\pdfoutput=1
\documentclass[11pt,twoside,a4paper,cmspaper,final,collab]{cms-tdr}
\def\svnVersion{00d9925-D}\def\svnDate{2021/05/03}\def\cmsCernNoTag{CERN-EP-2020-200}\def\cmsCernDate{\today}\def\cmsMessage{"Published in the European Physical Journal C as \href{http://dx.doi.org/10.1140/epjc/s10052-021-09014-x}{\doi{10.1140/epjc/s10052-021-09014-x}.}"}
\begin{document}\cmsNoteHeader{HIG-19-008}

\newcommand{\PHiggs}{\ensuremath{\PH}}
\newcommand{\Pgluon}{\ensuremath{\Pg}}
\newcommand{\Pquark}{\ensuremath{\PQq}}
\newcommand{\Pgamma}{\ensuremath{\PGg}}
\newcommand{\Pbottom}{\ensuremath{\PQb}}
\newcommand{\APbottom}{\ensuremath{\PAQb}}
\newcommand{\Ptop}{\ensuremath{\PQt}}
\newcommand{\APtop}{\ensuremath{\PAQt}}
\newcommand{\Plepton}{\ensuremath{\ell}}
\newcommand{\Pnu}{\ensuremath{\PGn}}
\newcommand{\APnu}{\ensuremath{\PAGn}}
\newcommand{\Pnut}{\ensuremath{\Pnu_{\Pgt}}}
\renewcommand{\Pgg}{\ensuremath{\PGg}\xspace}
\newcommand{\Pggx}{\ensuremath{\PGg^{*}}\xspace}
\newcommand{\PZggx}{\ensuremath{\PZ/\PGg^{*}}\xspace}
\newcommand{\Ppizero}{\ensuremath{\Pgpz}}
\newcommand{\tH}{\ensuremath{\Ptop\PHiggs}}
\newcommand{\ttH}{\ensuremath{\Ptop\APtop\PHiggs}}
\newcommand{\ggH}{\ensuremath{\Pgluon\Pgluon\PHiggs}}
\newcommand{\qqH}{\ensuremath{\Pquark\Pquark\PHiggs}}
\newcommand{\WH}{\ensuremath{\PW\PHiggs}}
\newcommand{\ZH}{\ensuremath{\PZ\PHiggs}}
\newcommand{\VH}{\ensuremath{\PV\PHiggs}}
\newcommand{\ttWH}{\ensuremath{\Ptop\APtop\PW\PHiggs}}
\newcommand{\ttZH}{\ensuremath{\Ptop\APtop\PZ\PHiggs}}
\newcommand{\ttVH}{\ensuremath{\Ptop\APtop\PV\PHiggs}}
\newcommand{\ttW}{\ensuremath{\Ptop\APtop\PW}}
\newcommand{\tW}{\ensuremath{\Ptop\PW}}
\newcommand{\ttWW}{\ensuremath{\Ptop\APtop\PW\PW}}
\newcommand{\ttWs}{\ensuremath{\Ptop\APtop\PW(\PW)}}
\newcommand{\ttZ}{\ensuremath{\Ptop\APtop\PZ}}
\newcommand{\tZ}{\ensuremath{\Ptop\PZ}}
\renewcommand{\ttbar}{\ensuremath{\Ptop\APtop}}
\newcommand{\mtop}{\ensuremath{m_{\Ptop}}}
\newcommand{\yt}{\ensuremath{y_{\Ptop}}}
\newcommand{\gW}{\ensuremath{g_{\PW}}}
\newcommand{\kappat}{\ensuremath{\kappa_{\Ptop}}}
\newcommand{\kappaV}{\ensuremath{\kappa_{\PV}}}
\renewcommand{\ss}{\ensuremath{\mkern 1mu\mathrm{SS}}}
\newcommand{\os}{\ensuremath{\mkern 1mu\mathrm{OS}}}
\newcommand{\fourFS}{\ensuremath{4\mathrm{\,FS}}}
\newcommand{\fiveFS}{\ensuremath{5\mathrm{\,FS}}}
\newcommand{\tHq}{\ensuremath{\tH\Pquark}}
\newcommand{\tHW}{\ensuremath{\tH\PW}}
\newcommand{\Zee}{\ensuremath{\PZggx \to \Pe\Pe}}
\newcommand{\Zmm}{\ensuremath{\PZggx \to \Pgm\Pgm}}
\newcommand{\Ztt}{\ensuremath{\PZggx \to \Pgt\Pgt}}
\newcommand{\vecMET}{\ptvecmiss}
\newcommand{\metLD}{\ensuremath{L_\mathrm{D}\xspace}}
\newcommand{\fix}{\ensuremath{\text{fix}}}
\newcommand{\mtfix}{\ensuremath{\mT^{\fix}}}
\newcommand{\pass}{\ensuremath{\text{pass}}}
\newcommand{\fail}{\ensuremath{\text{fail}}}
\newcommand{\fb}{\ensuremath{\unit{fb}}}
\renewcommand{\r}{\ensuremath{\mu}}
\newcommand{\rhat}{\ensuremath{\hat{\mu}}}
\newcommand{\vecr}{\ensuremath{\boldsymbol{\mu}}}
\newcommand{\vectheta}{\ensuremath{\boldsymbol{\theta}}}
\newcommand{\statsystlumi}{\ensuremath{\,\text{(stat+syst)}}}
\newcommand{\njets}{\ensuremath{N_{\mathrm{j}}}\xspace}
\newcommand{\zeroLeptonTwoTau}{\ensuremath{0\Plepton + 2\tauh}}
\newcommand{\oneLeptonOneTau}{\ensuremath{1\Plepton + 1\tauh}}
\newcommand{\oneLeptonTwoTau}{\ensuremath{1\Plepton + 2\tauh}}
\newcommand{\twoLeptonssZeroTau}{\ensuremath{2\Plepton\ss + 0\tauh}}
\newcommand{\twoLeptonssOneTau}{\ensuremath{2\Plepton\ss + 1\tauh}}
\newcommand{\twoLeptonosOneTau}{\ensuremath{2\Plepton\os + 1\tauh}}
\newcommand{\twoLeptonTwoTau}{\ensuremath{2\Plepton + 2\tauh}}
\newcommand{\threeLeptonZeroTau}{\ensuremath{3\Plepton + 0\tauh}}
\newcommand{\threeLeptonOneTau}{\ensuremath{3\Plepton + 1\tauh}}
\newcommand{\fourLeptonZeroTau}{\ensuremath{4\Plepton + 0\tauh}}

\newlength\cmsTabSkip\setlength{\cmsTabSkip}{1ex}
\providecommand{\cmsTable}[1]{\resizebox{\textwidth}{!}{#1}}

\providecommand{\NA}{\ensuremath{\text{---}}}
\providecommand{\CL}{CL\xspace}

\newcolumntype{C}[1]{>{\centering\let\newline\\\arraybackslash\hspace{0pt}}m{#1}}
\newcolumntype{L}[1]{>{\raggedright\let\newline\\\arraybackslash\hspace{0pt}}m{#1}}
\newcolumntype{R}[1]{>{\raggedleft\let\newline\\\arraybackslash\hspace{0pt}}m{#1}}

\setlength{\rotFPtop}{0pt plus 1fil}

\providecommand{\cmsLeft}{left\xspace}
\providecommand{\cmsRight}{right\xspace}
\providecommand{\cmsMid}{middle\xspace}
\providecommand{\cmsBottom}{lower\xspace}
\providecommand{\cmsTop}{upper\xspace}
\providecommand{\cmsCenter}{center\xspace}       

\cmsNoteHeader{HIG-19-008}
\title{
  Measurement of the Higgs boson production rate in association with top quarks in final states with electrons, muons, and hadronically decaying tau leptons at \texorpdfstring{$\sqrt{s} = 13\TeV$}{sqrt(s) = 13 TeV}
}
\titlerunning{Higgs boson production with top quarks}

\date{\today}

\abstract{
  The rate for Higgs ($\PHiggs$) bosons production in association with either one ($\Ptop\PHiggs$) or two ($\Ptop\APtop\PHiggs$) top quarks 
  is measured in final states containing multiple electrons, muons, or tau leptons decaying to hadrons and a neutrino,
  using proton-proton collisions recorded at a center-of-mass energy of $13\TeV$ by the CMS experiment.
  The analyzed data correspond to an integrated luminosity of 137\fbinv. 
  The analysis is aimed at events that contain $\PHiggs \to \PW\PW$, $\PHiggs \to \Pgt\Pgt$, or $\PHiggs \to \PZ\PZ$ decays
  and each of the top quark(s) decays either to lepton+jets or all-jet channels. 
  Sensitivity to signal is maximized by including ten signatures in the analysis, depending on the lepton multiplicity.
  The separation among $\Ptop\PHiggs$, $\Ptop\APtop\PHiggs$, and the backgrounds is enhanced through machine-learning techniques and matrix-element methods.
  The measured production rates for the $\Ptop\APtop\PHiggs$ and $\Ptop\PHiggs$ signals
  correspond to $0.92 \pm 0.19\stat \allowbreak^{+0.17}_{-0.13}\syst$ and $5.7 \pm 2.7\stat \pm 3.0\syst$ of their respective standard model (SM) expectations.
  The corresponding observed (expected) significance  amounts to $4.7$ ($5.2$) standard deviations for $\PQt\PAQt\PH$, and to $1.4$ ($0.3$) for $\PQt\PH$ production.
  Assuming that the Higgs boson coupling to the tau lepton is equal in strength to its expectation in the SM,
  the coupling $y_{\Ptop}$ of the Higgs boson to the top quark divided by its SM expectation, $\kappat=y_{\Ptop}/y_{\Ptop}^{\mathrm{SM}}$,  is constrained to be within $-0.9 < \kappat < -0.7$ or $0.7 < \kappat < 1.1$, at 95\% confidence level.
  This result is the most sensitive measurement of the $\Ptop\APtop\PHiggs$ production rate to date.
}

\hypersetup{
pdfauthor={CMS Collaboration},
pdftitle={Measurement of the Higgs boson production rate  in association with top quarks in final states containing electrons, muons, or hadronically decaying tau leptons at sqrt(s)=13 TeV},
pdfsubject={CMS},
pdfkeywords={CMS, Higgs, top, multilepton}}

\maketitle

\section{Introduction}
\label{sec:introduction}

The discovery of a Higgs ($\PHiggs$) boson by the ATLAS and CMS experiments at the CERN LHC~\cite{Aad:2012tfa,Chatrchyan:2012xdj,Chatrchyan:2013lba}
opened a new field for exploration in the realm of particle physics.
Detailed measurements of the properties of this new particle are important
to ascertain if the discovered resonance is indeed the Higgs boson predicted by the standard model (SM)~\cite{PhysRevLett.13.321,Higgs:1964ia,PhysRevLett.13.508,PhysRevLett.13.585}.
In the SM, the Yukawa coupling $y_{\mathrm{f}}$ of the Higgs boson to fermions is proportional to the mass $m_{\mathrm{f}}$ of the fermion,
namely $y_{\mathrm{f}} = m_{\mathrm{f}}/v$, where $v = 246\GeV$ denotes the vacuum expectation value of the Higgs field.
With a mass of $\mtop = 172.76 \pm 0.30\GeV$~\cite{Tanabashi:2018oca}, the top quark is by far the heaviest fermion known to date, and its Yukawa coupling is of order unity.
The large mass of the top quark may indicate that it plays a special role in the mechanism of electroweak symmetry breaking~\cite{Dobrescu:1997nm,Chivukula:1998wd,Delepine:1995qs}.
Deviations of $\yt$ from the SM prediction of $m_{\Ptop}/v$ would indicate the presence of physics beyond the SM.

\begin{figure}[h!]
  \centering\includegraphics[width=0.40\textwidth]{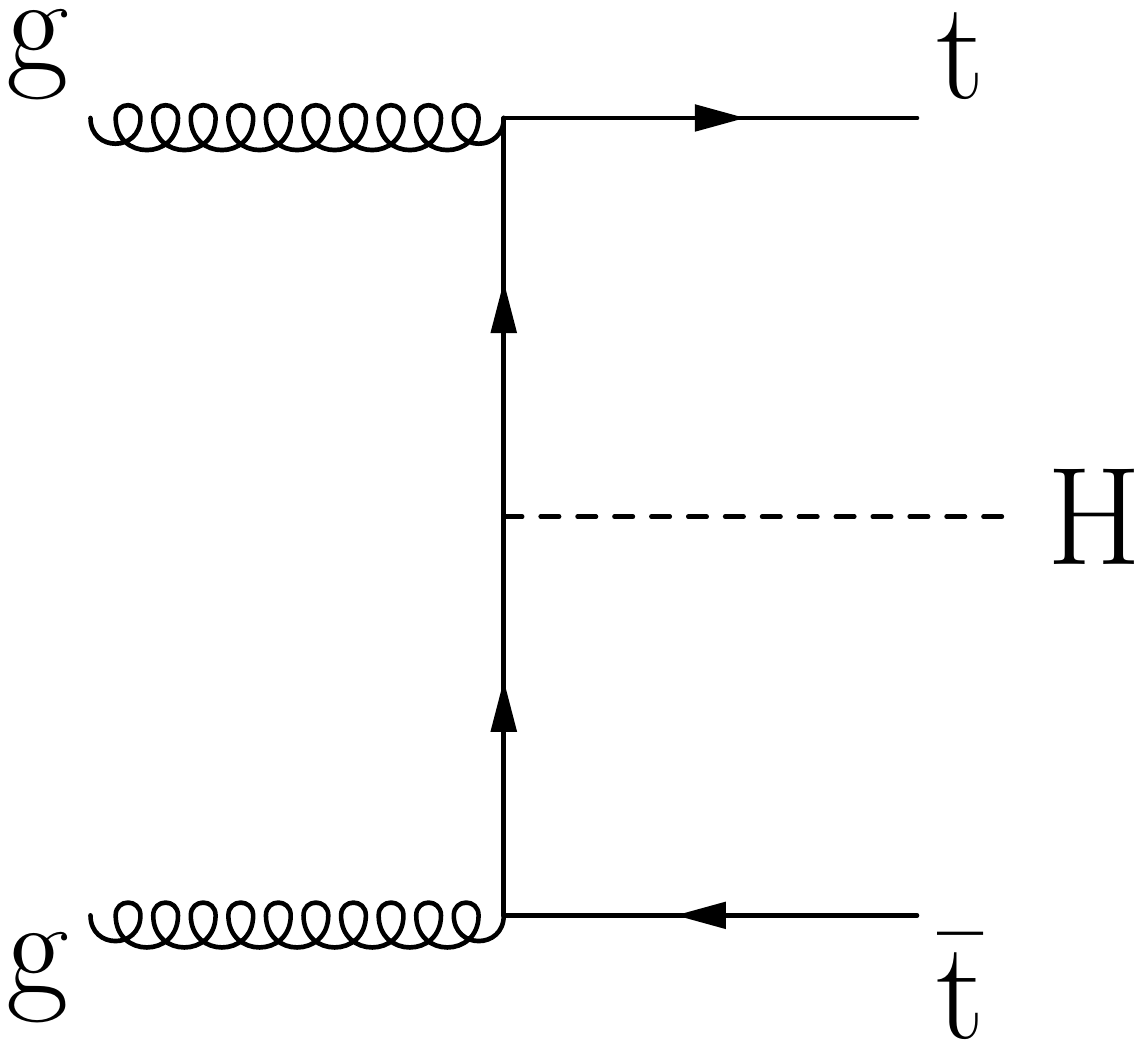}\hfill
  \centering\includegraphics[width=0.40\textwidth]{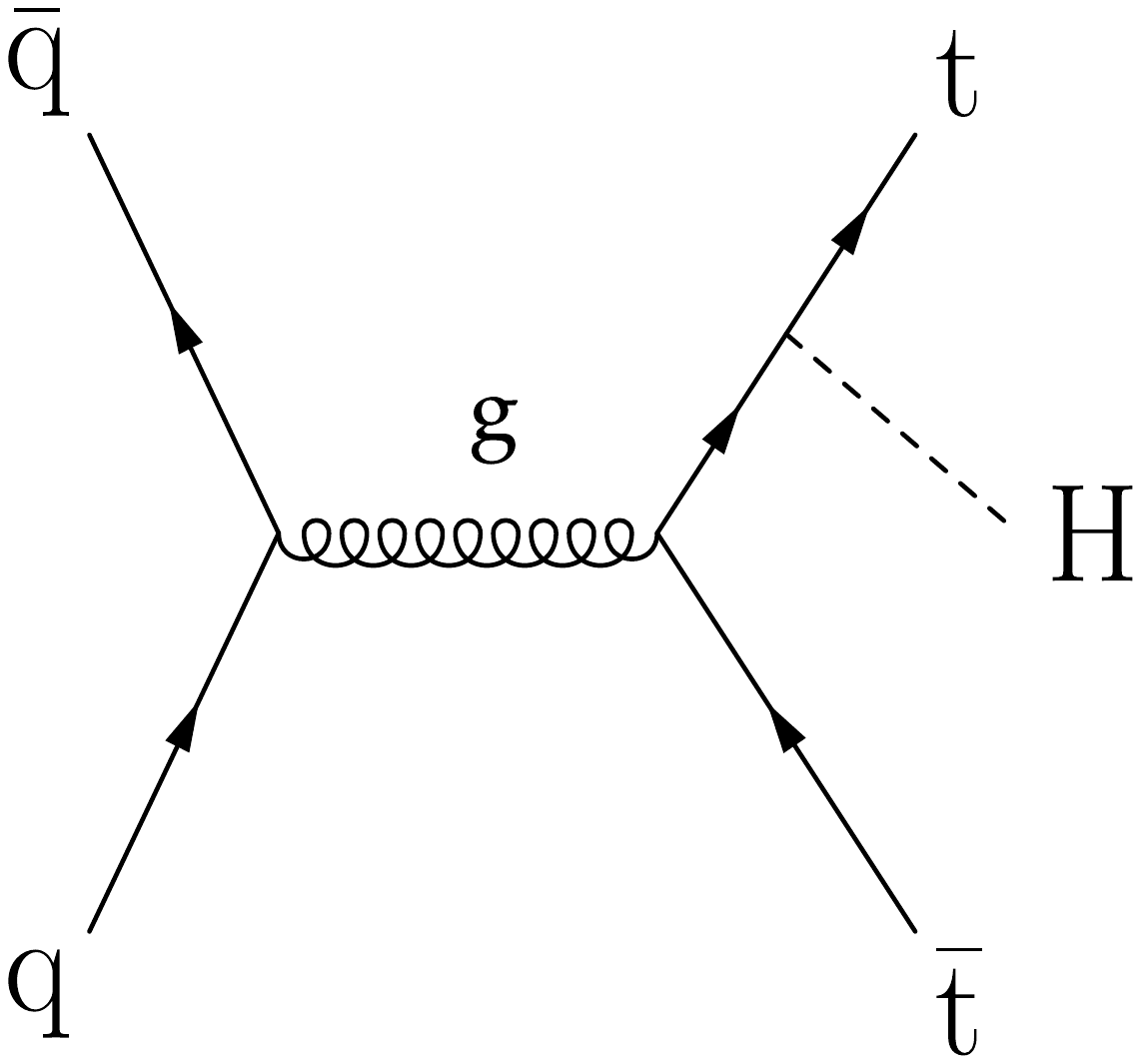}
  \caption{
    Feynman diagrams at LO for $\ttH$ production.
  }
  \label{fig:FeynmanDiagrams_ttH}
\end{figure}

\begin{figure*}[h!]
  \centering\includegraphics[width=0.35\textwidth]{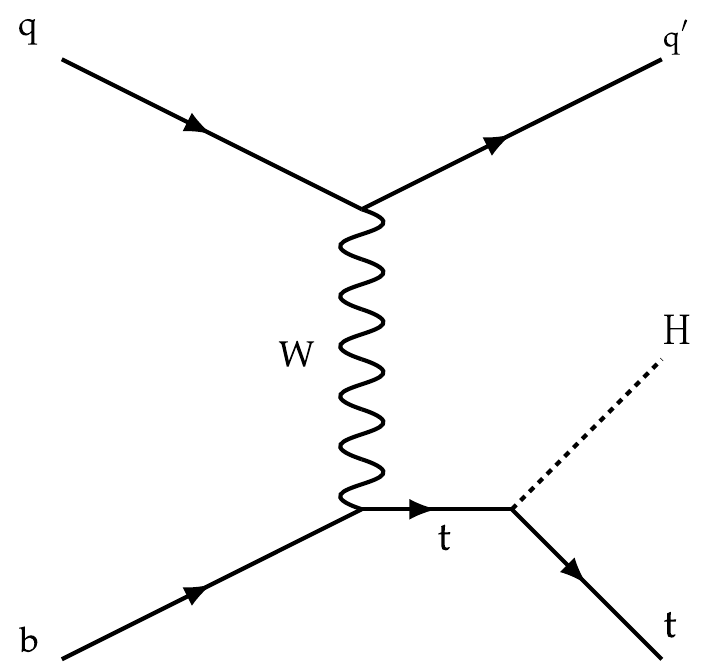}\hfill
  \centering\includegraphics[width=0.35\textwidth]{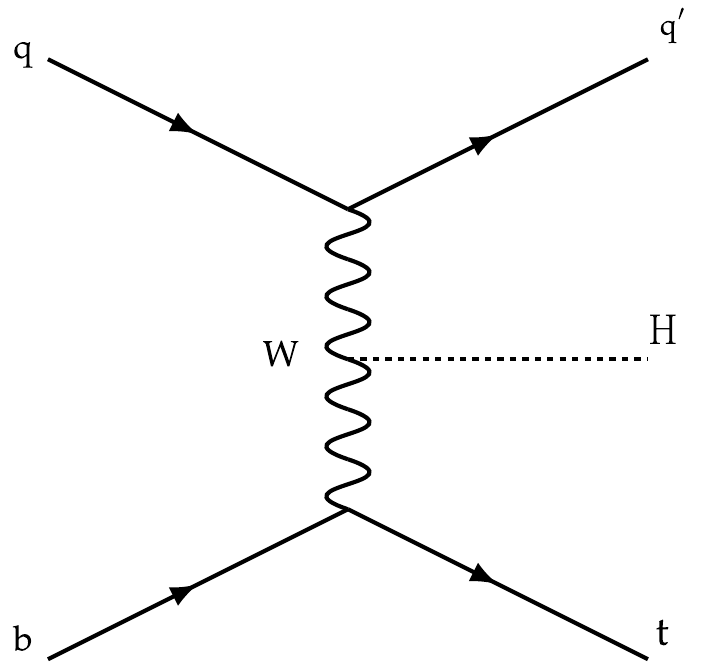}\\

  \vspace{1ex}

  \centering\includegraphics[width=0.35\textwidth]{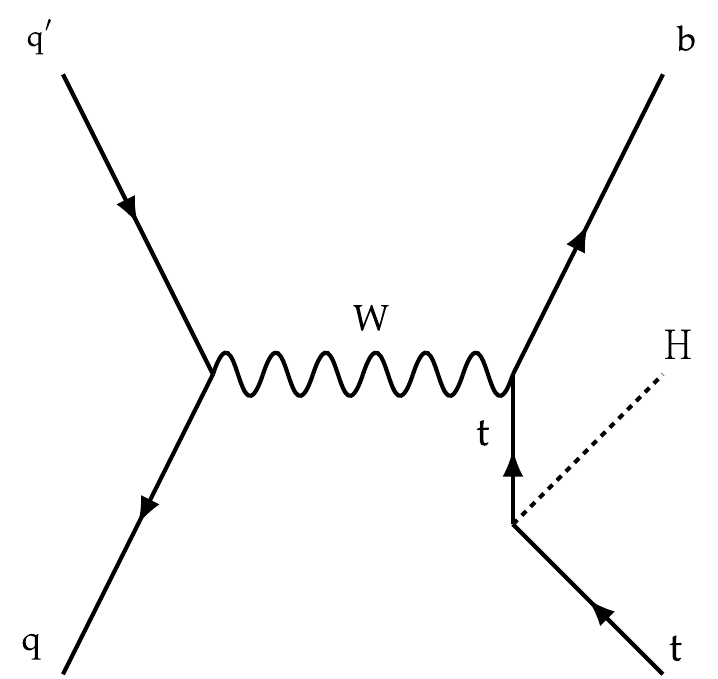}\\

  \vspace{1ex}

  \centering\includegraphics[width=0.35\textwidth]{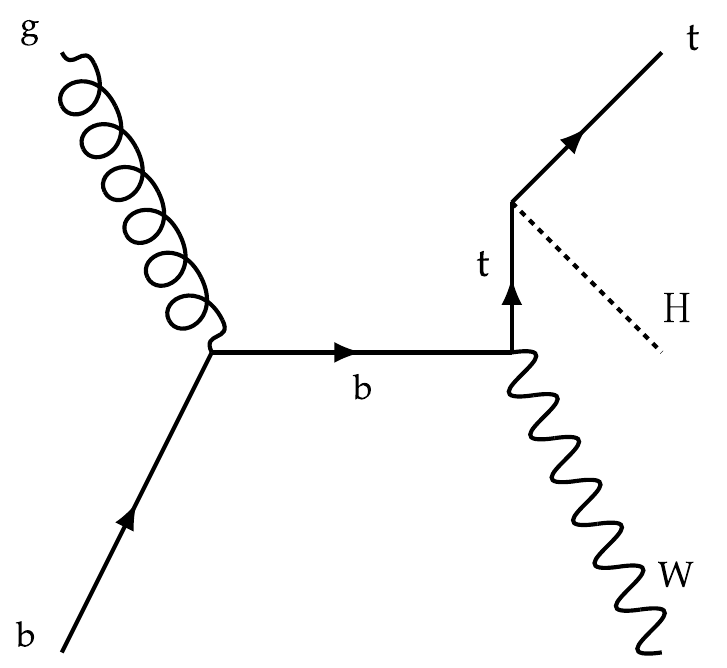}\hfill
  \centering\includegraphics[width=0.35\textwidth]{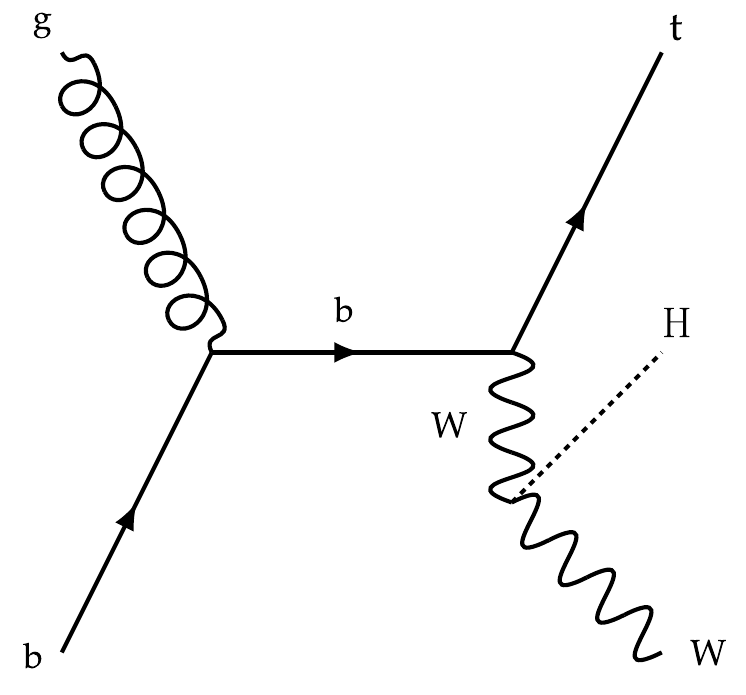}
  \caption{
    Feynman diagrams at LO for $\tH$ production via the $t$-channel ($\tHq$ in \cmsTop \cmsLeft and \cmsTop \cmsRight) and $s$-channel (\cmsMid) processes,
    and for associated production of a Higgs boson with a single top quark and a $\PW$ boson ($\tHW$ in \cmsBottom \cmsLeft and \cmsBottom \cmsRight).
    The $\tHq$ and $\tHW$ production processes are shown for the five-flavor scheme.
  }
  \label{fig:FeynmanDiagrams_tH}
\end{figure*}

The measurement of the Higgs boson production rate in association with a top quark pair ($\ttH$)
provides a model-independent determination of the magnitude of $\yt$, but not of its sign.
The sign of $\yt$ is determined from the associated production of a Higgs boson with a single top quark ($\tH$).
Leading-order (LO) Feynman diagrams for $\ttH$ and $\tH$ production are shown in Figs.~\ref{fig:FeynmanDiagrams_ttH} and~\ref{fig:FeynmanDiagrams_tH}, respectively.
The diagrams for $\tH$ production are separated into three contributions:
the $t$-channel ($\tHq$) and the $s$-channel, that proceed via the exchange of a virtual $\PW$ boson,
and the associated production of a Higgs boson with a single top quark and a $\PW$ boson ($\tHW$).
The interference between the diagrams where the Higgs boson couples to the top quark (Fig.~\ref{fig:FeynmanDiagrams_tH} \cmsTop and \cmsBottom \cmsLeft),
and those where the Higgs boson couples to the $\PW$ boson (Fig.~\ref{fig:FeynmanDiagrams_tH} \cmsTop and \cmsBottom \cmsRight)
is destructive when $\yt$ and $\gW$ have the same sign, 
where the latter denotes the coupling of the Higgs boson to the $\PW$ boson. This reduces the $\tH$  cross section and influences the kinematical properties of the event as a function of $\yt$ and $\gW$.
The interference becomes constructive when the coupling of the  $\gW$ and $\yt$ have opposite signs,
causing an increase in the cross section of up to one order of magnitude.
This is referred to as inverted top quark coupling.

Indirect constraints on the magnitude of $\yt$ are obtained from the rate of Higgs boson production via gluon fusion
and from the decay rate of Higgs bosons to photon pairs~\cite{Khachatryan:2016vau}, where in both cases, $\yt$ enters through top quark loops.
The $\PHiggs \to \Pgamma\Pgamma$ decay rate also provides sensitivity to the sign of $\yt$~\cite{Biswas:2013xva},
as does the rate for associated production of a Higgs boson with a $\PZ$ boson~\cite{Hespel:2015zea}.
The measured rates of these processes suggest that the Higgs boson coupling to top quarks is SM-like.
However, contributions from non-SM particles to these loops can compensate, and therefore mask, deviations of $\yt$ from its SM value.
A model-independent direct measurement of the top quark Yukawa coupling in $\ttH$ and $\tH$ production is therefore very important.
The comparison of the magnitude and sign of $\yt$ obtained from the measurement of the $\ttH$ and $\tH$ production rates,
where $\yt$ enters at lowest ``tree'' level,
with the value of $\yt$ obtained from processes where $\yt$ enters via loop contributions
can provide evidence about such contributions.

This manuscript presents the measurement of the $\ttH$ and $\tH$ production rates in final states containing multiple electrons, muons, or $\Pgt$ leptons that decay to hadrons and a neutrino (\tauh). In the following, we refer to \tauh as ``hadronically decaying \Pgt''. We also refer to electrons and muons collectively as ``leptons'' ($\Plepton$).
The measurement is based on data recorded by the CMS experiment in $\Pp\Pp$ collisions at $\sqrt{s} = 13\TeV$ during Run 2 of the LHC,
that corresponds to an integrated luminosity of 137\fbinv.

The associated production of Higgs bosons with top quark pairs was previously studied by the ATLAS and CMS experiments,
with up to 24.8\fbinv of data recorded at $\sqrt{s} = 7$ and $8\TeV$
during LHC Run 1~\cite{Khachatryan:2014qaa,Aad:2014lma,Aad:2015gra,Aad:2015iha,Aad:2016zqi},
and up to 79.8\fbinv of data recorded at $\sqrt{s} = 13\TeV$
during LHC Run 2~\cite{Sirunyan:2017exp,Aaboud:2017jvq,Aaboud:2017rss,Sirunyan:2018shy,Sirunyan:2018ygk,Sirunyan:2018ouh,Sirunyan:2018mvw}.
The combined analysis of data recorded at $\sqrt{s} = 7$, $8$, and $13\TeV$
resulted in the observation of $\ttH$ production by CMS and ATLAS~\cite{Sirunyan:2018hoz,Aaboud:2018urx}.
The production of Higgs bosons in association with a single top quark was also studied using the data recorded during LHC Run~1~\cite{Khachatryan:2015ota}
and Run 2~\cite{Sirunyan:2018lzm,Aad:2020ivc}.
These analyses covered Higgs boson decays to $\Pbottom\APbottom$, $\Pgamma\Pgamma$, $\PW\PW$, $\PZ\PZ$, and $\PGt\PGt$.

The measurement of the $\ttH$ and $\tH$ production rates presented in this manuscript constitutes their first simultaneous analysis in this channel. This approach is motivated by the high degree of overlap between the experimental signatures of both production processes
and takes into account the dependence of the $\ttH$ and $\tH$ production rates  as a function of $\yt$.
Compared to previous work~\cite{Sirunyan:2018shy}, the sensitivity of the present analysis is enhanced
by improvements in the identification of \tauh decays and of jets originating from the hadronization of bottom quarks,
as well as by performing the analysis in four additional experimental signatures, also referred to as analysis channels, that add up to a total of ten.
The signatures involve Higgs boson decays to $\PW\PW$, $\Pgt\Pgt$, and $\PZ\PZ$, and are defined according to the lepton and \tauh multiplicities in the events. Some of them require
leptons to have the same (opposite) sign of electrical charge and are therefore referred to as $\ss$ ($\os$).
The signatures 
$\twoLeptonssZeroTau$,
$\threeLeptonZeroTau$, $\twoLeptonssOneTau$, $\twoLeptonosOneTau$, $\oneLeptonTwoTau$,
$\fourLeptonZeroTau$, $\threeLeptonOneTau$, and $\twoLeptonTwoTau$
target events where at least one top quark decays via $\Ptop \to \Pbottom\PW^{+} \to \Pbottom\Plepton^{+}\Pnu_{\Plepton}$,
whereas the signatures
$\oneLeptonOneTau$ and $\zeroLeptonTwoTau$
target events where all top quarks decay via $\Ptop \to \Pbottom\PW^{+} \to \Pbottom\Pquark\PAQq'$. We refer to the first and latter top quark decay signatures as semi-leptonically and hadronically decaying top quarks, respectively.
Here and in the following, the term top quark includes the corresponding charge-conjugate decays of top antiquarks.
As in previous analyses, the separation of the $\ttH$ and $\tH$ signals from backgrounds is improved
through machine-learning techniques, specifically boosted decision trees (BDTs) and artificial neural networks (ANNs)~\cite{TMVA,2016arXiv160302754C,scikit-learn},
and through the matrix-element method~\cite{Kondo:1988yd,Kondo:1991dw}.
Machine-learning techniques are also employed to improve the separation between the $\ttH$ and $\tH$ signals.
We use the measured $\ttH$ and $\tH$ production rates to set limits on the magnitude and sign of $\yt$.

This paper is organized as follows. After briefly describing the CMS detector in Section~\ref{sec:detector}, we proceed to discuss the data and simulated events used in the measurement in Section~\ref{sec:datasamples_and_MonteCarloSimulation}. Section~\ref{sec:eventReconstruction} covers the object reconstruction and selection from signals recorded in the detector, while Section~\ref{sec:eventSelection} describes the selection criteria applied to events in the analysis. These events are grouped in categories, defined in Section~\ref{sec:eventClassification_and_signalExtraction}, while the estimation of background contributions in these categories is described  in Section~\ref{sec:backgroundEstimation}. The systematic uncertainties affecting the measurements are given in Section~\ref{sec:systematicUncertainties}, and the statistical analysis and the results of the measurements in Section~\ref{sec:results}. We end the paper with a brief summary in Section~\ref{sec:summary}.

\section{The CMS detector}
\label{sec:detector}

{\tolerance=800
The central feature of the CMS apparatus is a superconducting solenoid of 6\unit{m} internal diameter, providing a magnetic field of 3.8\unit{T}.
A silicon pixel and strip tracker, 
a lead tungstate crystal electromagnetic calorimeter (ECAL), and a brass and scintillator hadron calorimeter (HCAL), 
each composed of a barrel and two endcap sections,
are positioned within the solenoid volume.
The silicon tracker measures charged particles within the pseudorapidity range $\abs{\eta} < 2.5$. 
The ECAL is a fine-grained hermetic calorimeter with quasi-projective geometry,
and is segmented into the barrel region of $\abs{\eta} < 1.48$ and in two endcaps that extend up to $\abs{\eta} < 3.0$.
The HCAL barrel and endcaps similarly cover the region $\abs{\eta} < 3.0$.
Forward calorimeters extend the coverage up to $\abs{\eta} < 5.0$.
Muons are measured and identified in the range $\abs{\eta} < 2.4$ 
by gas-ionization detectors embedded in the steel flux-return yoke outside the solenoid. 
A two-level trigger system~\cite{Khachatryan:2016bia} is used to reduce the rate of recorded
events to a level suitable for data acquisition and storage.
The first level of the CMS trigger system, composed of custom hardware processors, 
uses information from the calorimeters and muon detectors to select the most interesting events with a latency of $4$\mus. 
The high-level trigger processor farm further decreases the event rate from around 100\unit{kHz} to about 1\unit{kHz}.
Details of the CMS detector and its performance, together with a definition of the coordinate system and the kinematic variables used in the analysis, are reported in Ref.~\cite{Chatrchyan:2008zzk}.
\par}

\section{Data samples and Monte Carlo simulation}
\label{sec:datasamples_and_MonteCarloSimulation}

The analysis uses $\Pp\Pp$ collision data recorded at $\sqrt{s} = 13\TeV$ at the LHC during 2016-2018.
Only the data-taking periods during which the CMS detector was fully operational are included in the analysis.
The total integrated luminosity of the analyzed data set amounts to 137\fbinv,
of which 35.9~\cite{LUM-17-001}, 41.5~\cite{LUM-17-004}, and 59.7~\cite{LUM-18-002}\fbinv have been recorded in 2016, 2017, and 2018, respectively.

The event samples produced via Monte Carlo (MC) simulation
are used for the purpose of calculating selection efficiencies for the $\ttH$ and $\tH$ signals,
estimating background contributions, and training machine-learning algorithms.
The contribution from $\ttH$ signal and the backgrounds arising from 
$\ttbar$ production in association with $\PW$ and $\PZ$ bosons ($\ttW$, $\ttZ$), 
from triboson ($\PW\PW\PW$, $\PW\PW\PZ$, $\PW\PZ\PZ$, $\PZ\PZ\PZ$, $\PW\PZ\Pgg$) production,
as well as from the production of four top quarks ($\Ptop\APtop\Ptop\APtop$)
are generated at next-to-LO (NLO) accuracy in perturbative quantum chromodynamics (pQCD) making use of the program \MGvATNLO $2.2.2$ or $2.3.3$~\cite{Alwall:2007fs,Artoisenet:2012st,Frederix:2012ps,Alwall:2014hca},
whereas the $\tH$ signal and the 
$\ttbar\Pgg$, $\ttbar\Pggx$, $\tZ$, $\ttWW$, $\PW$+jets, Drell--Yan (DY), $\PW\Pgg$, and $\PZ\Pgg$ backgrounds
are generated at LO accuracy using the same program.
The symbols $\Pggx$ and $\Pgg$ are employed to distinguish virtual photons from the real ones.
The event samples with virtual photons also include contributions from virtual $\PZ$ bosons.
The DY production of electron, muon, and $\Pgt$ lepton pairs are referred to as $\Zee$, $\Zmm$, and $\Ztt$, respectively.
The modeling of the $\ttW$ background includes additional $\alpS \alpha^3$ electroweak corrections~\cite{Frederix:2020jzp,Dror:2015nkp}, simulated using \MGvATNLO.
The NLO program \POWHEG v2.0~\cite{Nason:2004rx,Frixione:2007vw,Alioli:2010xd} is used to simulate 
the backgrounds arising from $\ttbar$+jets, $\Ptop\PW$, and diboson ($\PW^{\pm}\PW^{\mp}$, $\PW\PZ$, $\PZ\PZ$) production,
and from the production of single top quarks,
and from SM Higgs boson production via gluon fusion ($\ggH$) and vector boson fusion ($\qqH$) processes, 
and from the production of SM Higgs bosons in association with $\PW$ and $\PZ$ bosons ($\WH$, $\ZH$)
and with $\PW$ and $\PZ$ bosons along with a pair of top quarks ($\ttWH$, $\ttZH$).
The modeling of the top quark transverse momentum (\pt) distribution of $\ttbar$+jets events simulated with the program \POWHEG
is improved by reweighting the events to the differential cross section 
computed at next-to-NLO (NNLO) accuracy in pQCD, including electroweak corrections computed at NLO accuracy~\cite{Czakon:2017wor}.
We refer to the sum of $\WH$ plus $\ZH$ contributions by using the symbol $\VH$ 
and to the sum of $\ttWH$ plus $\ttZH$ contributions by using the symbol $\ttVH$.
The SM production of Higgs boson pairs or a Higgs boson in association with a pair of \PQb quarks is not considered as a background to this analysis,
because its impact on the event yields in all categories is found to be negligible.
The production of same-sign \PW pairs (\ss \PW) is simulated using the program \MGvATNLO in LO accuracy,
except for the contribution from double-parton interactions, which is simulated with \PYTHIA v$8.2$~\cite{Sjostrand:2014zea} (referred to as \PYTHIA hereafter).
The NNPDF3.0LO (NNPDF3.0NLO)~\cite{Ball:2011uy,Ball:2013hta,Ball:2014uwa} set of parton distribution functions (PDF) is used for the simulation of LO (NLO) 2016 samples, while \textsc{NNPDF3.1} NNLO~\cite{Ball:2017nwa} is used for 2017 and 2018 LO and NLO samples.

Different flavor schemes are chosen to simulate the $\tHq$ and $\tHW$ processes. In the five-flavor scheme ($\fiveFS$), bottom quarks are considered as sea quarks of the proton and may appear in the initial state of proton-proton ($\Pp\Pp$) scattering processes,
as opposed to the four-flavor scheme ($\fourFS$), where only up, down, strange, and charm quarks are considered as valence or sea quarks of the proton,
whereas bottom quarks are produced by gluon splitting at the matrix-element level, and therefore appear only in the final state~\cite{Maltoni:2012pa}.  In the $\fiveFS$ the distinction of $\tHq$, $s$-channel, and $\tHW$ contributions to $\tH$ production is well-defined up to NLO,
whereas at higher orders in perturbation theory the $\tHq$ and $s$-channel production processes start to interfere and can no longer be uniquely separated~\cite{Demartin:2015uha}.
Similarly, in the same regime the $\tHW$ process starts to interfere with $\ttH$ production at NLO.
In the $\fourFS$, the separation among the $\tHq$, $s$-channel, and $\tHW$ (if the $\PW$ boson decays hadronically) processes holds only up to LO,
and the $\tHW$ process starts to interfere with $\ttH$ production already at tree level~\cite{Demartin:2015uha}.

The $\tHq$ process is simulated at LO in the $\fourFS$ and the $\tHW$ process in the $\fiveFS$,  so that interference contributions of latter with $\ttH$ production are not present in the simulation. The contribution from $s$-channel $\tH$ production is negligible and is not considered in this analysis.

Parton showering, hadronization, and the underlying event are modeled using \PYTHIA with the tune CP5,
CUETP8M1, CUETP8M2, or CUETP8M2T4~\cite{Sirunyan:2019dfx,CMS-PAS-TOP-16-021,Khachatryan:2015pea}, depending on the dataset,
as are the decays of $\Pgt$ leptons, including polarization effects.
The matching of matrix elements to parton showers is done using the MLM scheme~\cite{Alwall:2007fs} for the LO samples
and the FxFx scheme~\cite{Frederix:2012ps} for the samples simulated at NLO accuracy.

The modeling of the $\ttH$ and $\tH$ signals, as well as of the backgrounds, is improved by normalizing the simulated event samples
to cross sections computed at higher order in pQCD.
The cross section for $\tH$ production is computed in the $\fiveFS$.
The SM cross section for $\tHq$ production has been computed at NLO accuracy in pQCD as $74.3\fb$~\cite{deFlorian:2016spz},
and the SM cross section for $\ttH$ production has been computed at NLO accuracy in pQCD as $506.5\fb$
with electroweak corrections calculated at the same order in perturbation theory~\cite{deFlorian:2016spz}.
Both cross sections are computed for $\Pp\Pp$ collisions at $\sqrt{s} = 13\TeV$.
The $\tHW$ cross section is computed to be $15.2\fb$ at NLO in the $\fiveFS$,
using the \textrm{DR2} scheme~\cite{Demartin:2016axk} to remove overlapping contributions between the $\tHW$ process and $\ttH$ production.
The cross sections for $\ttbar$+jets, $\PW$+jets, DY, and diboson production 
are computed at NNLO accuracy~\cite{Campbell:2011bn,Czakon:2011xx,Li:2012wna}.

Event samples containing Higgs bosons are normalized using the SM cross sections published in Ref.~\cite{deFlorian:2016spz}. Event samples of $\ttZ$ production are normalized to the cross
sections published in Ref.~\cite{deFlorian:2016spz}, while $\ttW$ simulated samples are normalized to the cross section published in the same reference increased by the contribution
from the $\alpS \alpha^3$ electroweak corrections~\cite{Frederix:2020jzp,Dror:2015nkp}. 
The SM cross sections for the $\ttH$ and $\tH$ signals
and for the most relevant background processes are given in Table~\ref{tab:backgroundXS}.

The $\ttH$ and $\tH$ samples are produced assuming all couplings of the Higgs boson have the values expected in the SM.
The variation in kinematical properties of $\tH$ signal events, which stem from the interference of the diagrams in Fig.~\ref{fig:FeynmanDiagrams_tH} described in Section~\ref{sec:introduction},
for values of $\yt$ and $\gW$ that differ from the SM expectation, is accounted for by applying weights calculated  for each $\tH$ signal event with \MGvATNLO, following the approach suggested in~\cite{Gainer:2014bta,Mattelaer:2016gcx}.
No such reweighting is necessary for the $\ttH$ signal,
because any variation of $\yt$ would only affect the inclusive cross section for $\ttH$ production, which increases proportional to $\yt^{2}$,
leaving the kinematical properties of $\ttH$ signal events unaltered.

The presence of simultaneous $\Pp\Pp$ collisions in the same or nearby bunch crossings, referred to as pileup (PU),
is modeled by superimposing inelastic $\Pp\Pp$ interactions, simulated using \PYTHIA, to all MC events.
Simulated events are weighed so the PU distribution of simulated samples matches the one observed in the data.

All MC events are passed through a detailed simulation of the CMS apparatus, based on \GEANTfour~\cite{Agostinelli:2002hh,Allison:2016lfl},
and are processed using the same version of the CMS event reconstruction software used for the data.

Simulated events are corrected by means of weights or by varying the relevant quantities to account for residual differences between data and simulation.
These differences arise in: trigger efficiencies; reconstruction and identification efficiencies for electrons, muons, and $\tauh$;
the energy scale of $\tauh$ and jets; the efficiency to identify jets originating from the hadronization of bottom quarks
and the corresponding misidentification rates for light-quark and gluon jets;
and the resolution in missing transverse momentum.
The corrections are typically at the level of a few percent~\cite{Khachatryan:2010xn,Khachatryan:2016kdb,Sirunyan:2017ezt,Sirunyan:2018pgf,Sirunyan:2019kia}.
They are measured using a variety of SM processes, 
such as $\Zee$, $\Zmm$, $\Ztt$, $\ttbar$+jets, and $\Pgamma$+jets production.

\begin{table*}[h!]
  \centering              
  \topcaption{
    Standard model cross sections for the $\ttH$ and $\tH$ signals as well as for the most relevant background processes.
    The cross sections are quoted for $\Pp\Pp$ collisions at $\sqrt{s} = 13\TeV$. The quoted value for DY production includes
    a generator-level requirement of $m_{\PZ/\PGg^*}>50\GeV$. 
  }
  \label{tab:backgroundXS}
  {
      \begin{tabular}{lc}
      \hline
      Process       & Cross section [$\text{fb}$] \\
      \hline
      $\ttH$        & $507$~\cite{deFlorian:2016spz} \\
      $\tHq$         & $74.3$~\cite{deFlorian:2016spz} \\
      $\tHW$         & $15.2$~\cite{Demartin:2016axk} \\
          [\cmsTabSkip]
          $\ggH$        &   $4.86 \times 10^{4}$~\cite{deFlorian:2016spz} \\
          $\qqH$        &   $3.78 \times 10^{3}$~\cite{deFlorian:2016spz} \\
          $\WH$         &   $1.37 \times 10^{3}$~\cite{deFlorian:2016spz} \\
          $\ZH$         & $884$~\cite{deFlorian:2016spz} \\
          \hline
    \end{tabular}
    \qquad
    \begin{tabular}{lc}
      \hline
      Process       & Cross section [$\text{fb}$] \\
      \hline
      $\ttZ$        & $839$~\cite{deFlorian:2016spz}     \\
      $\ttW$        & $650$~\cite{deFlorian:2016spz,Frederix:2020jzp,Dror:2015nkp}     \\
      $\ttWW$       &      $6.98$~\cite{Alwall:2014hca}\\
          $\ttbar$+jets & $8.33 \times 10^{5}$~\cite{Czakon:2011xx} \\
          DY            & $6.11 \times 10^{7}$~\cite{Campbell:2011bn} \\ 
          $\PW\PW$      & $1.19 \times 10^{5}$~\cite{Campbell:2011bn} \\
          $\PW\PZ$      & $4.50 \times 10^{4}$~\cite{Campbell:2011bn} \\
          $\PZ\PZ$      & $1.69 \times 10^{4}$~\cite{Campbell:2011bn} \\
          \hline
    \end{tabular}
  }
\end{table*}

\section{Event reconstruction}
\label{sec:eventReconstruction}

The CMS particle-flow (PF) algorithm~\cite{Sirunyan:2017ulk} provides a global event description
that optimally combines the information from all subdetectors, to reconstruct and identify all individual particles in the event.
The particles are subsequently classified into five mutually exclusive categories: 
electrons, muons, photons, and charged and neutral hadrons.

Electrons are reconstructed combining the information from tracker and ECAL~\cite{Khachatryan:2015hwa}
and are required to satisfy $\pt > 7\GeV$ and $\abs{\eta} < 2.5$.
Their identification is based on a multivariate (MVA) algorithm
that combines observables sensitive 
to: the matching of measurements of the electron energy and direction obtained from the tracker and the calorimeter;
the compactness of the electron cluster;
and the bremsstrahlung emitted along the electron trajectory.
Electron candidates resulting from photon conversions are removed by requiring 
that the track has no missing hits in the innermost layers of the silicon tracker and by vetoing candidates that are matched to a reconstructed conversion vertex.
In the $\twoLeptonssZeroTau$ and $\twoLeptonssOneTau$ channels (see Section~\ref{sec:eventSelection} for channel definitions),
we apply further electron selection criteria that demand the consistency among three independent measurements of the electron charge,
described as ``selective algorithm'' in Ref.~\cite{Khachatryan:2015hwa}.

The reconstruction of muons is based on linking track segments reconstructed in the silicon tracker 
to hits in the muon detectors that are embedded in the steel flux-return yoke~\cite{ Sirunyan:2018fpa}.
The quality of the spatial matching between the individual measurements in the tracker and in the muon detectors
is used to discriminate genuine muons from hadrons punching through the calorimeters and from muons produced by in-flight decays of kaons and pions.
Muons selected in the analysis are required to have $\pt > 5\GeV$ and $\abs{\eta} < 2.4$.
For events selected in the $\twoLeptonssZeroTau$ and $\twoLeptonssOneTau$ channels,
the relative uncertainty in the curvature of the muon track is required to be less than 20\% to ensure a high-quality charge measurement.

The electrons and muons satisfying the aforementioned selection criteria are referred to as ``loose leptons'' in the following.
Additional selection criteria are applied to discriminate electrons and muons produced in decays of $\PW$ and $\PZ$ bosons and leptonic $\Pgt$ decays (``prompt'')
from electrons and muons produced in decays of $\Pbottom$ hadrons (``nonprompt'').
The removal of nonprompt leptons reduces, in particular, the background arising from $\ttbar$+jets production.
To maximally exploit the information available in each event, 
we use MVA discriminants that
take as input the charged and neutral particles reconstructed in a cone around the lepton direction
besides the observables related to the lepton itself.
The jet reconstruction and $\Pbottom$ tagging algorithms are applied, and the resulting reconstructed jets are used as additional inputs to the MVA.
In particular, the ratio of the lepton \pt to the reconstructed jet \pt
and the component of the lepton momentum in a direction perpendicular to the jet direction are found to enhance the separation of prompt leptons from leptons originating from $\Pbottom$ hadron decays,
complementing more conventional observables such as the relative isolation of the lepton, 
calculated in a variable cone size depending on the lepton \pt~\cite{Rehermann:2010vq,Khachatryan:2016kod}, 
and the longitudinal and transverse impact parameters of the lepton trajectory with respect to the primary $\Pp\Pp$ interaction vertex.
Electrons and muons passing a selection on the MVA discriminants are referred to as ``tight leptons''.

Because of the presence of PU, the primary $\Pp\Pp$ interaction vertex typically needs to be chosen among the several vertex candidates that are reconstructed in each $\Pp\Pp$ collision event.
The candidate vertex with the largest value of summed physics-object $\pt^2$ is taken to be the primary $\Pp\Pp$ interaction vertex. The physics objects are the jets, clustered using the jet finding algorithm~\cite{Cacciari:2008gp,Cacciari:2011ma} with the tracks assigned to candidate vertices as inputs, and the associated missing transverse momentum, taken as the negative vector sum of the \pt of those jets.

While leptonic decay products of $\Pgt$ leptons are selected by the algorithms described above,
hadronic decays are reconstructed and identified by the ``hadrons-plus-strips'' (HPS) algorithm~\cite{Sirunyan:2018pgf}.
The algorithm is based on reconstructing individual hadronic decay modes of the $\Pgt$ lepton:
$\Pgt^{-} \to \Ph^{-}\Pnut$, $\Pgt^{-} \to \Ph^{-}\Ppizero\Pnut$, $\Pgt^{-} \to \Ph^{-}\Ppizero\Ppizero\Pnut$, 
$\Pgt^{-} \to \Ph^{-}\Ph^{+}\Ph^{-}\Pnut$, $\Pgt^{-} \to \Ph^{-}\Ph^{+}\Ph^{-}\Ppizero\Pnut$,
and all the charge-conjugate decays,
where the symbols $\Ph^{-}$ and $\Ph^{+}$ denotes either a charged pion or a charged kaon.
The photons resulting from the decay of neutral pions that are produced in the $\Pgt$ decay
have a sizeable probability to convert into an electron-positron pair when traversing the silicon tracker.
The conversions cause a broadening of energy deposits in the ECAL,
since the electrons and positrons produced in these conversions are bent in opposite azimuthal directions by the magnetic field
and may also emit bremsstrahlung photons.
The HPS algorithm accounts for this broadening when it reconstructs the neutral pions, 
by means of clustering photons and electrons in rectangular strips that are narrow in $\eta$ but wide in $\phi$.
The subsequent identification of $\tauh$ candidates is performed by the ``DeepTau'' algorithm~\cite{CMS-DP-2019-033}.
The algorithm is based on a convolutional ANN~\cite{lecun1989}, 
using as input a set of $42$ high-level observables in combination with low-level information obtained from the silicon tracker, the electromagnetic and hadronic calorimeters, and the muon detectors.
The high-level observables comprise the $\pt$, $\eta$, $\phi$, and mass of the $\tauh$ candidate; the reconstructed $\tauh$ decay mode;
observables that quantify the isolation of the $\tauh$ with respect to charged and neutral particles;
as well as observables that provide sensitivity to the small distance that a $\Pgt$ lepton typically traverses between its production and decay.
The low-level information quantifies the particle activity within two $\eta \times \phi$ grids,
an ``inner'' grid of size $0.2 \times 0.2$, filled with cells of size $0.02 \times 0.02$,
and an ``outer'' grid of size $0.5 \times 0.5$ (partially overlapping with the inner grid) and cells of size $0.05 \times 0.05$.
Both grids are centered on the direction of the $\tauh$ candidate.
The $\tauh$ considered in the analysis are required to have $\pt > 20\GeV$ and $\abs{\eta} < 2.3$ 
and to pass a selection on the output of the convolutional ANN.
The selection differs by analysis channel, targeting different efficiency and purity levels.
We refer to these as the very loose, loose, medium, and tight $\tauh$ selections, depending on the requirement imposed on the ANN output.

Jets are reconstructed using the anti-\kt algorithm~\cite{Cacciari:2008gp,Cacciari:2011ma} with a distance parameter of $0.4$
and with the particles reconstructed by the PF algorithm as inputs.
Charged hadrons associated with PU vertices are excluded from the clustering.
The energy of the reconstructed jets is corrected for residual PU effects using the method described in Refs.~\cite{Cacciari:2008gn, Cacciari:2007fd}
and calibrated as a function of jet \pt and $\eta$~\cite{Khachatryan:2016kdb}.
The jets considered in the analysis are required to: satisfy $\pt > 25\GeV$ and $\abs{\eta} < 5.0$;
pass identification criteria that reject spurious jets arising from calorimeter noise~\cite{JME-16-003};
and not overlap with any identified electron, muon or hadronic $\PGt$ within $\Delta R  = \sqrt{\smash[b]{(\Delta\eta)^2+(\Delta\phi)^2}} < 0.4$.
We tighten the requirement on the transverse momentum to the condition $\pt > 60\GeV$
for jets reconstructed within the range $2.7 < \abs{\eta} < 3.0$,
to further reduce the effect of calorimeter noise, which is sizeable in this detector region.
Jets passing these selection criteria are then categorized into central and forward jets,
the former satisfying the condition $\abs{\eta} < 2.4$ and the latter $2.4 < \abs{\eta} < 5.0$.
The presence of a high-\pt forward jet in the event is a characteristic signature of $\tH$ production in the $t$-channel
and is used to separate the $\ttH$ from the $\tH$ process in the signal extraction stage of the analysis.

Jets reconstructed within the region $\abs{\eta} < 2.4$ and originating from the hadronization of bottom quarks
are denoted as $\Pbottom$ jets and identified by the \textsc{DeepJet} algorithm~\cite{CMS-DP-2017-013}.
The algorithm exploits observables related to the long lifetime of $\Pbottom$ hadrons 
as well as to the higher particle multiplicity and mass of $\Pbottom$ jets compared to light-quark and gluon jets.
The properties of charged and neutral particle constituents of the jet, as well as of secondary vertices reconstructed within the jet,
are used as inputs to a convolutional ANN.
Two different selections on the output of the algorithm are employed in the analysis,
corresponding to $\Pbottom$ jet selection efficiencies of 84 (``loose'') and 70\% (``tight'').
The respective mistag rates for light-quark and gluon jets (\PQc jet) are 11 and 1.1\% (50\% and 15\%).

The missing transverse momentum vector, denoted by the symbol $\vecMET$,
is computed as the negative of the vector \pt sum of all particles reconstructed by the PF algorithm.
The magnitude of this vector is denoted by the symbol $\ptmiss$.
The analysis employs a linear discriminant, denoted by the symbol $\metLD$,
to remove backgrounds in which the reconstructed $\ptmiss$ arises from resolution effects.
The discriminant also reduces PU effects and is defined by the relation $\metLD = 0.6 \ptmiss + 0.4 \mht$,
where the observable $\mht$ corresponds to the magnitude of the vector \pt sum of electrons, muons, $\tauh$, and jets~\cite{Sirunyan:2018shy}.
The discriminant is constructed to combine the higher resolution of $\ptmiss$ with the robustness to PU of $\mht$.

\section{Event selection}
\label{sec:eventSelection}

The analysis targets $\ttH$ and $\tH$ production in events where the Higgs boson decays via
$\PHiggs \to \PW\PW$, $\PHiggs \to \Pgt\Pgt$, or $\PHiggs \to \PZ\PZ$,
with subsequent decays
$\PW\PW \to \Plepton^{+}\Pnu_{\Plepton}\Pquark\Pquark'$ or $\Plepton^{+}\Pnu_{\Plepton}\Plepton^{-}\APnu_{\Plepton}$; 
$\Pgt\Pgt \to \Plepton^{+}\Pnu_{\Plepton}\APnu_{\Pgt}\Plepton^{-}\APnu_{\Plepton}\Pnu_{\Pgt}$, $\Plepton^{+}\Pnu_{\Plepton}\APnu_{\Pgt}\tauh\Pnu_{\Pgt}$, or $\tauh\APnu_{\Pgt}\tauh\Pnu_{\Pgt}$;
$\PZ\PZ \to \Plepton^{+}\Plepton^{-}\Pquark\Pquark'$ or $\Plepton^{+}\Plepton^{-}\Pnu\APnu$;
and the corresponding charge-conjugate decays.
The decays $\PHiggs \to \PZ\PZ \to \Plepton^{+}\Plepton^{-}\Plepton^{+}\Plepton^{-}$ are covered by the analysis published in Ref.~\cite{Sirunyan:2017exp}.
The top quark may decay either semi-leptonically via $\Ptop \to \Pbottom\PW^{+} \to \Pbottom\Plepton^{+}\Pnu_{\Plepton}$ or hadronically via $\Ptop \to \Pbottom\PW^{+} \to \Pbottom\Pquark\Pquark'$,
and analogously for the top antiquarks.
The experimental signature of $\ttH$ and $\tH$ signal events consists of: multiple electrons, muons, and $\tauh$;
$\ptmiss$ caused by the neutrinos produced in the $\PW$ and $\PZ$ bosons, and tau lepton decays;
one ($\tH$) or two ($\ttH$) $\Pbottom$ jets from top quark decays;
and further light-quark jets, produced in the decays of either the Higgs boson or of the top quark(s).

The events considered in the analysis are selected in ten nonoverlapping channels,
targeting the signatures
$\twoLeptonssZeroTau$, $\threeLeptonZeroTau$, $\twoLeptonssOneTau$,
$\oneLeptonOneTau$, $\zeroLeptonTwoTau$,
$\twoLeptonosOneTau$, $\oneLeptonTwoTau$, 
$\fourLeptonZeroTau$, $\threeLeptonOneTau$, and $\twoLeptonTwoTau$, as stated earlier.
The channels $\oneLeptonOneTau$ and $\zeroLeptonTwoTau$ specifically target events in which the Higgs boson decays via $\PHiggs \to \Pgt\Pgt$ 
and the top quarks decay hadronically,
the other channels target a mixture of $\PHiggs \to \PW\PW$, $\PHiggs \to \Pgt\Pgt$, and $\PHiggs \to \PZ\PZ$ decays
in events with either one or two semi-leptonically decaying top quarks.

Events are selected at the trigger level using a combination of single-, double-, and triple-lepton triggers,
lepton$+\tauh$ triggers, and double-$\tauh$ triggers.
Spurious triggers are discarded by demanding that electrons, muons, and $\tauh$ reconstructed at the trigger level
match electrons, muons, and $\tauh$ reconstructed offline.
The \pt thresholds of the triggers typically vary by a few GeV during different data-taking periods, depending on the instantaneous luminosity.
For example, the threshold of the single-electron trigger ranges between 25 and 35\GeV in the analyzed data set,
and that of the single-muon trigger varies between 22 and 27\GeV.
The double-lepton (triple-lepton) triggers reduce the \pt threshold that is applied to the lepton of highest \pt 
to 23 (16)\GeV in case this lepton is an electron and to 17 (8)\GeV in case it is an muon.
The electron$+\tauh$ (muon$+\tauh$) trigger requires the presence of an electron of $\pt > 24\GeV$ (muon of $\pt > 19$ or 20\GeV) 
in combination with a $\tauh$ of $\pt > 20$ or 30\GeV ($\pt > 20$ or 27\GeV),
where the lower \pt thresholds were used in 2016 and the higher ones in 2017 and 2018.
The threshold of the double-$\tauh$ trigger ranges between 35 and 40\GeV and is applied to both $\tauh$.
In order to attain these \pt thresholds,
the geometric acceptance of the lepton$+\tauh$ and double-$\tauh$ triggers is restricted to the range $\abs{\eta} < 2.1$
for electrons, muons, and $\tauh$.
The \pt thresholds applied to electrons, muons, and $\tauh$ in the offline event selection 
are chosen above the trigger thresholds.

The charge of leptons and $\tauh$ is required to match the signature expected for the $\ttH$ and $\tH$ signals.
The $\zeroLeptonTwoTau$ and $\oneLeptonTwoTau$ channels target events where the Higgs boson decays to a $\Pgt$ lepton pair
and both $\Pgt$ leptons decay hadronically.
Consequently, the two $\tauh$ are required to have $\os$ charges in these channels.
In events selected in the channels $\fourLeptonZeroTau$, $\threeLeptonOneTau$, and $\twoLeptonTwoTau$,
the leptons and $\tauh$ are expected to originate from either the Higgs boson decay or from the decay of the top quark-antiquark pair
and the sum of their charges is required to be zero.
In the $\threeLeptonZeroTau$, $\twoLeptonssOneTau$, $\twoLeptonosOneTau$, and $\oneLeptonTwoTau$ channels
the charge-sum of leptons plus $\tauh$ is required to be either $+1$ or $-1$.
No requirement on the charge of the lepton and of the $\tauh$ is applied in the $\oneLeptonOneTau$ channel,
because studies performed with simulated samples of signal and background events indicate that the sensitivity of this channel is higher when no charge requirement is applied.
The $\twoLeptonssZeroTau$ channel targets events 
in which one lepton originates from the decay of the Higgs boson and the other lepton from a top quark decay.
Requiring $\ss$ leptons reduces the signal yield by about half,
but increases the signal-to-background ratio by a large factor
by removing in particular the large background arising from $\ttbar$+jets production with dileptonic decays of the top quarks.
The more favorable signal-to-background ratio for events with SS, rather than OS, lepton pairs
motivates the choice of analyzing the events containing two leptons and one $\tauh$ separately,
in the two channels $\twoLeptonssOneTau$ and $\twoLeptonosOneTau$.

The selection criteria on $\Pbottom$ jets are designed to maintain a high efficiency for the $\ttH$ signal:
one $\Pbottom$ jet can be outside of the \pt and $\eta$ acceptance of the jet selection or can fail the $\Pbottom$ tagging criteria,
provided that the other $\Pbottom$ jet passes the tight $\Pbottom$ tagging criteria.
This choice is motivated by the observation that the main background contributions,
arising from the associated production of single top quarks or top quark pairs with $\PW$ and $\PZ$ bosons, photons, and jets,
feature genuine $\Pbottom$ jets with a multiplicity resembling that of the $\ttH$ and $\tH$ signals.

The requirements on the overall multiplicity of jets, including $\Pbottom$ jets,
take advantage of the fact that the multiplicity of jets is typically higher in signal events compared to the background.
The total number of jets expected in $\ttH$ ($\tH$) signal events with the $\PH$ boson decaying into $\PW\PW$, $\PZ\PZ$, and $\PGt\PGt$ amounts to
$\njets = 10 - 2 N_{\Plepton} - 2 N_{\Pgt}$ ($\njets = 7 - 2 N_{\Plepton} - 2 N_{\Pgt}$),
where $\njets$,  $N_{\Plepton}$ and $N_{\Pgt}$ denote the total number of jets,  electrons or muons, and hadronic $\Pgt$ decays, respectively.
The requirements on $\njets$ applied in each channel permit up to two jets to be outside of the \pt and $\eta$ acceptance of the jet selection.
In the $\twoLeptonssZeroTau$ channel, the requirement on $\njets$ is relaxed further, to increase the signal efficiency in particular for the $\tH$ process.

Background contributions arising from $\ttZ$, $\tZ$, $\PW\PZ$, and DY production
are suppressed by vetoing events containing $\os$ pairs of leptons of the same flavor, referred to as SFOS lepton pairs,
passing the loose lepton selection criteria and having an invariant mass $m_{\Plepton\Plepton}$ within 10\GeV of the $\PZ$ boson mass, $m_{\PZ} = 91.19\GeV$~\cite{Tanabashi:2018oca}.
We refer to this selection criterion as ``$\PZ$ boson veto''.
In the $\twoLeptonssZeroTau$ and $\twoLeptonssOneTau$ channels,
the $\PZ$ boson veto is also applied to $\ss$ electron pairs,
because the probability to mismeasure the charge of electrons is significantly higher than the corresponding probability for muons. 

Background contributions arising from DY production in the $\twoLeptonssZeroTau$, $\threeLeptonZeroTau$, $\twoLeptonssOneTau$, $\fourLeptonZeroTau$, $\threeLeptonOneTau$, and $\twoLeptonTwoTau$ channels
are further reduced by imposing a requirement on the linear discriminant, $\metLD > 30\GeV$. 
The requirement on $\metLD$ is relaxed or tightened, depending on whether or not the event meets certain conditions,
in order to either increase the efficiency to select $\ttH$ and $\tH$ signal events or to reject more background.
In the $\twoLeptonssZeroTau$ and $\twoLeptonssOneTau$ channels,
the requirement on $\metLD$ is only applied to events where both reconstructed leptons are electrons,
to suppress the contribution of DY production entering the selection through a mismeasurement of the electron charge.
In the $\threeLeptonZeroTau$, $\fourLeptonZeroTau$, $\threeLeptonOneTau$, and $\twoLeptonTwoTau$ channels,
the distribution of $\njets$ is steeply falling for the DY background, 
thus rendering the expected contribution of this background small if the event contains a high number of jets;
we take advantage of this fact by applying the requirement on $\metLD$ only to events with three or fewer jets.
If events with $\njets \leq 3$ contain an SFOS lepton pair, 
the requirement on $\metLD$ is tightened to the condition $\metLD > 45\GeV$.
Events considered in the $\threeLeptonZeroTau$, $\fourLeptonZeroTau$, $\threeLeptonOneTau$, and $\twoLeptonTwoTau$ channels containing three or fewer jets
and no SFOS lepton pair are required to satisfy the nominal condition $\metLD > 30\GeV$.

Events containing a pair of leptons passing the loose selection criteria and having an invariant mass $m_{\Plepton\Plepton}$ of less than 12\GeV are vetoed,
to remove events in which the leptons originate from quarkonium decays, cascade decays of heavy-flavor hadrons,
and low-mass DY production, because such events are not well modeled by the MC simulation.

In the $\threeLeptonZeroTau$ and $\fourLeptonZeroTau$ channels,
events containing four leptons passing the loose selection criteria
and having an invariant mass of $m_{4\Plepton}$ of the four-lepton system of less than 140\GeV are vetoed,
to remove $\ttH$ and $\tH$ signal events in which the Higgs boson decays via 
$\PHiggs \to \PZ\PZ \to \Plepton^{+}\Plepton^{-}\Plepton^{+}\Plepton^{-}$,
thereby avoiding overlap with the analysis published in Ref.~\cite{Sirunyan:2017exp}.

A summary of the event selection criteria applied in the different channels is given in Tables~\ref{tab:eventSelection1}-\ref{tab:eventSelection3}.

\begin{table*}[!htbp]
  {\centering
  \topcaption{
    Event selections applied in the $\twoLeptonssZeroTau$, $\twoLeptonssOneTau$, $\threeLeptonZeroTau$, and $\threeLeptonOneTau$ channels.
    The \pt thresholds applied to the lepton of highest, second-highest, and third-highest \pt are separated by slashes. The symbol ``\NA'' indicates that no requirement is applied.
  }
  \label{tab:eventSelection1}
  \cmsTable{
    \begin{tabular}{lC{6cm}C{6cm}}
      \hline
      Selection step & $\twoLeptonssZeroTau$ & $\twoLeptonssOneTau$ \\
      \hline
      Targeted $\ttH$ decay & $\Ptop \to \Pbottom\Plepton\Pnu$, $\Ptop \to \Pbottom\Pquark\Pquark'$ with  & $\Ptop \to \Pbottom\Plepton\Pnu$, $\Ptop \to \Pbottom\Pquark\Pquark'$ with  \\
      & $\PHiggs \to \PW\PW \to \Plepton\Pnu\Pquark\Pquark'$ & $\PHiggs \to \Pgt\Pgt \to \Plepton\Pnu\Pnu\tauh\Pnu$ \\ 
      [\cmsTabSkip]
      Targeted $\tH$ decays & $\cPqt\to\cPqb\Plepton\Pnu$, & $\cPqt\to\cPqb\Plepton\Pnu$, \\
      & $\PHiggs\to\PW\PW\to\Plepton\Pnu\cPq\cPq'$ & $\PHiggs\to\PGt\PGt\to\Plepton\tauh+\Pnu'\text{s}$ \\
      [\cmsTabSkip]
      Trigger & \multicolumn{2}{c}{Single- and double-lepton triggers} \\
      [\cmsTabSkip]
      Lepton \pt & $\pt > 25$ / $15\GeV$ & $\pt > 25$ / $15\GeV$ ($\Pe$) or $10\GeV$ ($\Pgm$) \\
      Lepton $\eta$ & \multicolumn{2}{c}{$\abs{\eta} < 2.5$ ($\Pe$) or $2.4$ ($\Pgm$)} \\
      $\tauh$ \pt & \NA & $\pt > 20\GeV$ \\
      $\tauh$ $\eta$ & \NA & $\abs{\eta} < 2.3$ \\
      $\tauh$ identification & \NA & very loose \\
      Charge requirements & 2 $\ss$ leptons & 2 $\ss$ leptons \\
      & and charge quality requirements & and charge quality requirements \\
      & & $\sum\limits_{\Plepton,\tauh} q = \pm 1$ \\
      [\cmsTabSkip]
      Multiplicity of central jets & $\geq$3 jets & $\geq$3 jets \\
      $\Pbottom$ tagging requirements & \multicolumn{2}{c}{$\geq$1 tight $\Pbottom$-tagged jet or $\geq$2 loose $\Pbottom$-tagged jets} \\
      [\cmsTabSkip]
      Missing transverse & \multicolumn{2}{c}{$\metLD > 30\GeV^{\dagger}$} \\
      momentum & & \\
      [\cmsTabSkip]
      Dilepton invariant mass & \multicolumn{2}{c}{$\abs{m_{\Plepton\Plepton} - m_{\PZ}} > 10\GeV^{\ddagger}$ and $m_{\Plepton\Plepton} > 12\GeV$} \\
      \hline
    \end{tabular}
  }
  \vspace*{0.2 cm}
  
  \cmsTable{
    \begin{tabular}{lC{6cm}C{6cm}}
      \hline
      Selection step &  $\threeLeptonZeroTau$ & $\threeLeptonOneTau$ \\
      \hline
      Targeted $\ttH$ decays & $\Ptop \to \Pbottom\Plepton\Pnu$, $\Ptop \to \Pbottom\Plepton\Pnu$ with & $\Ptop \to \Pbottom\Plepton\Pnu$, $\Ptop \to \Pbottom\Plepton\Pnu$ with \\ 
      & $\PHiggs \to \PW\PW \to \Plepton\Pnu\Pquark\Pquark'$ & $\PHiggs \to \Pgt\Pgt \to \Plepton\Pnu\Pnu\tauh\Pnu$ \\
      & $\Ptop \to \Pbottom\Plepton\Pnu$, $\Ptop \to \Pbottom\Pquark\Pquark'$ with $\PHiggs \to \PW\PW \to \Plepton\Pnu\Plepton\Pnu$ & \\
      & $\Ptop \to \Pbottom\Plepton\Pnu$, $\Ptop \to \Pbottom\Pquark\Pquark'$ with & \\
      &  $\PHiggs \to \PZ\PZ \to \Plepton\Plepton\Pquark\Pquark'$ or $\Plepton\Plepton\Pnu\Pnu$ & \\
      [\cmsTabSkip]
      Targeted $\tH$ decays & $\cPqt\to\cPqb\Plepton\Pnu$, $\PHiggs\to\PW\PW\to\Plepton\Pnu\Plepton\Pnu$ & \NA \\
      [\cmsTabSkip]
      Trigger & \multicolumn{2}{c}{Single-, double- and triple-lepton triggers} \\
      [\cmsTabSkip]
      Lepton \pt & \multicolumn{2}{c}{$\pt > 25$ / $15$ / $10\GeV$} \\
      Lepton $\eta$ & \multicolumn{2}{c}{$\abs{\eta} < 2.5$ ($\Pe$) or $2.4$ ($\Pgm$)} \\
      $\tauh$ \pt & \NA & $\pt > 20\GeV$ \\
      $\tauh$ $\eta$ & \NA & $\abs{\eta} < 2.3$ \\
      $\tauh$ identification & \NA & very loose \\
      Charge requirements & $\sum\limits_{\Plepton} q = \pm 1$ & $\sum\limits_{\Plepton,\tauh} q = 0$ \\
      [\cmsTabSkip]
      Multiplicity of central jets & \multicolumn{2}{c}{$\geq$2 jets} \\
      $\Pbottom$ tagging requirements & \multicolumn{2}{c}{$\geq$1 tight $\Pbottom$-tagged jet or $\geq$2 loose $\Pbottom$-tagged jets} \\
      [\cmsTabSkip]
      Missing transverse & \multicolumn{2}{c}{$\metLD > 0$ / $30$ / $45\GeV^{\ddagger}$} \\
      momentum & & \\
      [\cmsTabSkip]
      Dilepton invariant mass & \multicolumn{2}{c}{$m_{\Plepton\Plepton} > 12\GeV$ and $\abs{m_{\Plepton\Plepton} - m_{\PZ}} > 10\GeV^{\mathsection}$} \\
      Four-lepton invariant mass & $m_{4\Plepton} > 140\GeV^{\mathparagraph}$ & \NA \\
      \hline
    \end{tabular}
    \par}
  }
    $^{\dagger}$ A complete description of this requirement can be found in the main text. \\
    $^{\ddagger}$ Applied to all SFOS lepton pairs and to pairs of electrons of SS charge. \\
    $^{\mathsection}$ Applied to all SFOS lepton pairs. \\
    $^{\mathparagraph}$ If the event contains two SFOS pairs of leptons that pass the loose lepton selection criteria.
\end{table*}

\begin{table*}[h!]
  {\centering
  \topcaption{
    Event selections applied in the $\zeroLeptonTwoTau$, $\oneLeptonOneTau$, $\oneLeptonTwoTau$, and $\twoLeptonTwoTau$ channels.
    The \pt thresholds applied to the lepton and to the $\tauh$ of highest and second-highest \pt are separated by slashes. The symbol ``\NA'' indicates that no requirement is applied.
  }
  \label{tab:eventSelection2}
  \cmsTable{
    \begin{tabular}{lC{6cm}C{6cm}}
      \hline
      Selection step &  $\zeroLeptonTwoTau$ & $\oneLeptonOneTau$ \\
      \hline
      Targeted $\ttH$ decays & $\Ptop \to \Pbottom\Pquark\Pquark'$, $\Ptop \to \Pbottom\Pquark\Pquark'$ with  & $\Ptop \to \Pbottom\Pquark\Pquark'$, $\Ptop \to \Pbottom\Pquark\Pquark'$ with  \\
      & $\PHiggs \to \Pgt\Pgt \to \tauh\Pnu\tauh\Pnu$ & $\PHiggs \to \Pgt\Pgt \to \Plepton\Pnu\Pnu\tauh\Pnu$ \\ 
      [\cmsTabSkip]
      Trigger & Double-$\tauh$ trigger & Single-lepton \\
      & & and lepton$+\tauh$ triggers \\
      [\cmsTabSkip]
      Lepton \pt & \NA & $\pt > 30$ ($\Pe$) or $25\GeV$ ($\Pgm$)  \\
      Lepton $\eta$ & \NA & $\abs{\eta} < 2.1$ \\
      $\tauh$ \pt & $\pt > 40\GeV$ & $\pt > 30\GeV$ \\
      $\tauh$ $\eta$ & \multicolumn{2}{c}{$\abs{\eta} < 2.1$} \\
      $\tauh$ identification & loose & medium \\
      Charge requirements & $\sum\limits_{\tauh} q = 0$ & $\sum\limits_{\Plepton,\tauh} q = 0$ \\
      [\cmsTabSkip]
      Multiplicity of central jets & \multicolumn{2}{c}{$\geq$4 jets} \\
      $\Pbottom$ tagging requirements & \multicolumn{2}{c}{$\geq$1 tight $\Pbottom$-tagged jet or $\geq$2 loose $\Pbottom$-tagged jets} \\
      [\cmsTabSkip]
      Dilepton invariant mass & \multicolumn{2}{c}{$m_{\Plepton\Plepton} > 12\GeV$} \\
      \hline
    \end{tabular}
  }
  \vspace*{0.2 cm}

  \cmsTable{
    \begin{tabular}{lC{6cm}C{6cm}}
      \hline
      Selection step & $\oneLeptonTwoTau$ & $\twoLeptonTwoTau$ \\
      \hline
      Targeted $\ttH$ decays & $\Ptop \to \Pbottom\Plepton\Pnu$, $\Ptop \to \Pbottom\Pquark\Pquark'$ with  & $\Ptop \to \Pbottom\Plepton\Pnu$, $\Ptop \to \Pbottom\Plepton\Pnu$ with  \\
      & $\PHiggs \to \Pgt^{+}\Pgt^{-} \to \tauh\Pnu\tauh\Pnu$ & $\PHiggs \to \Pgt^{+}\Pgt^{-} \to \tauh\Pnu\tauh\Pnu$ \\
      [\cmsTabSkip]
      Trigger & Single-lepton & Single- \\ 
      & and lepton+$\tauh$ triggers & and double-lepton triggers \\
      [\cmsTabSkip]
      Lepton \pt & $\pt > 30$ ($\Pe$) or $25\GeV$ ($\Pgm$) & $\pt > 25$ / 10 $(15)\GeV$ ($\Pe$) \\
      Lepton $\eta$ & $\abs{\eta} < 2.1$ & $\abs{\eta} < 2.5$ ($\Pe$) or $2.4$ ($\Pgm$) \\
      $\tauh$ \pt & $\pt > 30$ / $20\GeV$ & $\pt > 20\GeV$ \\
      $\tauh$ $\eta$ & $\abs{\eta} < 2.1$ & $\abs{\eta} < 2.3$ \\
      $\tauh$ identification & medium & medium \\
      Charge requirements & $\sum\limits_{\Plepton,\tauh} q = \pm 1$ & $\sum\limits_{\Plepton,\tauh} q = 0$ \\
      [\cmsTabSkip]
      Multiplicity of central jets & $\geq$3 jets & $\geq$2 jets \\
      $\Pbottom$ tagging requirements & \multicolumn{2}{c}{$\geq$1 tight $\Pbottom$-tagged jet or $\geq$2 loose $\Pbottom$-tagged jets} \\
      [\cmsTabSkip]
      Missing transverse & \NA & $\metLD > 0$ / $30$ / $45\GeV^{\dagger}$ \\
      momentum & & \\
      [\cmsTabSkip]
      Dilepton invariant mass & \multicolumn{2}{c}{$m_{\Plepton\Plepton} > 12\GeV$} \\
      \hline
    \end{tabular}
    \par}}
    $^{\dagger}$ A complete description of this requirement can be found in the main text.\\

\end{table*}

\begin{table*}[h!]
  {\centering
  \topcaption{
    Event selections applied in the $\twoLeptonosOneTau$ and $\fourLeptonZeroTau$ channels. The symbol ``\NA'' indicates that no requirement is applied.
  }
  \label{tab:eventSelection3}
  \cmsTable{
    \begin{tabular}{lC{6cm}C{6cm}}
      \hline
      Selection step & $\twoLeptonosOneTau$ & $\fourLeptonZeroTau$ \\
      \hline
Targeted $\ttH$ decays & $\Ptop \to \Pbottom\Plepton\Pnu$, $\Ptop \to \Pbottom\Pquark\Pquark'$ with  & $\Ptop \to \Pbottom\Plepton\Pnu$, $\Ptop \to \Pbottom\Plepton\Pnu$ with  \\
      & $\PHiggs \to \Pgt^{+}\Pgt^{-} \to \Plepton\Pnu\Pnu\tauh\Pnu$ &  $\PHiggs \to \PW\PW \to \Plepton\Pnu\Plepton\Pnu$ \\
      & & $\Ptop \to \Pbottom\Plepton\Pnu$, $\Ptop \to \Pbottom\Plepton\Pnu$ with \\
      & &  $\PHiggs \to \PZ\PZ \to \Plepton\Plepton\Pquark\Pquark'$ or $\Plepton\Plepton\Pnu\Pnu$ \\
      [\cmsTabSkip]
      Trigger & Single- & Single-, double-  \\
      & and double-lepton triggers & and triple-lepton triggers \\
      [\cmsTabSkip]
      Lepton \pt & $\pt > 25$ / $15\GeV$ ($\Pe$) or $10\GeV$ ($\Pgm$) & $\pt > 25$ / $15$ / $15$ / $10\GeV$ \\
      Lepton $\eta$ & \multicolumn{2}{c}{$\abs{\eta} < 2.5$ ($\Pe$) or $2.4$ ($\Pgm$)} \\
      $\tauh$ \pt & $\pt > 20\GeV$ & \NA \\
$\tauh$ $\eta$ & $\abs{\eta} < 2.3$ & \NA \\
      $\tauh$ identification & tight & \NA \\
      Charge requirements & $\sum\limits_{\Plepton} q = 0$ and $\sum\limits_{\Plepton,\tauh} q = \pm 1$ & $\sum\limits_{\Plepton} q = 0$ \\
      [\cmsTabSkip]
      Multiplicity of central jets & $\geq$3 jets & $\geq$2 jets \\
      $\Pbottom$ tagging requirements & \multicolumn{2}{c}{$\geq$1 tight $\Pbottom$-tagged jet or $\geq$2 loose $\Pbottom$-tagged jets} \\
      [\cmsTabSkip]
      Missing transverse & $\metLD > 30\GeV^{\dagger}$ & $\metLD > 0$ / $30$ / $45\GeV^{\ddagger}$ \\
      momentum & & \\
      [\cmsTabSkip]
      Dilepton invariant mass & $m_{\Plepton\Plepton} > 12\GeV$ & $\abs{m_{\Plepton\Plepton} - m_{\PZ}} > 10\GeV^{\mathsection}$ \\
      & & and $m_{\Plepton\Plepton} > 12\GeV$ \\
      Four-lepton invariant mass & \NA & $m_{4\Plepton} > 140\GeV^{\mathparagraph}$ \\
      \hline
    \end{tabular}
    \par}}
    $^{\dagger}$ Only applied to events containing two electrons. \\
    $^{\ddagger}$ A complete description of this requirement can be found in the main text.\\
    $^{\mathsection}$ Applied to all SFOS lepton pairs. \\
    $^{\mathparagraph}$ If the event contains two SFOS pairs of leptons passing the loose lepton selection criteria.
\end{table*}

\section{Event classification, signal extraction, and analysis strategy}
\label{sec:eventClassification_and_signalExtraction}

\begin{figure*}[h!]
  \centering\includegraphics[width=\textwidth]{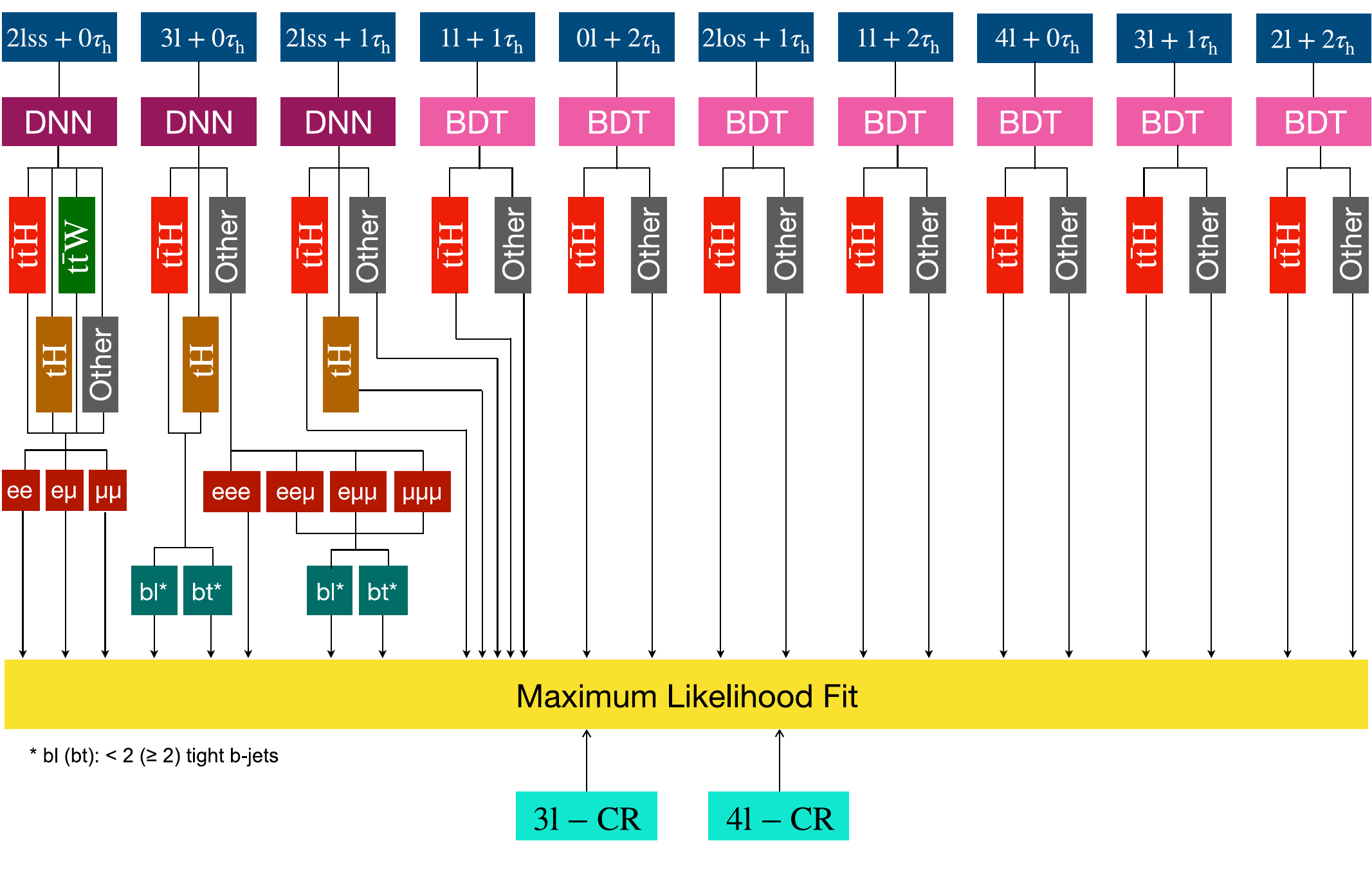}
  \caption{
    Diagram showing the categorization strategy used for the signal extraction, making use of
    MVA-based algorithms and topological variables. In addition to the ten channels, the ML fit receives input
    from two control regions (CRs) defined in Section~\ref{sec:backgroundEstimation_control_regions}.
  }
  \label{fig:diagram}
\end{figure*}

Contributions from background processes that pass the event selection criteria detailed in Section~\ref{sec:eventSelection}, significantly exceed
the expected $\ttH$ and $\tH$ signal rates.
The ratio of expected signal to background yields is particularly unfavorable in channels with a low multiplicity of leptons and $\tauh$,
notwithstanding that these channels also provide the highest acceptance for the $\ttH$ and $\tH$ signals.
In order to separate the $\ttH$ and $\tH$ signals from the background contributions,
we employ a maximum-likelihood (ML) fit to the distributions of a number of discriminating observables.
The choice of these observables is based on studies, performed with simulated samples of signal and background events, that aim at maximizing the expected sensitivity of the analysis. 
Compared to the alternative of reducing the background by applying more stringent event selection criteria,
the chosen strategy has the advantage
of retaining events reconstructed in kinematic regions of low signal-to-background ratio for analysis.
Even though these events enter the ML fit with a lower ``weight'' compared to the signal events reconstructed in kinematic regions where the signal-to-background ratio is high,
the retained events increase the overall sensitivity of the statistical analysis,
firstly by increasing the overall $\ttH$ and $\tH$ signal yield
and secondly by simultaneously constraining the background contributions.
The likelihood function used in the ML fit is described in Section~\ref{sec:results}.
The diagram displayed in Fig.~\ref{fig:diagram} describes the classification employed
in each of the categories, which defines the regions that are fitted in the signal extraction fit.

The chosen discriminating observables are the outputs of machine-learning algorithms
that are trained using simulated samples of $\ttH$ and $\tH$ signal events as well as $\ttW$, $\ttZ$, $\ttbar$+jets, and diboson background samples.
For the purpose of separating the $\ttH$ and $\tH$ signals from backgrounds, the $\twoLeptonssZeroTau$, $\threeLeptonZeroTau$, and $\twoLeptonssOneTau$ channels employ ANNs,
which allows to discriminate among the two signals and background simultaneously, while the other channels use BDTs.

The observables used as input to the ANNs and BDTs are outlined in Table~\ref{tab:vblesinputmba}. These
are chosen to maximize the discrimination power of the discriminators,
with the objective of maximizing the expected sensitivity of the analysis.
The optimization is performed separately for each of the ten analysis channels.
Typical observables used are:
the number of leptons, $\tauh$, and jets that are reconstructed in the event,
where electrons and muons, as well as forward jets, central jets, and jets passing the loose and the tight $\Pbottom$ tagging criteria are counted separately;
the 3-momentum of leptons, $\tauh$, and jets;
the magnitude of the missing transverse momentum, quantified by the linear discriminant $\metLD$;
the angular separation between leptons, $\tauh$, and jets; 
the average $\Delta R$ separation between pairs of jets;
the sum of charges for different combinations of leptons and $\tauh$;
observables related to the reconstruction of specific top quark and Higgs boson decay modes;
as well as a few other observables that provide discrimination between the $\ttH$ and $\tH$ signals.
A boolean variable that indicates whether the event has an SFOS lepton pair passing looser isolation criteria is included
in regions with at least three leptons in the final state.

\begin{sidewaystable*}[htbp!]
  \topcaption{Input variables to the multivariate discriminants in each of the ten analysis channels. The symbol ``\NA'' indicates that the variable is not used. For all objects, the three-momentum is constituted by the \pt, $\eta$, and $\phi$ components of the object momentum.  }
  \label{tab:vblesinputmba}
\scriptsize
\begin{tabular}{lcccccccccc}
\hline
                                                                 & $\twoLeptonssZeroTau$ & $\twoLeptonssOneTau$ & $\threeLeptonZeroTau$ & $\oneLeptonOneTau$ & $\zeroLeptonTwoTau$ & $\twoLeptonosOneTau$ & $\oneLeptonTwoTau$ & $\fourLeptonZeroTau$ & $\threeLeptonOneTau$ & $\twoLeptonTwoTau$ \\
\hline                                                                                                                                                                                                                                                                                             
Electron multiplicity                                            & \checkmark            & \checkmark           & \checkmark            & \NA                & \NA                 & \NA                  & \NA                & \NA                  & \NA                  & \NA                \\
Three-momenta of leptons and/or $\tauh$s                         & \checkmark            & \checkmark           & \checkmark            & \checkmark         & \checkmark          & \checkmark           & \checkmark         & \NA                  & \checkmark           & \checkmark         \\
\pt of leptons and/or $\tauh$s                                   & \NA                   & \NA                  & \NA                   & \NA                & \NA                 & \NA                  & \NA                & \checkmark           & \NA                  & \NA             \\
Transverse mass of leptons and/or $\tauh$s                       & \checkmark            & \checkmark           & \NA                   & \checkmark         & \checkmark          & \checkmark           & \checkmark         & \NA                  & \NA                  & \NA                \\
Invariant mass of leptons and/or $\tauh$s                        & \checkmark            & \NA                  & \NA                   & \checkmark         & \checkmark          & \checkmark           & \checkmark         & \checkmark           & \checkmark           & \checkmark         \\
SVFit mass of leptons and/or $\tauh$s                            & \NA                   & \NA                  & \NA                   & \checkmark         & \checkmark          & \NA                  & \NA                & \NA                  & \NA                  & \NA                \\
$\Delta R$ between leptons and/or $\tauh$s                       & \checkmark            & \checkmark           & \checkmark            & \checkmark         & \checkmark          & \checkmark           & \checkmark         & \NA                  & \NA                  & \checkmark         \\
cos$\theta^{*}$ of leptons and $\tauh$s                          & \NA                   & \NA                  & \NA                   & \checkmark         & \checkmark          & \NA                  & \checkmark         & \NA                  & \NA                  & \checkmark         \\
Charge of leptons and/or $\tauh$s                                & \checkmark            & \checkmark           & \checkmark            & \checkmark         & \NA                 & \NA                  & \NA                & \NA                  & \NA                  & \NA                \\
Has SFOS lepton pairs                                            & \NA                   & \NA                  & \checkmark            & \NA                & \NA                 & \NA                  & \NA                & \checkmark           & \checkmark           & \NA                \\
Jet multiplicity                                                 & \checkmark            & \checkmark           & \checkmark            & \NA                & \NA                 & \NA                  & \NA                & \NA                  & \NA                  & \NA                \\
Jets three-momenta                                               & \checkmark            & \checkmark           & \checkmark            & \NA                & \NA                 & \NA                  & \NA                & \NA                  & \NA                  & \NA                \\
Average $\Delta R$ between jets                                  & \checkmark            & \checkmark           & \checkmark            & \checkmark         & \checkmark          & \checkmark           & \checkmark         & \NA                  & \NA                  & \checkmark         \\
Forward jet multiplicity                                         & \checkmark            & \checkmark           & \checkmark            & \NA                & \NA                 & \NA                  & \NA                & \NA                  & \NA                  & \NA                \\
Leading forward jet three-momenta                                & \checkmark            & \checkmark           & \checkmark            & \NA                & \NA                 & \NA                  & \NA                & \NA                  & \NA                  & \NA                \\
Minimum $\abs{\Delta \eta}$ between leading forward jet and jets & \NA                   & \checkmark           & \checkmark            & \NA                & \NA                 & \NA                  & \NA                & \NA                  & \NA                  & \NA                \\
$\Pbottom$ jet multiplicity                                      & \checkmark            & \checkmark           & \checkmark            & \NA                & \NA                 & \NA                  & \NA                & \NA                  & \NA                  & \NA                \\
Invariant mass of $\Pbottom$ jets                                & \checkmark            & \checkmark           & \checkmark            & \checkmark         & \checkmark          & \checkmark           & \checkmark         & \NA                  & \NA                  & \checkmark         \\
Linear discriminant $\metLD$                                     & \checkmark            & \checkmark           & \checkmark            & \checkmark         & \checkmark          & \checkmark           & \checkmark         & \checkmark           & \checkmark           & \checkmark         \\
Hadronic top quark tagger                                        & \checkmark            & \checkmark           & \checkmark            & \checkmark         & \checkmark          & \checkmark           & \checkmark         & \NA                  & \NA                  & \NA                \\
Hadronic top \pt                                                 & \NA                   & \checkmark           & \checkmark            & \NA                & \NA                 & \checkmark           & \checkmark         & \NA                  & \NA                  & \NA                \\
Higgs boson jet tagger                                           & \checkmark            & \NA                  & \NA                   & \NA                & \NA                 & \NA                  & \NA                & \NA                  & \NA                  & \NA                \\
[\cmsTabSkip]                                                                                                                                                                                                                                                                                      
Number of variables                                              & 36                    & 41                   & 37                    & 16                 & 15                  & 18                   & 17                 & 7                    & 9                    & 9 \\
\hline
\end{tabular}
\end{sidewaystable*}

Input variables are included related to the reconstruction of specific top quark and Higgs boson decay modes comprise the transverse mass of a given lepton, 
$\mT = \allowbreak\sqrt{\smash[b]{2 \pt^{\Plepton} \ptmiss \left( 1 - \cos\Delta\phi \right)}}$,
where $\Delta\phi$ refers to the angle in the transverse plane between the lepton momentum and the $\vecMET$ vector;
the invariant masses of different combinations of leptons and $\tauh$; 
and the invariant mass of the pair of jets with the highest and second-highest values of the $\Pbottom$ tagging discriminant.
These observables are complemented by the outputs of MVA-based algorithms, documented in Ref.~\cite{Sirunyan:2018shy},
that reconstruct hadronic top quark decays
and identify the jets originating from $\PHiggs \to \PW\PW \to \Plepton^{+}\Pnu_{\Plepton}\Pquark\PAQq'$ decays.

In the $\zeroLeptonTwoTau$ channel,
we use as additional inputs the invariant mass of the $\Pgt$ lepton pair, 
which is expected to be close to the Higgs boson mass in signal events and is reconstructed using the algorithm documented in Ref.~\cite{Bianchini:2016yrt} (SVFit),
in conjunction with  the decay angle, denoted by $\cos\theta^{*}$, of the two tau leptons in the Higgs boson rest frame.

In the $\twoLeptonssZeroTau$, $\threeLeptonZeroTau$, and $\twoLeptonssOneTau$ channels, 
the \pt and $\eta$ of the forward jet of highest \pt, as well as the distance $\Delta\eta$ of this jet to the jet nearest in pseudorapidity,
are used as additional inputs to the ANN, in order to improve the separation of the $\tH$ from the $\ttH$ signal.
The presence of such a jet is a characteristic signature of $\tH$ production in the $t$-channel.
The forward jet in such $\tH$ signal events is expected to be separated from other jets in the event by a pseudorapidity gap,
since there is no color flow at tree level between this jet and the jets originating from the top quark and Higgs boson decays.

The number of simulated signal and background events that pass the event selection criteria described in Section~\ref{sec:eventSelection}
and are available for training the BDTs and ANNs typically amount to a few thousand.
In order to increase the number of events in the training samples,
in particular for the channels with a high multiplicity of leptons and $\tauh$ where the amount of available events is most limited,
we relax the identification criteria for electrons, muons, and hadronically decaying tau leptons.
The resulting increase in the ratio of misidentified to genuine leptons and $\tauh$
is corrected.
We have checked that the distributions of the observables used for the BDT and ANN training are compatible, within statistical uncertainties,
between events selected with relaxed and with nominal lepton and $\tauh$ selection criteria,
provided that these corrections are applied.

The ANNs used in the $\twoLeptonssZeroTau$, $\threeLeptonZeroTau$, and $\twoLeptonssOneTau$ channels are of the multiclass type.
Such ANNs have multiple output nodes that, besides discriminating the $\ttH$ and $\tH$ signals from backgrounds,
accomplish both the separation of the $\tH$ from the $\ttH$ signal and the distinction between individual types of backgrounds.
In the $\twoLeptonssZeroTau$ channel, we use four output nodes, 
to distinguish between $\ttH$ signal, $\tH$ signal, $\ttW$ background, and other backgrounds.
No attempt is made to distinguish between individual types of backgrounds in the $\threeLeptonZeroTau$ and $\twoLeptonssOneTau$ channels,
which therefore use three output nodes. The ANNs in the $\twoLeptonssZeroTau$, $\threeLeptonZeroTau$, and $\twoLeptonssOneTau$ channels implement 16, 5 and 3 hidden layers, respectively, each one of them containing  8 to 32 neurons.
The \textrm{softmax}~\cite{softmax} function is chosen as an activation function for all output nodes,
permitting the interpretation of their activation values as probability for a given event 
to be either $\ttH$ signal, $\tH$ signal, $\ttW$ background, or other background ($\ttH$ signal, $\tH$ signal, or background)
in the $\twoLeptonssZeroTau$ channel (in the $\threeLeptonZeroTau$ and $\twoLeptonssOneTau$ channels). 
The events selected in the $\twoLeptonssZeroTau$ channel ($\threeLeptonZeroTau$ and $\twoLeptonssOneTau$ channels)
are classified into four (three) categories, corresponding to the 
$\ttH$ signal, $\tH$ signal, $\ttW$ background, or other background ($\ttH$ signal, $\tH$ signal, or background),
according to the output node that has the highest such probability value.
We refer to these categories as ANN output node categories.
The four (three) distributions of the probability values of the output nodes in the $\twoLeptonssZeroTau$ channel 
(in the $\threeLeptonZeroTau$ and $\twoLeptonssOneTau$ channels) are used as input to the ML fit.
Events are prevented from entering more than one of these distributions
by assigning each event only to the distribution corresponding to the output node that has the highest activation value.
The rectified linear activation function~\cite{ReLu} is used for the hidden layers.
The training is performed using the \textsc{TensorFlow}~\cite{tensorflow} package with the \textsc{Keras}~\cite{keras} interface.
The objective of the training is to minimize the cross-entropy loss function~\cite{cross-entropy}.
Batch gradient descent is used to update the weights of the ANN during the training.
Overtraining is minimized by using Tikhonov regularization~\cite{Tikhonov-regularization} and dropout~\cite{dropout}.

\begin{figure*}[h!]
  \centering\includegraphics[width=0.32\textwidth]{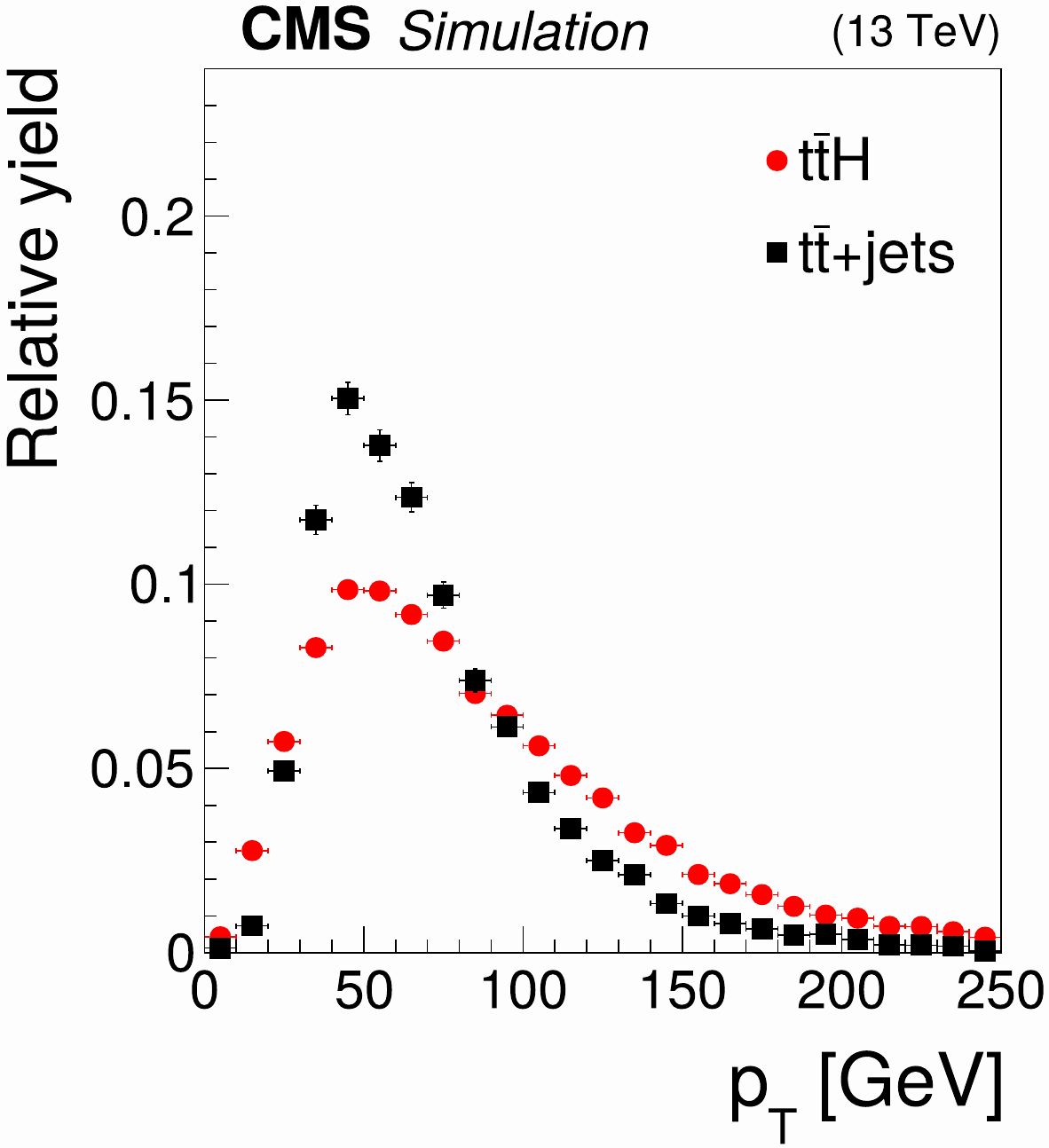}
  \centering\includegraphics[width=0.32\textwidth]{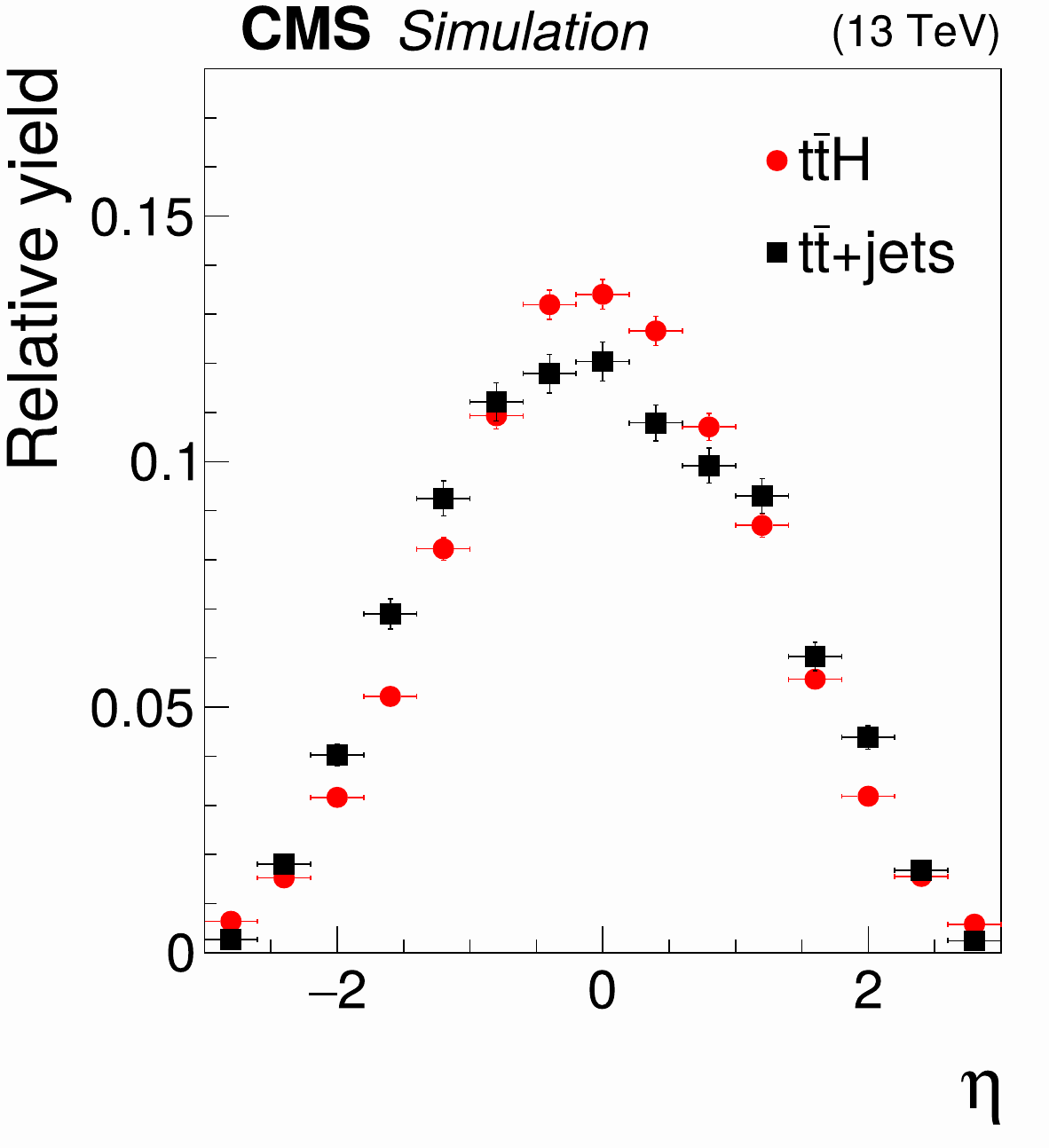}
  \centering\includegraphics[width=0.32\textwidth]{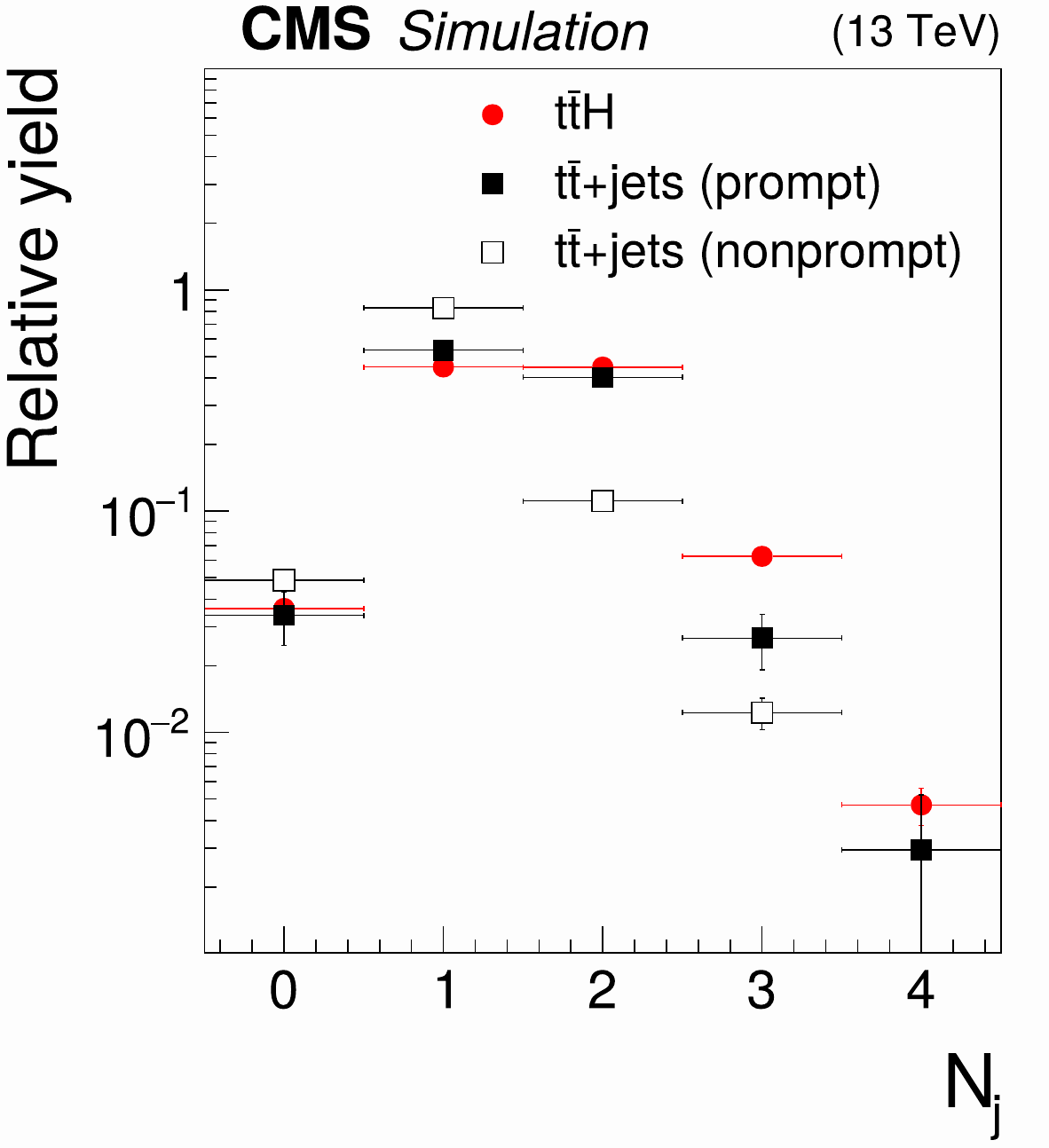}
  \caption{
  Transverse momentum (\cmsLeft) and pseudorapidity (\cmsMid) distributions of bottom quarks produced in top quark decays in $\ttH$ signal events compared to $\ttbar$+jets background events,
  and multiplicity of jets passing tight $\Pbottom$ jet identification criteria (\cmsRight).
  The latter distribution is shown separately for $\ttbar$+jets background events in which a nonprompt lepton is misidentified as a prompt lepton
  and for those background events in which all reconstructed leptons are prompt leptons.
  The events are selected in the $\twoLeptonssZeroTau$ channel.
}
\label{fig:eventClassification_motivation}
\end{figure*}

The sensitivity of the $\twoLeptonssZeroTau$ and $\threeLeptonZeroTau$ channels,
which are the channels with the largest event yields out of the three using multiclass ANN,
is further improved by analyzing selected events in subcategories based on the flavor (electron or muon) of the leptons 
and on the number of jets passing the tight $\Pbottom$ tagging criteria.
The motivation for distinguishing events by lepton flavor is that the rate for misidentifying nonprompt leptons as prompt ones
and, in the $\twoLeptonssZeroTau$ channel, also the probability for mismeasuring the lepton charge 
is significantly higher for electrons compared to muons.
Distinguishing events by the multiplicity of $\Pbottom$ jets improves in particular the separation of the $\ttH$ signal from the $\ttbar$+jets background.
This occurs because if a nonprompt lepton produced in the decay of a $\Pbottom$ hadron gets misidentified as a prompt lepton,
the remaining particles resulting from the hadronization of the bottom quark are less likely to pass the $\Pbottom$ jet identification criteria,
thereby reducing the number of $\Pbottom$ jets in such $\ttbar$+jets background events.
The distribution of the multiplicity of $\Pbottom$ jets in $\ttbar$+jets background events 
in which a nonprompt lepton is misidentified as prompt lepton (``nonprompt'')
and in $\ttbar$+jets background events in which this is not the case (``prompt'') is shown in Fig.~\ref{fig:eventClassification_motivation}.
The figure also shows the distributions of \pt and $\eta$ of bottom quarks produced in top quark decays in $\ttH$ signal events 
compared to in $\ttbar$+jets background events.
The $\ttH$ signal features more bottom quarks of high \pt, 
whereas the distribution of $\eta$ is similar for the $\ttH$ signal and for the $\ttbar$+jets background.

The number of subcategories is optimized for each of the four (three) ANN output categories of the
$\twoLeptonssZeroTau$ ($\threeLeptonZeroTau$) channel individually.
In the $\twoLeptonssZeroTau$ channel, each of the $4$ ANN output node categories is subdivided into three subcategories, 
based on the flavor of the two leptons ($\Pe\Pe$, $\Pe\Pgm$, $\Pgm\Pgm$).
In the $\threeLeptonZeroTau$ channel, the ANN output node categories corresponding to the $\ttH$ signal and to the $\tH$ signal are subdivided into two subcategories,
based on the multiplicity of jets passing tight $\Pbottom$ tagging criteria
(\textrm{bl}: $<$2 tight $\Pbottom$-tagged jets, \textrm{bt}: $\geq$2 tight $\Pbottom$-tagged jets),
while the output node category corresponding to the backgrounds is subdivided into seven subcategories,
based on the flavor of the three leptons and on the multiplicity of jets passing tight $\Pbottom$ tagging criteria
($\Pe\Pe\Pe$; $\Pe\Pe\Pgm$~\textrm{bl}, $\Pe\Pe\Pgm$~\textrm{bt}; $\Pe\Pgm\Pgm$~\textrm{bl}, $\Pe\Pgm\Pgm$~\textrm{bt}; $\Pgm\Pgm\Pgm$~\textrm{bl}, $\Pgm\Pgm\Pgm$~\textrm{bt}),
where \textrm{bl} (\textrm{bt}) again corresponds to the condition of $<$2 ($\geq$2) tight $\Pbottom$-tagged jets.
The $\Pe\Pe\Pe$ subcategory is not further subdivided by the number of $\Pbottom$-tagged jets, 
because of the lower number of events containing three electrons compared to events in other categories.
The aforementioned event categories are constructed based on the output of the BDTs and ANNs with the goal of enhancing the analysis sensitivity, while
keeping a sufficiently high rate of background events for a precise estimation.

The BDTs used in the $\oneLeptonOneTau$, $\zeroLeptonTwoTau$, $\twoLeptonosOneTau$, $\oneLeptonTwoTau$, 
$\fourLeptonZeroTau$, $\threeLeptonOneTau$, and $\twoLeptonTwoTau$ channels address the binary classification problem
of separating the sum of $\ttH$ and $\tH$ signals from the aggregate of all backgrounds.
The training is performed using the \textsc{scikit-learn}~\cite{scikit-learn} package with the \textsc{XGBoost}~\cite{2016arXiv160302754C} algorithm.
The training parameters are chosen to maximize the integral, or area-under-the-curve, 
of the receiver-operating-characteristic curve of the BDT output.

\section{Background estimation}
\label{sec:backgroundEstimation}

The dominant background in most channels comes from the production of top quarks in association with $\PW$ and $\PZ$ bosons.
We collectively refer to the sum of $\ttW$ and $\ttWW$ backgrounds using the notation $\ttWs$.
In $\ttWs$ and $\ttZ$ background events selected in the signal regions (SRs),
reconstructed leptons typically originate from genuine prompt leptons or reconstructed $\Pbottom$ jets arising from the hadronization of bottom quarks,
whereas reconstructed $\tauh$ are a mixture of genuine hadronic $\Pgt$ decays and misidentified quark or gluon jets.
Background events from $\ttZ$ production may pass the $\PZ$ boson veto
applied in the $\twoLeptonssZeroTau$, $\threeLeptonZeroTau$, $\twoLeptonssOneTau$, $\twoLeptonosOneTau$, $\fourLeptonZeroTau$, and $\threeLeptonOneTau$ channels
in the case that the $\PZ$ boson either decays to leptons and one of the leptons fails to get selected,
or the $\PZ$ boson decays to $\Pgt$ leptons and the $\Pgt$ leptons subsequently decay to electrons or muons.
In the latter case, the invariant mass $m_{\Plepton\Plepton}$ of the lepton pair is shifted to lower values because of the neutrinos produced in the $\Pgt$ decays.
Additional background contributions arise from off-shell $\Ptop\APtop\Pggx$ and $\Ptop\Pggx$ production:
we include them in the $\ttZ$ background.
The $\ttbar$+jets production cross section is about three orders of magnitude larger than the cross section for associated production of top quarks with $\PW$ and $\PZ$ bosons,
but in most channels the $\ttbar$+jets background is strongly reduced by the lepton and $\tauh$ identification criteria.
Except for the channels $\oneLeptonOneTau$ and $\zeroLeptonTwoTau$,
the $\ttbar$+jets background contributes solely in the cases that a nonprompt lepton (or a jet) is misidentified as a prompt lepton,
a quark or gluon jet is misidentified as $\tauh$, or the charge of a genuine prompt lepton is mismeasured.
Photon conversions are a relevant background in the event categories with one or more reconstructed electrons in the $\twoLeptonssZeroTau$ and $\threeLeptonZeroTau$ channels.
The production of $\PW\PZ$ and $\PZ\PZ$ pairs in events with two or more jets
constitutes another relevant background in most channels.
In the $\oneLeptonOneTau$ and $\zeroLeptonTwoTau$ channels,
an additional background arises from DY production of $\Pgt$ lepton pairs.

We categorize the contributions of background processes into reducible and irreducible ones.
A background is considered irreducible
if all reconstructed electrons and muons are genuine prompt leptons and all reconstructed $\tauh$ are genuine hadronic $\Pgt$ decays;
in the $\twoLeptonssZeroTau$ and $\twoLeptonssOneTau$ channels,
we further require that the measured charge of reconstructed electrons and muons matches their true charge.
The irreducible background contributions are modeled using simulated events fulfilling the above criteria
to avoid double-counting of all the other background contributions, which are considered to be reducible and
are mostly determined from data.

Throughout the analysis, we distinguish three sources of reducible background contributions: misidentified leptons and $\tauh$ (``misidentified leptons''),
asymmetric conversions of a photon into electrons (``conversions''), and mismeasurement of the lepton charge (``flips'').

The background from misidentified leptons and $\tauh$ refers to events
in which at least one reconstructed electron or muon is caused by the misidentification of a nonprompt lepton or hadron,
or at least one reconstructed $\tauh$ arises from the misidentification of a quark or gluon jet.
The main contribution to this background stems from $\ttbar$+jets production, reflecting the large cross section for this background process.

The conversions background consists of events
in which one or more reconstructed electrons are due to the conversion of a photon.
The conversions background is typically caused by $\ttbar\Pgg$ events
in which one electron or positron produced in the photon conversion carries most of the energy of the converted photon,
whereas the other electron or positron is of low energy and fails to get reconstructed.
We refer to such photon conversions as asymmetric conversions.

The flips background is specific to the $\twoLeptonssZeroTau$ and $\twoLeptonssOneTau$ channels
and consists in events where the charge of a reconstructed lepton is mismeasured.
The main contribution to the flips background stems from $\ttbar$+jets events in which both top quarks decay semi-leptonically.
In case of the $\twoLeptonssOneTau$ channel, a quark or gluon jet is additionally misidentified as $\tauh$.
The mismeasurement of the electron charge typically results from the emission of a hard bremsstrahlung photon,
followed by an asymmetric conversion of this photon.
The reconstructed electron is typically the electron or positron that carries most of the energy of the converted photon,
resulting in an equal probability for the reconstructed electron to have either the same or opposite charge
compared to the charge of the electron or positron that emitted the bremsstrahlung photon~\cite{Khachatryan:2015hwa}.
The probability of mismeasuring the charge of muons is negligible in this analysis.

The three types of reducible background are made mutually exclusive
by giving preference to the misidentified leptons type over the flips and conversions types and by giving preference to the flips type over the conversions type
when an event qualifies for more than one type of reducible background.
The misidentified leptons and flips backgrounds are determined from data,
whereas the conversions background is modeled using the MC simulation.
The procedures for estimating the misidentified leptons and flips backgrounds
are described in Sections~\ref{sec:backgroundEstimation_fakes} and~\ref{sec:backgroundEstimation_flips}, respectively.
We performed dedicated studies in the data to ascertain that photon conversions are adequately modeled by the MC simulation
similar to the ones performed in Ref.~\cite{Sirunyan:2017lae}.
To avoid potential double-counting of the background estimates obtained from data
with background contributions modeled using the MC simulation, we match reconstructed electrons, muons, and $\tauh$
to their generator-level equivalents and veto simulated signal and background events selected in the SR that qualify as misidentified leptons or flips backgrounds.

Concerning the irreducible backgrounds,
we refer to the aggregate of background contributions
other than those arising from $\ttWs$, $\ttZ$, $\ttbar$+jets, DY, and diboson backgrounds,
or from SM Higgs boson production via the processes $\ggH$, $\qqH$, $\WH$, $\ZH$, $\ttWH$, and $\ttZH$ 
as ``rare'' backgrounds.
The rare backgrounds typically yield a minor background contribution to each of the ten analysis channels
and include such processes as $\tW$ and $\tZ$ production,
the production of \ss $\PW$ boson pairs, triboson, and $\Ptop\APtop\Ptop\APtop$ production.

We validate the modeling of the $\ttWs$, $\ttZ$, $\PW\PZ$, and $\PZ\PZ$ backgrounds
in dedicated control regions (CRs) whose definitions are detailed in Section~\ref{sec:backgroundEstimation_control_regions}.

\subsection{Estimation of the ``misidentified leptons'' background}
\label{sec:backgroundEstimation_fakes}

The background from misidentified leptons and $\tauh$ is estimated using the misidentification probability (MP) method~\cite{Sirunyan:2018shy}.
The method is based on selecting a sample of events satisfying all selection criteria of the SR, detailed in Section~\ref{sec:eventSelection},
except that the electrons, muons, and $\tauh$ used to construct the signal regions are
required to pass relaxed selections instead of the nominal ones.
We refer to this sample of events as the application region (AR) of the MP method.
Events in which all leptons and $\tauh$ satisfy the nominal selections are vetoed, to avoid overlap with the SR.

An estimate of the background from misidentified leptons and $\tauh$ in the SR is obtained by applying suitably chosen weights to the events selected in the AR.
The weights, denoted by the symbol $w$, are given by the expression:
\begin{equation}
w = (-1)^{n+1} \, \prod_{i=1}^{n} \, \frac{f_{i}}{1 - f_{i}}\,
\label{eq:FF_weights}
\end{equation}
where the product extends over all electrons, muons, and $\tauh$
that pass the relaxed, but fail the nominal selection criteria,
and $n$ refers to the total number of such leptons and $\tauh$.
The symbol $f_{i}$ denotes the probability for an electron, muon, or $\tauh$ passing the relaxed selection
to also satisfy the nominal one.
The contributions of irreducible backgrounds to the AR are subtracted based on the MC expectation of such contributions.
The $\ttH$ and $\tH$ signal yields in the AR are found to be negligible.

\begin{figure*}[htbp!]
  \centering\includegraphics[width=0.49\textwidth]{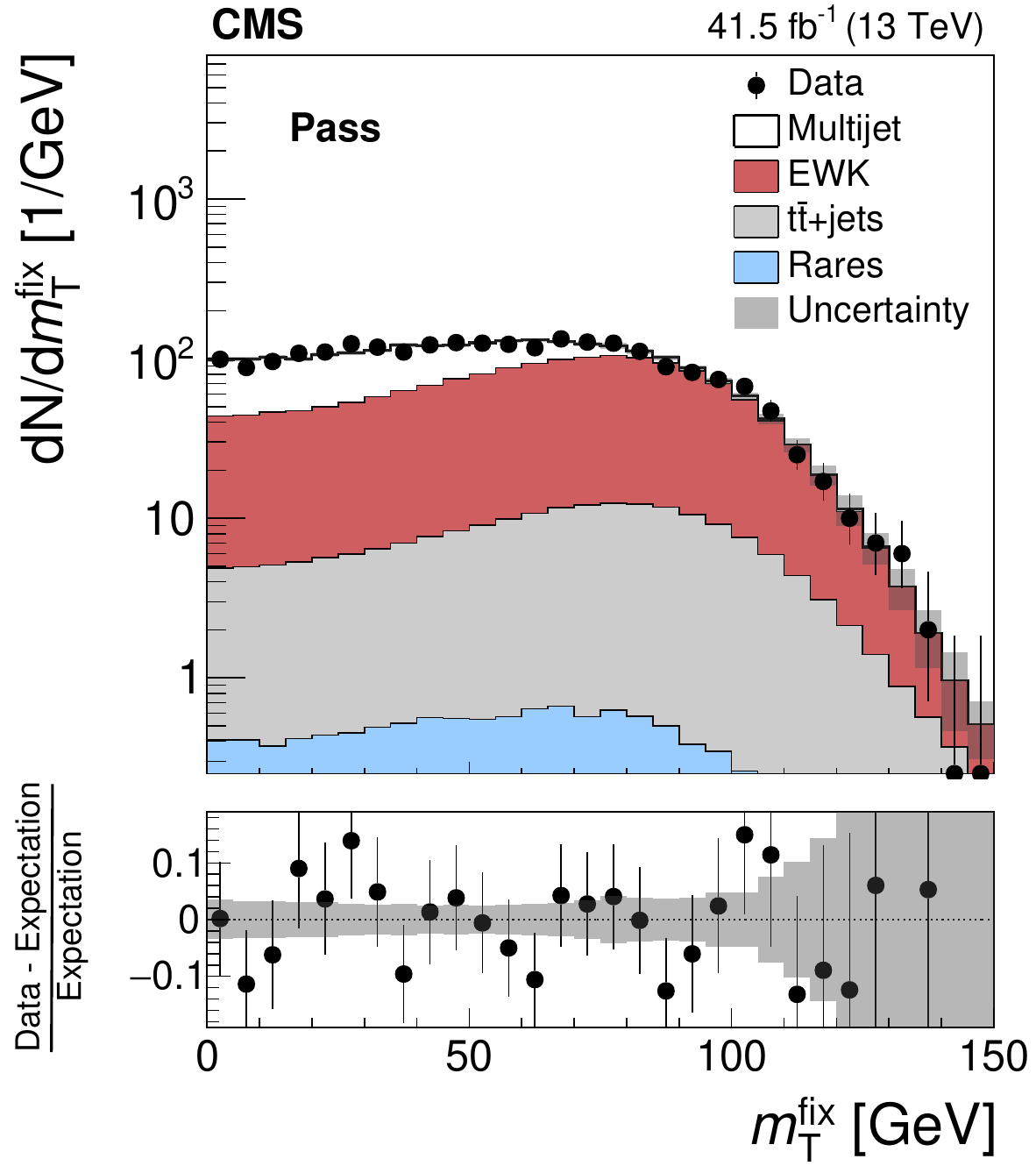}
  \centering\includegraphics[width=0.49\textwidth]{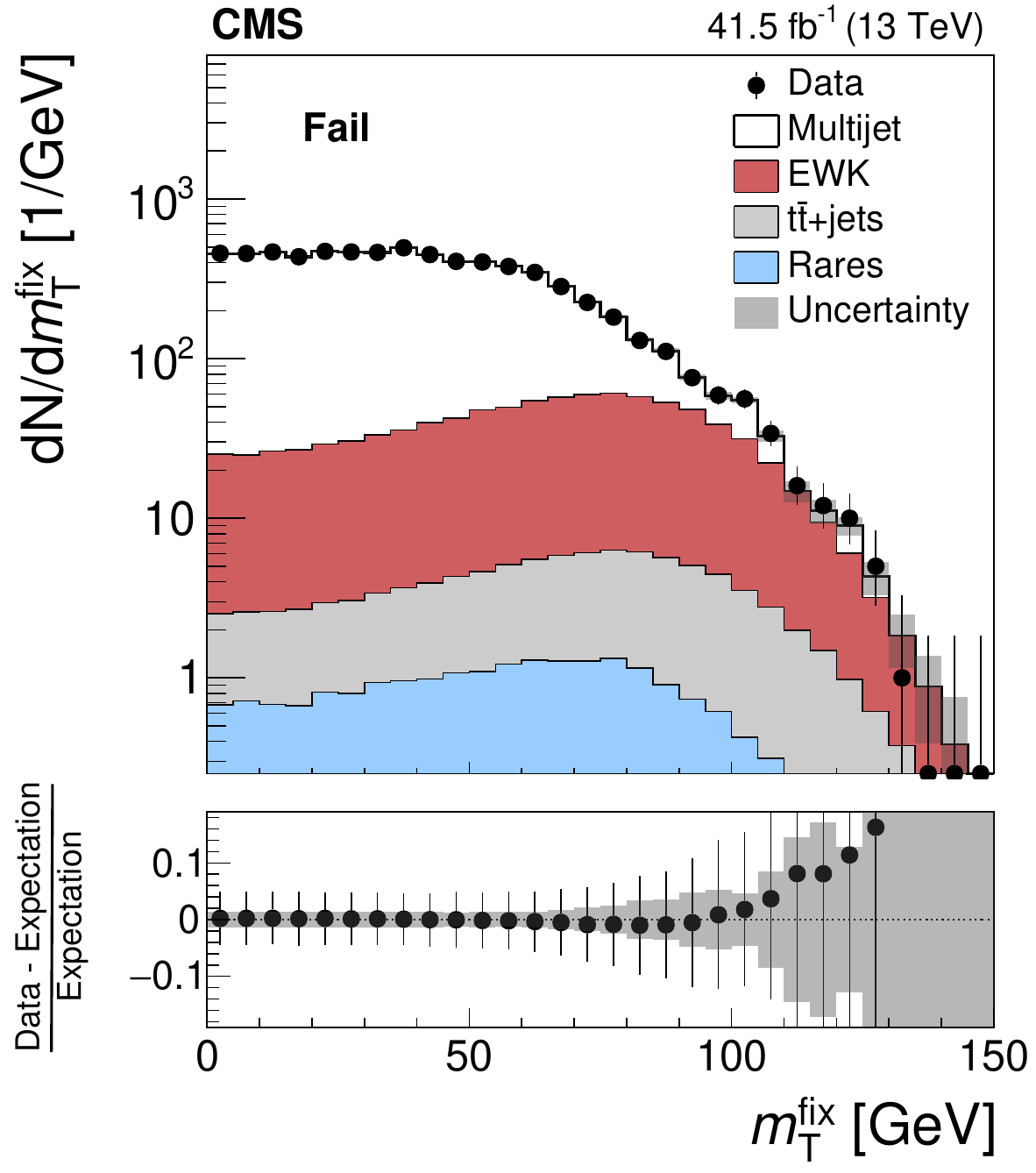}
  \caption{
    Distributions of $\mtfix$ for events containing an electron candidate of $25 < \pt < 35\GeV$ in the ECAL barrel,
    which (\cmsLeft) passes the nominal selection and (\cmsRight) passes the relaxed, but fails the nominal selection. 
    The ``electroweak'' (EWK) background refers to the sum of $\PW$+jets, DY, and diboson production.
    The ``rare'' backgrounds are defined in the text.
    The data in the fail sample agrees with the sum of multijet, EWK, $\ttbar$+jets, and rare backgrounds by construction,
    as the number of multijet events in the fail sample is computed by subtracting the sum of EWK, $\ttbar$+jets, and rare background contributions
    from the data. The misidentification probabilities are derived separately for each era: this figure shows, as an example, the results obtained with the 2017 data set.
    The uncertainty band represents the total uncertainty after the fit has been performed.
  }
  \label{fig:controlPlots_fakes}
\end{figure*}

The probabilities $f_{i}$ for leptons are measured in multijet events, separately for electrons and muons, 
and are binned in \pt and $\eta$ of the lepton candidate.
The measurement is based on selecting events containing exactly one electron or muon that passes the relaxed selection
and at least one jet separated from the lepton by $\Delta R > 0.7$.
Selected events are then subdivided into ``pass'' and ``fail'' samples,
depending on whether the lepton candidate passes the nominal selection or not.
The fail sample is dominated by the contribution of multijet events.
The contributions of other processes, predominantly arising from $\PW$+jets, DY, diboson, and $\ttbar$+jets production,
are subtracted based on MC estimates of these contributions.
The number of multijet events in the pass sample is obtained by an ML fit to the distribution of the observable:
\begin{equation}
\mtfix = \sqrt{2 \, \pt^{\fix} \, \ptmiss \, \left( 1 - \cos \Delta\phi \right)}\,,
\label{eq:mTfix}
\end{equation}
where $\pt^{\fix}$ is a constant value set to $35\GeV$, and the symbol $\Delta\phi$ refers to the angle in the transverse plane between the lepton momentum and the $\vecMET$ vector.
$\pt^{\fix}$ is used instead of the lepton \pt to reduce the correlation between \mtfix and the lepton \pt.
The ML fit is similar to the one used in the measurement of the $\ttH$ and $\tH$ signal rates, described in Section~\ref{sec:results}.
The distribution of $\PW$+jets, DY, diboson, $\ttbar$+jets, and rare backgrounds in the observable $\mtfix$ is modeled using the MC simulation,
whereas the distribution of multijet events in the pass sample is obtained from data in the fail region,
from which the $\PW$+jets, DY, diboson, and $\ttbar$+jets contributions are subtracted based on their MC estimate.
The observable $\mtfix$ exploits the fact that the $\ptmiss$ reconstructed in multijet events 
is mainly caused by resolution effects and is typically small, resulting in a falling distribution of $\mtfix$,
whereas $\PW$+jets and $\ttbar$+jets events exhibit a broad maximum around $m_{\PW} \approx 80\GeV$.
Compared to the usual transverse mass, 
the observable $\mtfix$ has the advantage of not depending on the \pt of the lepton,
and is therefore better suited for the purpose of measuring the probabilities $f_{i}$ in bins of lepton \pt.
For illustration, the distributions of $\mtfix$ in the pass and fail samples are shown in Fig.~\ref{fig:controlPlots_fakes}
for events containing an electron of $25 < \pt < 35\GeV$ in the ECAL barrel.
The contributions from $\PW$+jets, DY, and diboson production 
are assumed to scale by a common factor with respect to their MC expectation in the fit;
we refer to their sum as ``electroweak'' (EWK) background.
Finally, denoting the number of multijet events in the pass and fail samples by the symbols $N_{\pass}$ and $N_{\fail}$,
the probabilities $f_{i}$ are given by $f_{i} = N_{\pass} / (N_{\pass} + N_{\fail})$.

The $f_{i}$ for $\tauh$ are determined as a function of \pt and $\eta$ of the \tauh
candidate in a region enriched in $\ttbar$+jets events containing a reconstructed opposite-sign
electron-muon pair and at least two loose \PQb-tagged jets in addition to the \tauh candidate.
Contributions of genuine $\tauh$ are modeled using the MC simulation and subtracted.

\begin{figure*}[htp]
  \centering\includegraphics[width=0.49\textwidth]{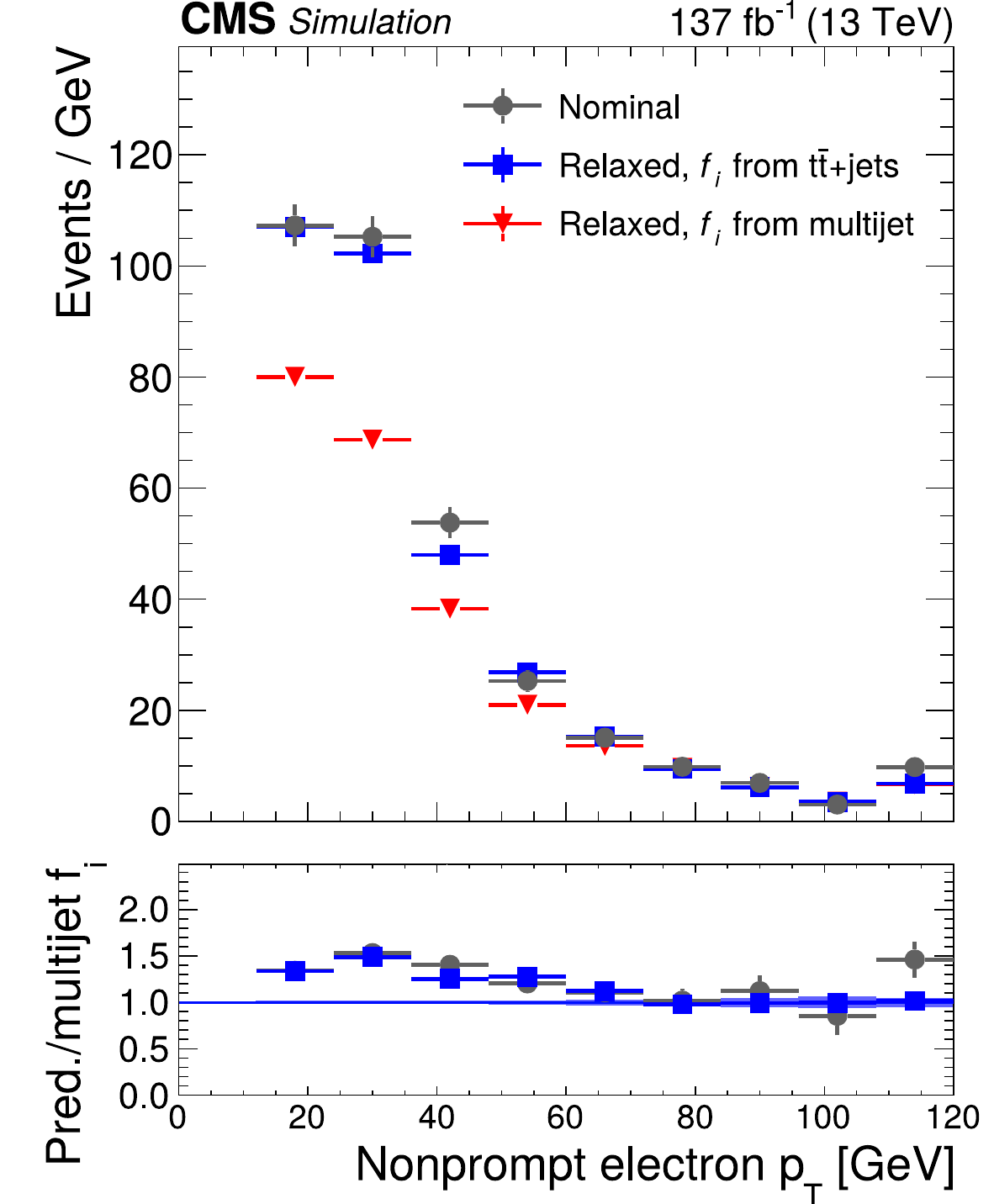}
  \centering\includegraphics[width=0.49\textwidth]{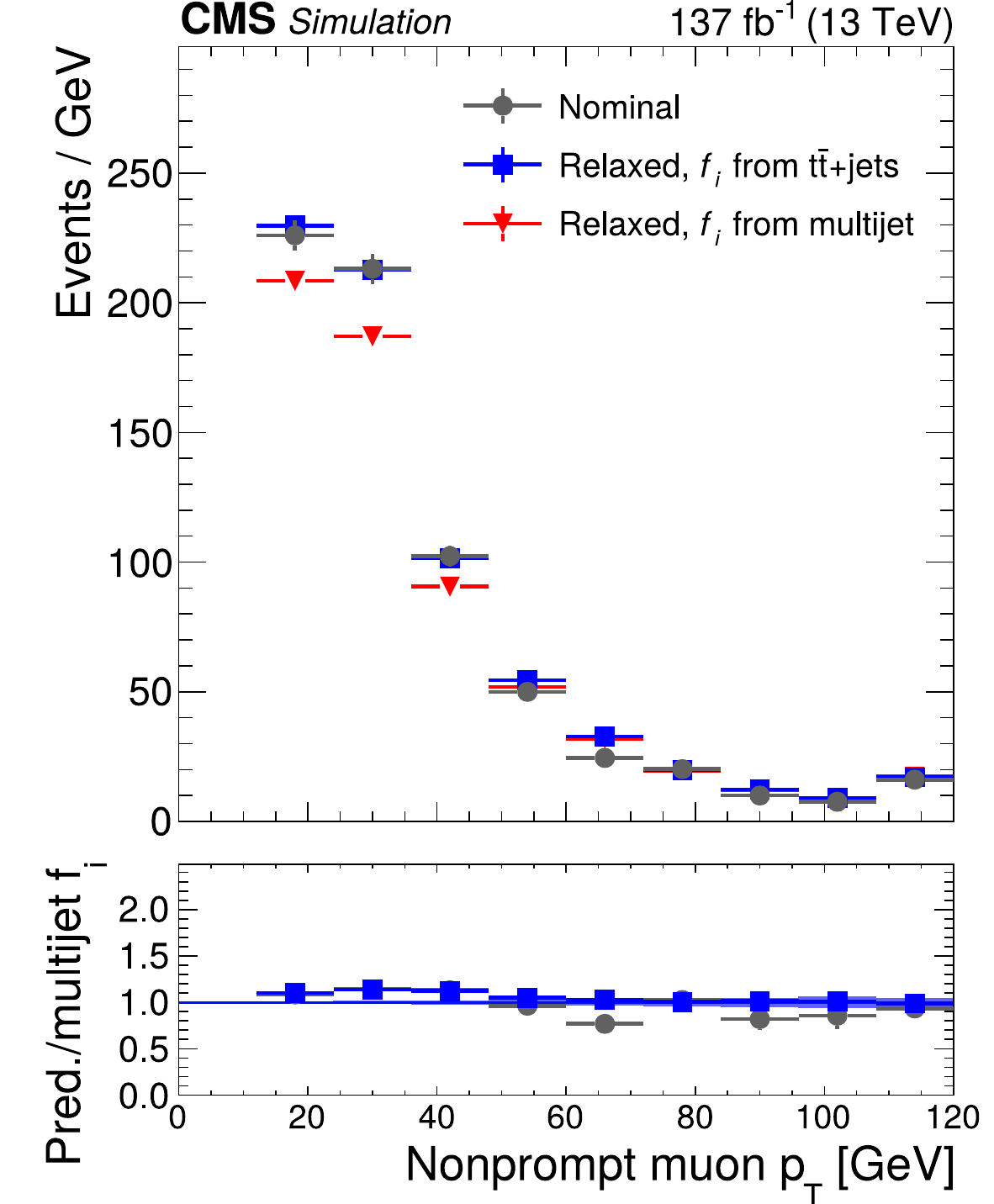}
  \caption{
    Transverse momentum distributions of nonprompt (\cmsLeft) electrons and (\cmsRight) muons in simulated $\ttbar$+jets events,
    for the three cases 
    ``nominal'', ``relaxed, $f_{i}$ from $\ttbar$+jets'', and ``relaxed, $f_{i}$ from multijet''
    discussed in text.
    The figure illustrates that a nonclosure correction needs to be applied to the probabilities $f_{i}$ measured for electrons in data,
    while no such correction is needed for muons.
  }
  \label{fig:closureTest_fakes}
\end{figure*}

The event samples used to measure the $f_{i}$ are referred to as measurement regions (MRs) of the MP method.
Potential biases in the estimate of the background from misidentified leptons and $\tauh$,
arising from differences between AR and MR in the \pt spectrum of the lepton and $\tauh$ candidates
and in the mixture of nonprompt leptons and hadrons that are misidentified as prompt leptons,
are mitigated as detailed in Ref.~\cite{Khachatryan:2016kod}.
A closure test performed using simulated $\ttbar$+jets and multijet events reveals a residual difference 
between the probabilities $f_{i}$ for electrons in $\ttbar$+jets and those in multijet events.
The test is illustrated in Fig.~\ref{fig:closureTest_fakes},
which compares the distributions of \pt of nonprompt electrons in simulated $\ttbar$+jets events for three cases:
nonprompt electrons passing the nominal selection criteria (``nominal'');
nonprompt electrons passing the relaxed, but failing the nominal selection criteria, 
weighted by probabilities $f_{i}$ determined in simulated $\ttbar$+jets events (``relaxed, $f_{i}$ from $\ttbar$+jets'');
and nonprompt electrons passing the relaxed, but failing the nominal selection criteria,
weighted by probabilities $f_{i}$ determined in simulated multijet events (``relaxed, $f_{i}$ from multijet'').
The electron and muon \pt distributions obtained in the first and second cases are in agreement,
demonstrating the performance of the MP method.
The ratio of the distributions obtained in the second and third cases
is fitted by a linear function in \pt of the lepton and is applied
as a multiplicative correction to the $f_{i}$ measured in data, that accounts for the different
flavor composition of jets between AR and MR.
For the lepton and $\tauh$ selections used in this analysis,
the probabilities $f_{i}$ range from $0.04$ to $0.13$,
$0.02$ to $0.20$, and $0.10$ to $0.50$ for electrons, muons, and $\tauh$, respectively.

The probabilities $f_{i}$ for electrons and muons obtained as described above
are validated in a CR dominated by semileptonic $\ttbar$+jets events.
The events are selected by requiring the presence of two $\ss$ leptons
and exactly three jets, one of which exactly passes the tight $\Pbottom$ tagging criteria.
The three jets are interpreted as originating from the hadronic decay of one of the top quarks,
while the other top quark decays semi-leptonically.
One of the two reconstructed leptons is assumed to arise from the misidentification of a $\Pbottom$ hadron
originating from the semi-leptonically decaying top quark.
A kinematic fit using the constraints from kinematic relations between the top quark decay products is employed to increase the purity of semileptonic $\ttbar$+jets events that are correctly reconstructed in this CR.
The level of compatibility of selected events with the aforementioned experimental signature
is quantified using a $\chi^{2}$ criterion; events with a high value of $\chi^{2}$,
corresponding to a poor-quality fit, are discarded.
Good agreement is observed between semileptonic $\ttbar$+jets events where both leptons pass the nominal selection
and semileptonic $\ttbar$+jets events where both leptons pass the relaxed selection, but one or both leptons fail the nominal selection,
provided that the weights given by Eq.~(\ref{eq:FF_weights}) are applied to the latter events by
using the probabilities $f_{i}$ measured in multijet events 
and corrected (for electrons) as described in the previous paragraph.

The MP method is applied in all channels except for $\twoLeptonssOneTau$ and $\threeLeptonOneTau$,
where a modified version of the method is used,
in which only the selections for the leptons are relaxed in the AR, while the $\tauh$ is required to satisfy the nominal selection.
Correspondingly, only the leptons are considered when computing the weights $w$, given by Eq.~(\ref{eq:FF_weights}),
that are applied to events in the AR of the $\twoLeptonssOneTau$ and $\threeLeptonOneTau$ channels.
Background contributions where the reconstructed leptons are genuine prompt leptons
and the reconstructed $\tauh$ is due to the misidentification of a quark or gluon jet are modeled using the MC simulation.
Weights are applied to these simulated events to correct for differences in the $\tauh$ misidentification rates between data and simulation.
Using a modified version of the MP method in the $\twoLeptonssOneTau$ and $\threeLeptonOneTau$ channels
permits the retention as signal of those $\ttH$ and $\tH$ signal events in which the reconstructed $\tauh$ is not a genuine hadronic $\Pgt$ decay,
but arises instead from the misidentification of a quark or gluon jet.
The fraction of $\ttH$ and $\tH$ signal events retained as signal amounts to approximately 30\% of the total $\ttH$ and $\tH$ signal yield
in the $\twoLeptonssOneTau$ and $\threeLeptonOneTau$ channels.

\subsection{Estimation of the ``flips'' background}
\label{sec:backgroundEstimation_flips}

The flips background, relevant for events containing either one or two reconstructed electrons in the $\twoLeptonssZeroTau$ and $\twoLeptonssOneTau$ channels,
is estimated using a procedure similar to the MP method.
A sample of events passing all selection criteria of the SR, except that both leptons are required to be of $\os$ instead of $\ss$, are selected
and assigned appropriately chosen weights.
In the $\twoLeptonssZeroTau$ channel, the weight is given by the sum of the probabilities for the charge of either lepton to be mismeasured,
whereas in the $\twoLeptonssOneTau$ channel, only the lepton that has the same charge as the $\tauh$ is considered,
since only those events in which the charge of this lepton is mismeasured satisfy the condition $\sum\limits_{\Plepton,\tauh} q = \pm 1$ that is applied in the SR of this channel.

\begin{figure*}
  \centering\includegraphics[width=0.49\textwidth]{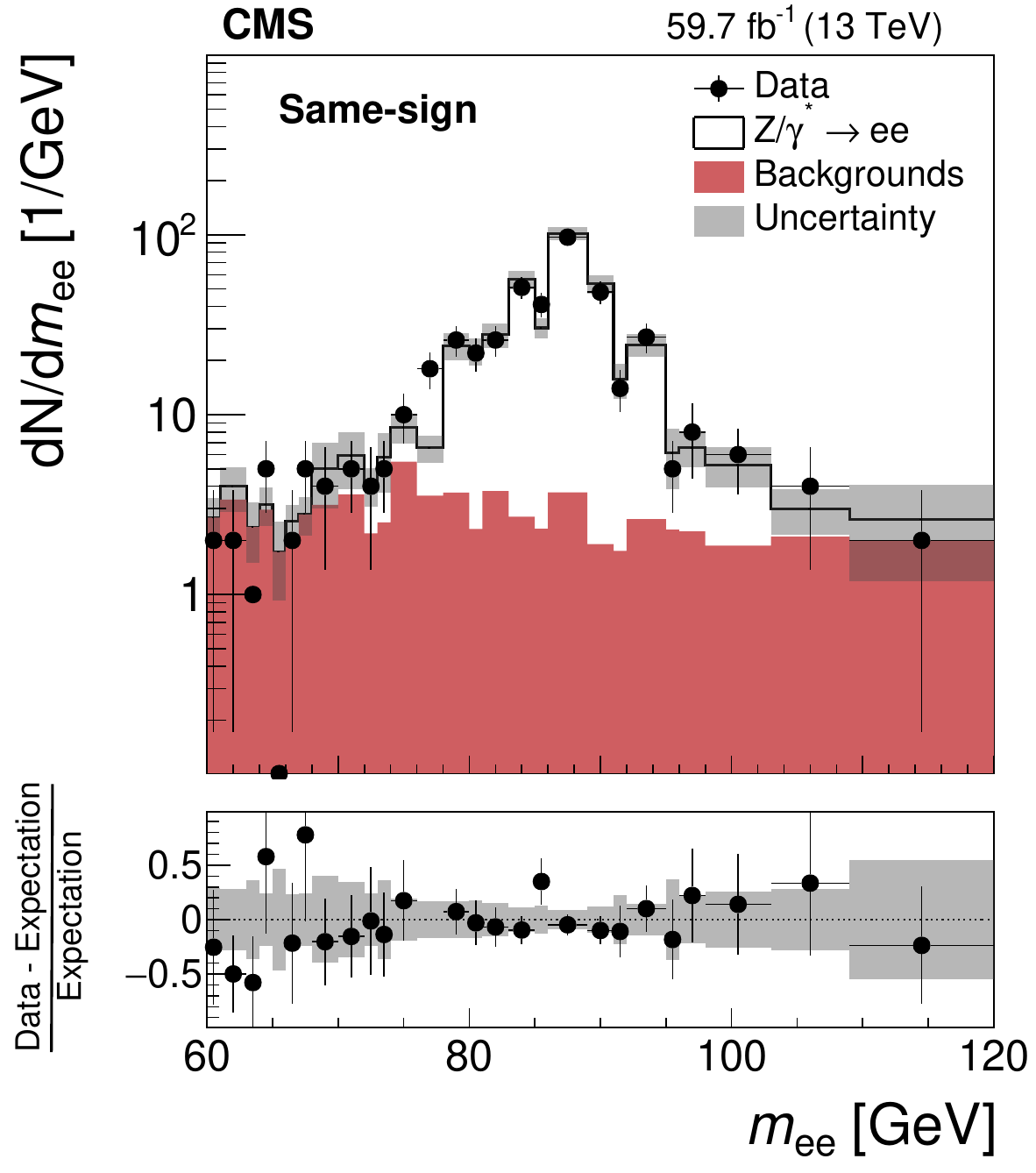}
  \centering\includegraphics[width=0.49\textwidth]{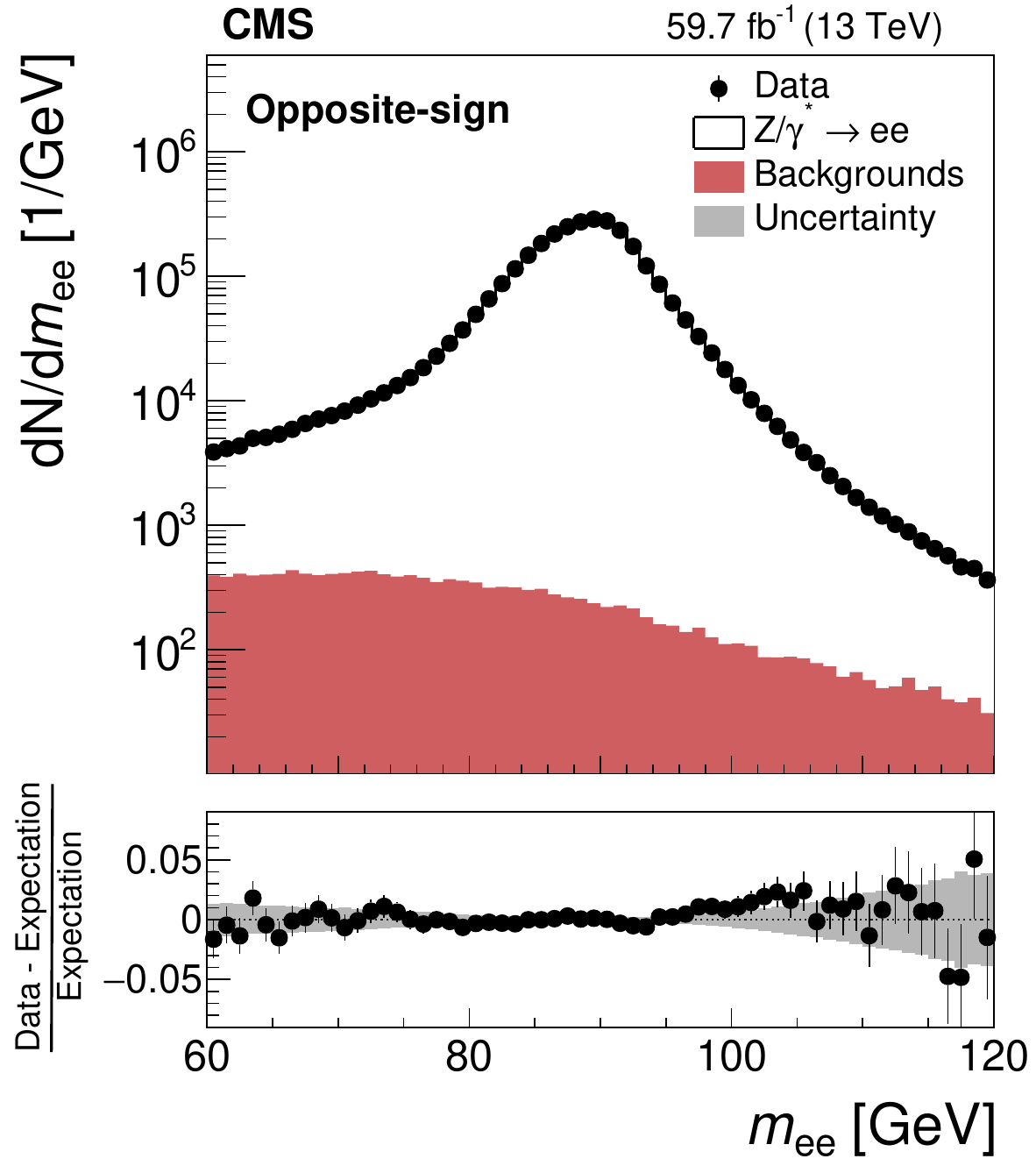}
  \caption{
    Distributions of $m_{\Pe\Pe}$ for (\cmsLeft) SS and (\cmsRight) OS electron pairs
    in $\Zee$ candidate events in which both electrons are in the ECAL barrel and have transverse momenta within the range $25 < \pt < 50\GeV$,
    for data recorded in 2018, compared to the expectation. Uncertainties shown are statistical only. A similar level of agreement is
    present in all the other momentum ranges and data-taking periods.
  }
  \label{fig:controlPlots_flips}
\end{figure*}

The probability for the charge of electrons to be mismeasured, referred to as the electron charge misidentification rate,
is determined using $\Zee$ events.
The events are selected by requiring the presence of an electron pair of invariant mass $m_{\Pe\Pe}$ within the range $60 < m_{\Pe\Pe} < 120\GeV$. No requirement is imposed on the charge of the electron pair.
Contributions to the selected event sample arising from processes other than DY production of electron pairs
are determined by performing an ML fit to the $m_{\Pe\Pe}$ distribution.
Referring to the number of $\Zee$ events containing reconstructed $\ss$ and $\os$ electron pairs, respectively, by the symbols $N_{\ss}$ and $N_{\os}$,
the electron charge misidentification rate is given by the ratio $N_{\ss}/(N_{\os} + N_{\ss})$.
The ratio is measured as a function of electron \pt and $\eta$ and varies between $5.1 \times 10^{-5}$ for electrons of low \pt in the ECAL barrel
and $1.6 \times 10^{-3}$ for electrons of high \pt in the ECAL endcap.
For illustration, the $m_{\Pe\Pe}$ distributions for SS and OS electron pairs are shown in Fig.~\ref{fig:controlPlots_flips}
for events in which both electrons are reconstructed in the ECAL barrel and have \pt within the range $25 < \pt < 50\GeV$.

\subsection{Control regions for irreducible backgrounds}
\label{sec:backgroundEstimation_control_regions}

The accuracy of the simulation-based modeling of the main irreducible backgrounds, 
arising from $\ttWs$, $\ttZ$, $\PW\PZ$, and $\PZ\PZ$ production,
is validated in three CRs.
The first CR is based on the SR for the $\threeLeptonZeroTau$ channel and targets the $\ttZ$ and $\PW\PZ$ backgrounds.
We refer to this CR as the $3\Plepton$-CR.
The selection criteria applied in the $3\Plepton$-CR differ from those applied in the SR of the $\threeLeptonZeroTau$ channel
in that:
no $\PZ$ boson veto is applied in the $3\Plepton$-CR;
the presence of at least one SFOS lepton pair
of invariant mass $m_{\Plepton\Plepton}$ with $\abs{m_{\Plepton\Plepton} - m_{\PZ}} < 10\GeV$ is demanded instead;
the requirement on the multiplicity of jets is relaxed to demanding the presence of at least one jet;
and no requirement on the presence of $\Pbottom$-tagged jets is applied.
The contributions arising from $\ttZ$ and from $\PW\PZ$ production 
are separated by binning the events selected in the $3\Plepton$-CR
in the flavor of the three leptons ($\Pe\Pe\Pe$, $\Pe\Pe\Pgm$, $\Pe\Pgm\Pgm$, $\Pgm\Pgm\Pgm$) 
and in the multiplicity of jets and of $\Pbottom$-tagged jets.
The second CR targets the $\PZ\PZ$ background. 
We refer to it as the $4\Plepton$-CR, since it is based on the SR for the $\fourLeptonZeroTau$ channel.
Compared to the latter, the event selection criteria applied in the $4\Plepton$-CR
are modified by applying no $\PZ$ veto, instead requiring the presence of at least one SFOS lepton pair
of invariant mass $m_{\Plepton\Plepton}$ with $\abs{m_{\Plepton\Plepton} - m_{\PZ}} < 10\GeV$,
and applying no requirements on the multiplicity of jets and of $\Pbottom$-tagged jets.
To separate the $\PZ\PZ$ background from other backgrounds, predominantly arising from $\ttZ$ production,
the events selected in the $4\Plepton$-CR are binned in the multiplicity of SFOS lepton pairs of invariant mass $\abs{m_{\Plepton\Plepton} - m_{\PZ}} < 10\GeV$
and in the number of jets passing tight $\Pbottom$ tagging criteria.
The third CR targets the $\ttWs$ background and is identical to the SR of the $\twoLeptonssZeroTau$ channel,
except that the output node of the ANN that has the highest activation value is required to be the output node corresponding to the $\ttW$ background.

The numbers of events observed in the $3\Plepton$- and $4\Plepton$-CRs and in the CR for the $\ttWs$ background
are given in Table~\ref{tab:eventYields_control_regions}.
The contributions arising from the misidentified leptons and flips backgrounds
are estimated using the methods described in Sections~\ref{sec:backgroundEstimation_fakes} and~\ref{sec:backgroundEstimation_flips}, respectively.
The uncertainties include both statistical and systematic sources, added in quadrature.
The systematic uncertainties that are relevant for the CRs are similar to the ones applied to the SR. 
The latter are detailed in Section~\ref{sec:systematicUncertainties}.

\begin{table*}[h!]
  \centering
  \topcaption{
    Number of events selected in the $3\Plepton$- and $4\Plepton$-CRs and in the CR for the $\ttWs$ background,
    compared to the event yields expected from different types of background and from the $\ttH$ and $\tH$ signals, after the fit to data
    is performed as described in Section~\ref{sec:results}. Uncertainties shown include all systematic components.
    The symbol ``\NA'' indicates that the corresponding background does not apply.
  }
  \label{tab:eventYields_control_regions}
    \begin{tabular}{lr@{ $\pm$ }lr@{ $\pm$ }lr@{ $\pm$ }l}
      \hline
      Process                          & \multicolumn{2}{c}{$3\Plepton$-CR} & \multicolumn{2}{c}{$4\Plepton$-CR} & \multicolumn{2}{c}{$\ttWs$ CR}                           \\
      \hline
      $\ttH$                           & $15.9 $                            & $  4.4 $                           & $ 1.4 $                 & $  0.4 $ & $62 $    & $ 14 $   \\
      $\tH$                            & $ 4.4 $                            & $  3.0 $                           & \multicolumn{2}{c}{\NA} & $22 $    & $ 18 $              \\
      $\ttZ + \ttbar\Pggx$             & $550 $                             & $ 43 $                             & $41.5 $                 & $  3.0 $ & $100.3 $ & $  8.1 $ \\
      $\ttW + \ttWW$                   & $26.8 $                            & $  1.7 $                           & \multicolumn{2}{c}{\NA} & $588 $   & $ 35 $              \\
      $\PW\PZ$                         & $4320 $                            & $ 120 $                            & \multicolumn{2}{c}{\NA} & $51.6 $  & $  7.5 $            \\
      $\PZ\PZ$                         & $298 $                             & $ 18 $                             & $1030 $                 & $ 32 $   & $ 0.2 $  & $  0.1 $ \\
      Nonprompt leptons                & $210 $                             & $ 20 $                             & \multicolumn{2}{c}{\NA} & $102 $   & $ 14 $              \\
      Flips                            & \multicolumn{2}{c}{\NA}            & \multicolumn{2}{c}{\NA}            & $24.9 $                 & $  4.0 $                       \\
      Rare backgrounds                 & $311 $                             & $ 61 $                             & $17.0 $                 & $  3.4 $ & $58 $    & $ 13 $   \\
      Conversions                      & $ 1.0 $                            & $  0.3 $                           & $ 0.1 $                 & $  0.1 $ & $ 1.4 $  & $  0.6 $ \\
      $\ggH + \qqH + \VH + 	\ttVH$ & $42.8 $                            & $  3.1 $                           & $ 5.8 $                 & $  0.4 $ & $ 1.6 $  & $  0.3 $ \\
      Total expected background        & $5761 $                            & $ 99 $                             & $1094 $                 & $ 33 $   & $949 $   & $ 33 $   \\
      Data                             & \multicolumn{2}{c}{5778}           & \multicolumn{2}{c}{1089}           & \multicolumn{2}{c}{986}                                  \\
      \hline
    \end{tabular}
\end{table*}

Figure~\ref{fig:controlPlots_3leptonCR_and_4leptonCR}, discussed in Section~\ref{sec:results}, shows the distributions of events selected in the $3\Plepton$- and $4\Plepton$-CRs 
in the binning scheme employed to separate the $\PW\PZ$ and $\PZ\PZ$ backgrounds from the $\ttZ$ backgrounds.
The events selected in the $3\Plepton$-CR are first subdivided by lepton flavor and then by the multiplicity of jets and $\Pbottom$-tagged jets.
For each lepton flavor, $12$ bins are used, defined as follows (in order of increasing bin number):
$0$ jets passing the tight $\Pbottom$ tagging criteria with $1$, $2$, $3$, or $\geq$4 jets in total; 
$1$ jet passing the tight $\Pbottom$ tagging criteria  with $2$, $3$, $4$, or $\geq$5 jets in total;
$\geq$2 jets passing the tight $\Pbottom$ tagging criteria with $2$, $3$, $4$, or $\geq$5 jets in total. 
In the $4\Plepton$-CR, $4$ bins are used in total, defined as (again in order of increasing bin number):
$2$ SFOS lepton pairs of invariant mass $\abs{m_{\Plepton\Plepton} - m_{\PZ}} < 10\GeV$;
$1$ such SFOS lepton pair with $0$, $1$, or $\geq$2 jets passing the tight $\Pbottom$ tagging criteria.

The data in the $3\Plepton$- and $4\Plepton$-CRs and in the CR for the $\ttWs$ background
are in agreement with the background estimates within the quoted uncertainties.

\section{Systematic uncertainties}
\label{sec:systematicUncertainties}

The event rates and the distributions of the discriminating observables used for signal extraction may
be altered by several experiment- or theory-related effects, referred to as systematic uncertainties.
Experimental sources comprise the uncertainties in auxiliary measurements, 
performed to validate and, if necessary, correct the modeling of the data by the MC simulation,
and the uncertainties in the data-driven estimates of the misidentified leptons and flips backgrounds.
The latter are largely unaffected by potential inaccuracies of the MC simulation.
Theoretical uncertainties mainly arise from missing higher-order corrections to the perturbative expansions employed for the computation of cross sections
and from uncertainties in the PDFs.

The efficiencies of triggers based on the presence of one, two, or three electrons or muons
are measured as a function of the lepton multiplicity with an uncertainty ranging from 1 to 2\%,
using samples of $\ttbar$+jets and diboson events that have been recorded using triggers based on $\ptmiss$.

The efficiencies for electrons and muons to pass the offline reconstruction and identification criteria 
are measured as a function of the lepton \pt and $\eta$ by applying the ``tag-and-probe'' method detailed in Ref.~\cite{Khachatryan:2010xn} to $\Zee$ and $\Zmm$ events. 
Additionally, we cross-check these efficiencies in a CR enriched in $\ttbar$+jets events
to account for differences in event topology between DY events and the events in the SR of this analysis,
which may cause a change in the efficiencies for electrons and muons to pass isolation requirements. 
Events in the $\ttbar$+jets CR are selected by requiring the presence of an OS $\Pe$+$\Pgm$ pair and at least two jets.
Nonprompt-lepton backgrounds in the CR are subtracted using a sideband region SS $\Pe$+$\Pgm$ events.
The difference between the efficiency measured in the $\ttbar$+jets CR and the one measured in DY events
is included as a systematic uncertainty, amounting to 1--2\%.
The $\tauh$ identification efficiency and energy scale are measured with respective uncertainties of 5 and 1.2\%
using $\Ztt$ events~\cite{Sirunyan:2018pgf}.

The energy scale of jets is measured with an uncertainty amounting to a few percent, depending on the jet \pt and $\eta$,
using the \pt-balance method, which is applied to $\Zee$, $\Zmm$, $\Pgg$+jets, dijet, and multijet events~\cite{Khachatryan:2016kdb}.
The resulting effect on signal and background expectations is evaluated by varying the energies of jets in simulated events within their uncertainties,
recalculating all kinematic observables, and reapplying the event selection criteria.
The effect of uncertainties in the jet energy resolution is evaluated in a similar way, but is smaller than the effect of the uncertainties in the jet energy scale.

The $\Pbottom$ tagging efficiency is measured with an uncertainty of a few per cent in $\ttbar$+jets and multijet events as a function of jet \pt and $\eta$.
The heavy-flavor content of the multijet events is enriched by requiring the presence of a muon in the event.
The mistag rates for light-quark and gluon jets are measured in multijet events
yielding an uncertainty of 5--10\% for the loose and 20--30\% for the tight $\Pbottom$ tagging criteria,
depending on \pt and $\eta$~\cite{Sirunyan:2017ezt}.

The integrated luminosities of the 2016, 2017, and 2018 data-taking periods are individually known with uncertainties in the 2.3--2.5\% range~\cite{LUM-17-001,LUM-17-004,LUM-18-002},
while the total Run~2 (2016--2018) integrated luminosity has an uncertainty of 1.8\%, the improvement in precision reflecting the (uncorrelated) time evolution of some systematic effects.

The uncertainties related to the number of PU interactions are evaluated 
by varying the number of inelastic $\Pp\Pp$ interactions that are superimposed on simulated events by 4.6\%~\cite{Sirunyan:2018nqx}.
The resulting effect on the $\ttH$ and $\tH$ signal yields 
and on the yields of background contributions modeled using the MC simulation amounts to less than 1\%.

The effect of theory-related uncertainties on the event yields and on the distributions of the BDTs and ANNs classifier outputs that are used for the signal extraction
is assessed for the $\ttH$ and $\tH$ signals, as well as for the main irreducible backgrounds that arise from $\ttW$, $\ttWW$, and $\ttZ$ production.
The uncertainties in the production cross sections amount to $^{+6.8}_{-9.9}$ and $^{+5.1}_{-7.3}\%$ for the $\ttH$ and $\tH$ signals,
and to $^{+13.5}_{-12.2}$, $^{+8.6}_{-11.3}$, and $^{+11.7}_{-10.2}\%$ for the $\ttW$, $\ttWW$, and $\ttZ$ backgrounds, respectively.
These uncertainties are taken from Ref.~\cite{deFlorian:2016spz} and consist of the sum in quadrature of three sources:
missing higher-order corrections in the perturbative expansion, different choices of PDFs, 
and uncertainties in the value of the strong coupling constant $\alpS$.
The uncertainties in the cross sections are relevant 
for the purpose of quoting the measured production rates with respect to their SM expectations for these rates.
In addition, the uncertainty in the $\ttH$ and $\tH$ production cross sections is relevant
for setting limits on the coupling of the Higgs boson to the top quark.
The effect of missing higher-order corrections on the distributions of the discriminating observables
is estimated by varying the renormalization and factorization scales up and down by a factor of two with respect to their nominal value,
following the recommendations of Refs.~\cite{Cacciari:2003fi,Catani:2003zt,Frederix:2011ss}, avoiding cases in which the two variations are done in opposite directions.
The effect of uncertainties in the PDFs on these distributions is evaluated following the recommendations given in Ref.~\cite{Butterworth:2015oua}.
The uncertainties in the branching fractions of the Higgs boson decay modes
$\PHiggs \to \PW\PW$, $\PHiggs \to \Pgt\Pgt$, and $\PHiggs \to \PZ\PZ$ are taken from Ref.~\cite{deFlorian:2016spz}
and amount to 1.5, 1.7, and 1.5\%, respectively.

In the $\oneLeptonOneTau$ and $\zeroLeptonTwoTau$ channels,
the $\ttbar$+jets and DY production may contribute as irreducible backgrounds
and are modeled using the MC simulation.
The $\ttbar$+jets and DY production cross sections are known to an uncertainty of 5~\cite{Czakon:2011xx} and 4\%~\cite{Gavin:2010az}, respectively.
An additional uncertainty on the modeling of top quark \pt distribution of $\ttbar$+jets events is 
considered, defined as the difference between the nominal \POWHEG sample and that sample reweighed to improve the quality of the top quark \pt modeling, as described in Section~\ref{sec:datasamples_and_MonteCarloSimulation}.
The modeling of the multiplicity of jets and of $\Pbottom$-tagged jets in simulated DY events
is improved by comparing these multiplicities between MC simulation and data using $\Zee$ and $\Zmm$ events.
The average ratio of data and MC simulation in the $\Zee$ and $\Zmm$ event samples is taken as a correction,
while the difference between the ratios measured in $\Zee$ and $\Zmm$ events is taken as the systematic uncertainty
and added in quadrature to the statistical uncertainties in these ratios.
The $\Zee$ and $\Zmm$ event samples used to determine this correction 
have little overlap with the SRs of the $\oneLeptonOneTau$ and $\zeroLeptonTwoTau$ channels,
since most of the DY background in these channels arises from $\Ztt$ events.

Other background processes, notably the conversions and rare backgrounds, are modeled using the MC simulation;
the uncertainty in their event yields is conservatively taken to be 50\%.
This choice accounts for the extrapolation from the inclusive phase space to the phase space relevant for this analysis,
in particular to events with a high multiplicity of jets and $\Pbottom$-tagged jets, as required to pass
the event selection criteria detailed in Section~\ref{sec:eventSelection}.
The inclusive cross sections for most of these background processes have been measured with uncertainties amounting to significantly less than 50\%
by previous analyses of the LHC data.

The extrapolation of the $\PW\PZ$ and $\PZ\PZ$ background rates from the $3\Plepton$- and $4\Plepton$-CRs to the SR depends on the heavy-flavor content of $\PW\PZ$ and $\PZ\PZ$ background events.
According to the MC simulation, most of the $\Pbottom$ jets reconstructed in $\PW\PZ$ and $\PZ\PZ$ background events
arise from the misidentification of light-quark or gluon jets rather than from charm or bottom quarks.
We assign an uncertainty of 40\% to the modeling of the heavy-flavor content in $\PW\PZ$ and $\PZ\PZ$ background events,
accounting for the differences in the jet multiplicity distribution between data and simulation in the $3\Plepton$ CR.
The misidentification of light quark or gluon jets as $\Pbottom$ jets is covered by a separate systematic uncertainty.

The uncertainties in the rate and in the distribution of the discriminating observables for the background from misidentified leptons and $\tauh$
stem from statistical uncertainties in the events selected in the MR and AR
as well as from systematic uncertainties related to the subtraction of the prompt-lepton contributions from the data selected in the MR and AR of the MP method.
The effect of these uncertainties on the analysis 
is evaluated by applying independent variations of the probabilities $f_{i}$ for electrons and muons in different bins of lepton-candidate \pt and $\eta$
and determining the resulting change in the yield and distribution of the misidentified leptons background estimate.
We introduce an additional uncertainty in the nonclosure correction to the $f_{i}$ for electrons and muons,
accounting for differences between the probabilities $f_{i}$ in $\ttbar$+jets and multijet events shown in Fig.~\ref{fig:closureTest_fakes}.
The size of this uncertainty is equal to the magnitude of the correction.
In case of $\tauh$, the misidentification rates $f_{i}$ measured in each bin in $\eta$ and reconstructed $\tauh$ decay mode
are fitted by a linear function in \pt of the $\tauh$ candidate and the uncertainty in the slope and offset of this fit is propagated to the final result.
The uncertainty in the rate of the misidentified leptons background is, in general, higher for channels with $\tauh$.
The uncertainty varies between 10\% in the $\twoLeptonssZeroTau$ channel and 60\% in the $\twoLeptonTwoTau$ channel.
The resulting uncertainty in the distribution of the discriminating observables is of moderate size.
Additional nonclosure uncertainties account for small differences between the misidentified leptons background estimate
obtained by computing the probabilities $f_{i}$ for simulated events and applying the weights $w$ given by Eq.~(\ref{eq:FF_weights}) to simulated events selected in the AR,
and the background estimates obtained by modeling the background from misidentified leptons and $\tauh$ in the SR using the MC simulation directly.

The uncertainty in the flips background in the $\twoLeptonssZeroTau$ and $\twoLeptonssOneTau$ channels is evaluated in a similar way:
it amounts to 30\% in each channel.

The effects of systematic uncertainties representing the same source are treated as fully correlated between all ten analysis channels.
Theoretical uncertainties are furthermore treated as fully correlated among all data-taking periods,
whereas the uncertainties arising from experimental sources are treated as uncorrelated between the data recorded in each of the years 2016, 2017, and 2018.
The latter treatment is justified by the fact that the uncertainties related to the auxiliary measurements that are performed to validate, and if necessary correct,
the modeling of the data by the MC simulation,
are mainly of statistical origin and hence independent for measurements
that are performed independently for each of the three data-taking periods
because of the changes in the detector conditions from one period to another.

The impact of the systematic and statistical uncertainties on the measurement of the $\ttH$ and $\tH$ signal rates 
is summarized in Table~\ref{tab:systematicUncertainties_impact}.
The largest impacts are due to: the statistical uncertainty of observed data; the uncertainty in the efficiency to reconstruct and identify \tauh;
the uncertainties related to the estimation of the misidentified leptons and flips backgrounds;
and the theoretical uncertainties, which affect the yield and the distribution of the discriminating observables for the $\ttH$ and $\tH$ signals as well as for the main irreducible backgrounds,
arising from $\ttW$, $\ttWW$, $\tW$, $\ttZ$, and $\tZ$ production.

\begin{table*}[h!]
  \centering
  \topcaption{
    Summary of the sources of systematic and statistical uncertainties and their impact on the measurement of the $\ttH$ and $\tH$ signal rates, and the measured value of the unconstrained nuisance parameters.
    The quantity $\Delta\r_{x}/\r_{x}$ corresponds to the change in uncertainty when fixing the nuisance parameters associated 
    with that uncertainty in the fit. Under the label ``MC and sideband statistical uncertainty'' are the uncertainties associated
    with the limited number of simulated MC events and the amount of data events in the application region of the 
    MP method.
  }
  \label{tab:systematicUncertainties_impact}
  \cmsTable{
    \renewcommand{\arraystretch}{1.5}
    \begin{tabular}{lcccc}
      \hline
      Source                                                                 & $\Delta\r_{\ttH}/\r_{\ttH}$  [\%] & $\Delta\r_{\tH}/\r_{\tH}$ [\%] & $\Delta\r_{\ttW}/\r_{\ttW}$  [\%] & $\Delta\r_{\ttZ}/\r_{\ttZ}$  [\%] \\ \hline 
      Trigger efficiency                                                     & 2.3                               & 8.1                            & 1.2                               & 1.9                               \\  
      \Pe, \Pgm reconstruction and identification efficiency                 & 2.9                               & 7.1                            & 1.7                               & 3.2                               \\
      \tauh identification efficiency                                        & 4.6                               & 9.1                            & 1.7                               & 1.3                               \\
      $\Pbottom$ tagging efficiency and mistag rate                          & 3.6                               & 13.6                           & 1.3                               & 2.9                               \\
      Misidentified leptons and flips                                        & 6.0                               & 36.8                           & 2.6                               & 1.4                               \\
      Jet energy scale and resolution                                        & 3.4                               & 8.3                            & 1.1                               & 1.2                               \\
      MC sample and sideband statistical uncertainty                         & 7.1                               & 27.2                           & 2.4                               & 2.3                               \\
      Theory-related sources affecting acceptance                            & \multirow{2}{*}{4.6}              & \multirow{2}{*}{18.2}          & \multirow{2}{*}{2.0}              & \multirow{2}{*}{4.2}              \\
      and shape of distributions                                             &                                   &                                &                                   &                                   \\
      Normalization of MC-estimated processes                                & 13.3                              & 12.3                           & 13.9                              & 11.3                              \\
      Integrated luminosity                                                  & 2.2                               & 4.6                            & 1.8                               & 3.1                               \\ [\cmsTabSkip]
      Statistical uncertainty                                                & 20.9                              & 48.0                           & 5.9                               & 5.8                               \\
      \hline
    \end{tabular}
  }
\end{table*}

\subsection{Additional checks}
\label{sub:addchecks}
As a cross-check, and to highlight the enhancement in sensitivity provided by machine-learning techniques,
a complementary measurement of the $\ttH$ signal rate is performed using a set of alternative observables in the ML fit.
We refer to this cross-check as the control analysis, as distinguished from the analysis previously discussed, which we refer to as the main analysis.
The control analysis (CA) is restricted to the $\twoLeptonssZeroTau$, $\threeLeptonZeroTau$, $\twoLeptonssOneTau$, and $\fourLeptonZeroTau$ channels.
The production rate of the $\tH$ signal is fixed to its SM expectation in the CA.
In the $\twoLeptonssZeroTau$ channel, the invariant mass of the lepton pair is used as the discriminating observable.
The event selection criteria applied in the CA in this channel are modified to the condition $\njets \geq 4$
and the events are analyzed in subcategories based on lepton flavor, the charge-sum of the leptons ($+2$ or $-2$), and the multiplicity of jets.
In the $\threeLeptonZeroTau$ channel, the invariant mass of the three-lepton system is used as discriminating observable
and the events are analyzed in subcategories based on the multiplicity of jets and on the charge-sum of the leptons ($+1$ or $-1$).
A discriminant based on the matrix-element method~\cite{Kondo:1988yd,Kondo:1991dw} is used as discriminating observable in the $\twoLeptonssOneTau$ channel
and the events are analyzed in two subcategories based on the multiplicity of jets, defined by the conditions $\njets = 3$ and $\njets \geq 4$,
and referred to as the ``missing-jet'' and ``no-missing-jet'' subcategories.
The computation of the discriminant exploits the fact that the differential cross sections for the $\ttH$ signal, 
as well as for the dominant background processes in the $\twoLeptonssOneTau$ channel, are well known;
this permits the computation of the probabilities for a given event to be either signal or background,
given the measured values of kinematic observables in the event and taking into account the experimental resolution of the detector. 
The probabilities are computed for the $\ttH$ signal hypothesis and for three types of background hypotheses:
$\ttZ$ events in which the $\PZ$ boson decays into a pair of $\Pgt$ leptons;
$\ttZ$ events in which the $\PZ$ boson decays into a pair of electrons or muons and one lepton is misidentified as $\tauh$;
and $\Ptop\APtop \to \Pbottom\Plepton\Pnu \, \APbottom\Pgt\Pnu$ events with one additional nonprompt lepton originating from a $\Pbottom$ hadron decay.
Details on the computation of these probabilities are given in Ref.~\cite{Sirunyan:2018shy}.
The ratio of the probability for a given event to be $\ttH$ signal to the sum of the probabilities for the event to be one of the three backgrounds
constitutes, according to the Neyman-Pearson lemma~\cite{Neyman:1937uhy}, an optimal observable for the purpose of separating the $\ttH$ signal from backgrounds
and is taken as the discriminant used for the signal extraction.
In the $\fourLeptonZeroTau$ channel, the invariant mass of the four-lepton system, $m_{4\Plepton}$, is used as the discriminating observable.

\section{Statistical analysis and results}
\label{sec:results}

The production rates of the $\ttH$ and $\tH$ signals are determined through a binned simultaneous ML fit
to the total of $105$ distributions:
the outputs of the BDTs in each of the seven channels
$\oneLeptonOneTau$, $\zeroLeptonTwoTau$,
$\twoLeptonosOneTau$, $\oneLeptonTwoTau$,
$\fourLeptonZeroTau$, $\threeLeptonOneTau$, and $\twoLeptonTwoTau$;
the distributions of the 10 output nodes of the ANNs in the $\twoLeptonssZeroTau$, $\threeLeptonZeroTau$, and $\twoLeptonssOneTau$ channels in the categories described in Fig.~\ref{fig:diagram};
and the distributions of the observables that discriminate the $\ttZ$ background from each of the $\PW\PZ$ and $\PZ\PZ$ backgrounds in the $3\Plepton$- and $4\Plepton$-CRs, respectively;
separately for the three data-taking periods considered in the analysis.
The $\twoLeptonssZeroTau$ ($\threeLeptonZeroTau$) channel contributes a total of $12$ ($11$) distributions per data-taking period to the ML fit,
reflecting the subdivision of these channels into event categories based on lepton flavor and on the multiplicity of $\Pbottom$-tagged jets.

The production rates of the $\ttH$ and $\tH$ signals constitute the parameters of interest (POI) in the fit.
We denote by the symbols $\r_{\ttH}$ and $\r_{\tH}$ the ratio of these production rates to their SM expectation
and use the notation $\vecr$ to refer to the set of both POIs.

The likelihood function is denoted by the symbol $\mathcal{L}$ and is given by the expression:
\begin{equation}
\mathcal{L}\left(\text{data} \, \vert \, \vecr, \vectheta \right)
= \prod_{i} \, \mathcal{P}\left(n_{i} \vert \, \vecr, \vectheta \right) \, \prod_{k} \, \text{p}\left(\tilde{\theta}_{k} \vert \theta_{k}\right)\,,
\label{eq:likelihoodFunction}
\end{equation}
where the index $i$ refers to individual bins of the $105$ distributions of the discriminating observables that are included in the fit,
and the factor $\mathcal{P}\left(n_{i} \vert \, \vecr, \vectheta \right)$ represents the probability
to observe $n_{i}$ events in a given bin $i$, where $\nu_{i}(\vecr, \vectheta)$ events are expected from the sum of signal and background contributions in that bin.
The number of expected events is a linear function of the two POIs indicated by $\r_{\ttH}$ and  $\r_{\tH}$
\begin{equation}
\nu_{i}(\vecr, \vectheta) = \r_{\ttH} \nu_{i}^{\ttH}(\vectheta) + \r_{\tH} \nu_{i}^{\tH}(\vectheta) + \nu_{i}^{\mathrm{B}}(\vectheta)\,,
\end{equation}
where the symbols $\nu_{i}^{\ttH}$, $\nu_{i}^{\tH}$, and $\nu_{i}^{\mathrm{B}}$ denote, respectively,
the SM expectation for the $\ttH$ and $\tH$ signal contributions and the aggregate of contributions expected from background processes in bin $i$.
We use the notation $\nu_{i}(\vecr, \vectheta)$ to indicate that the number of events expected from signal and background processes in each bin $i$
depends on a set of parameters, denoted by the symbol $\vectheta$,
that represent the systematic uncertainties detailed in Section~\ref{sec:systematicUncertainties}
and are referred to as nuisance parameters.
Via the dependency of the $\nu_{i}(\vecr, \vectheta)$ on $\vectheta$,
the nuisance parameters accommodate for variations of the event yields as well as of the distributions of the discriminating observables during the fit.
The probability $\mathcal{P}\left(n_{i} \vert \, \vecr, \vectheta \right)$ is given by the Poisson distribution:
\begin{equation}
\mathcal{P}\left(n_{i} \vert \, \vecr, \vectheta \right) = \frac{\left(\nu_{i}(\vecr, \vectheta)\right)^{n_{i}}}{n_{i}!} \, \exp\left(-\nu_{i}(\vecr, \vectheta)\right)\,.
\end{equation}
Individual elements of the set of nuisance parameters $\vectheta$ are denoted by the symbol $\theta_{k}$,
where each $\theta_{k}$ represents a specific source of systematic uncertainty.
The function $\mathrm{p}\left(\tilde{\theta}_{k} \vert \theta_{k}\right)$ represents the probability to
observe a value $\tilde{\theta}_{k}$ in an auxiliary measurement of the nuisance parameter,
given that its true value is $\theta_{k}$.
Systematic uncertainties that affect only the normalization, but not the shape of the distribution of the discriminating observables,
are represented by a Gamma probability density function if they are statistical in origin,
\eg if they correspond to the number of events observed in a CR,
and otherwise by a log-normal probability density function;
systematic uncertainties that also affect the shape of distributions of the discriminating observables
are incorporated into the ML fit via the technique detailed in Ref.~\cite{Conway:2011in}
and represented by a Gaussian probability density function.

The rates of the $\ttW$ and $\ttZ$ backgrounds are separately left unconstrained  in the fit.
The rate of the small $\ttWW$ background is constrained to scale by the same factor
with respect to its SM expectation as the rate of the $\ttW$ background.

Statistical fluctuations in the background predictions arise because of a limited number of events in the MC simulation
as well as in the ARs that are used to estimate the misidentified leptons and flips backgrounds from data.
These fluctuations are incorporated into the likelihood function via the approach described in Ref.~\cite{Barlow:1993dm}.

Further details concerning the treatment of systematic uncertainties and concerning the choice of the functions $\mathrm{p}(\tilde{\theta}_{k} \vert \theta_{k})$
are given in Refs.~\cite{ATL-PHYS-PUB-2011-011,HIG-11-032,Conway:2011in}.

A complication in the signal extraction arises from the fact that a deviation in the top quark Yukawa coupling $\yt$ with respect to the SM expectation $m_{\Ptop}/v$
would change the distribution of kinematic observables for the $\tH$ signal
and alter the proportion between the $\tH$ and $\ttH$ signal rates.
We address this complication by first determining the production rates for the $\tH$ and $\ttH$ signals,
assuming that the distributions of kinematic observables for the $\tH$ signal conform to the distributions expected in the SM;
we then determine the Yukawa coupling $\yt$ of the Higgs boson to the top quark,
accounting for modifications in the interference effects for the $\tH$ signal. These studies assume a Higgs boson mass of
125 \GeV.

\begin{figure*}[h!]
  \centering\includegraphics[width=0.4\textwidth]{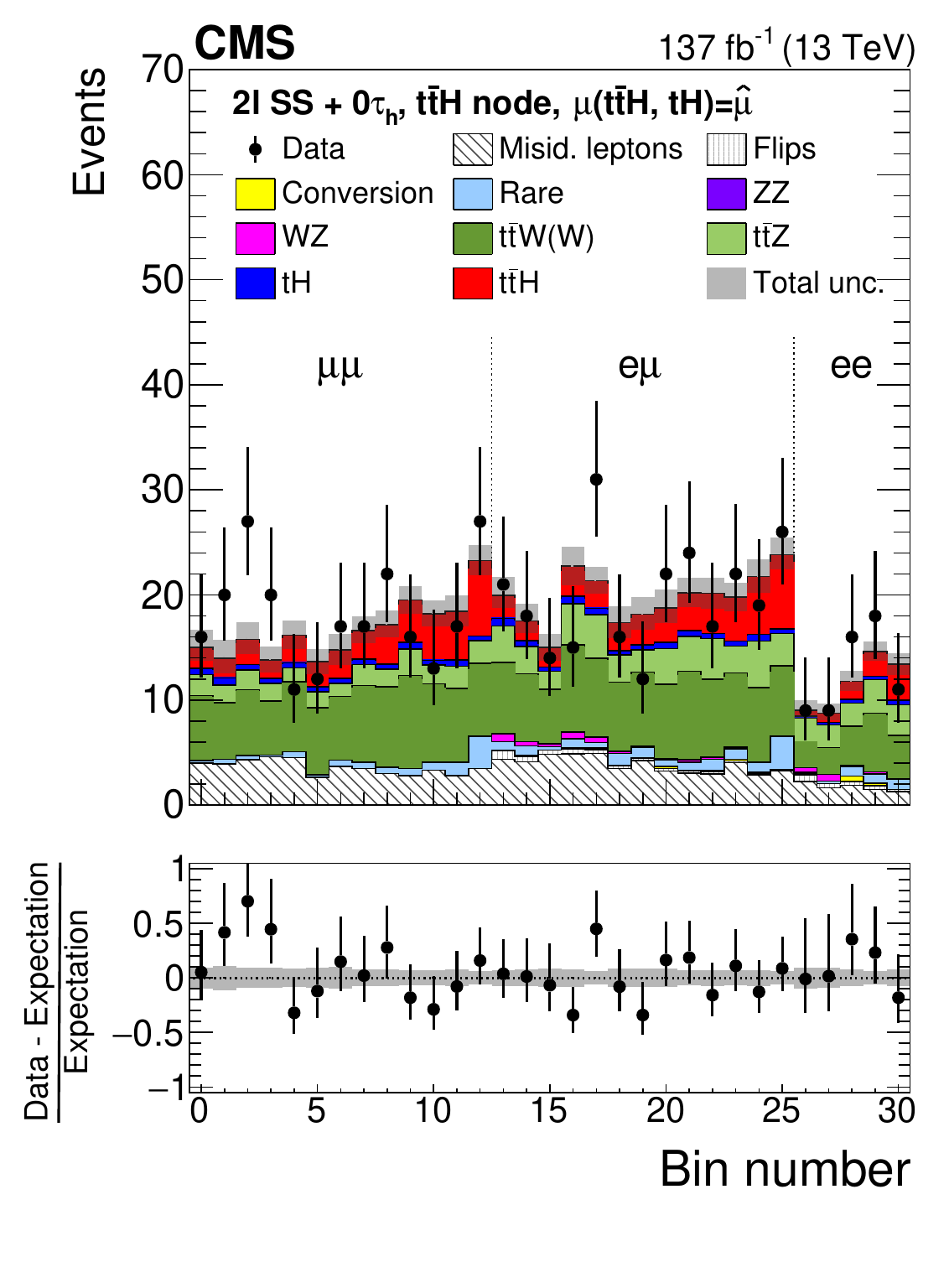}
  \centering\includegraphics[width=0.4\textwidth]{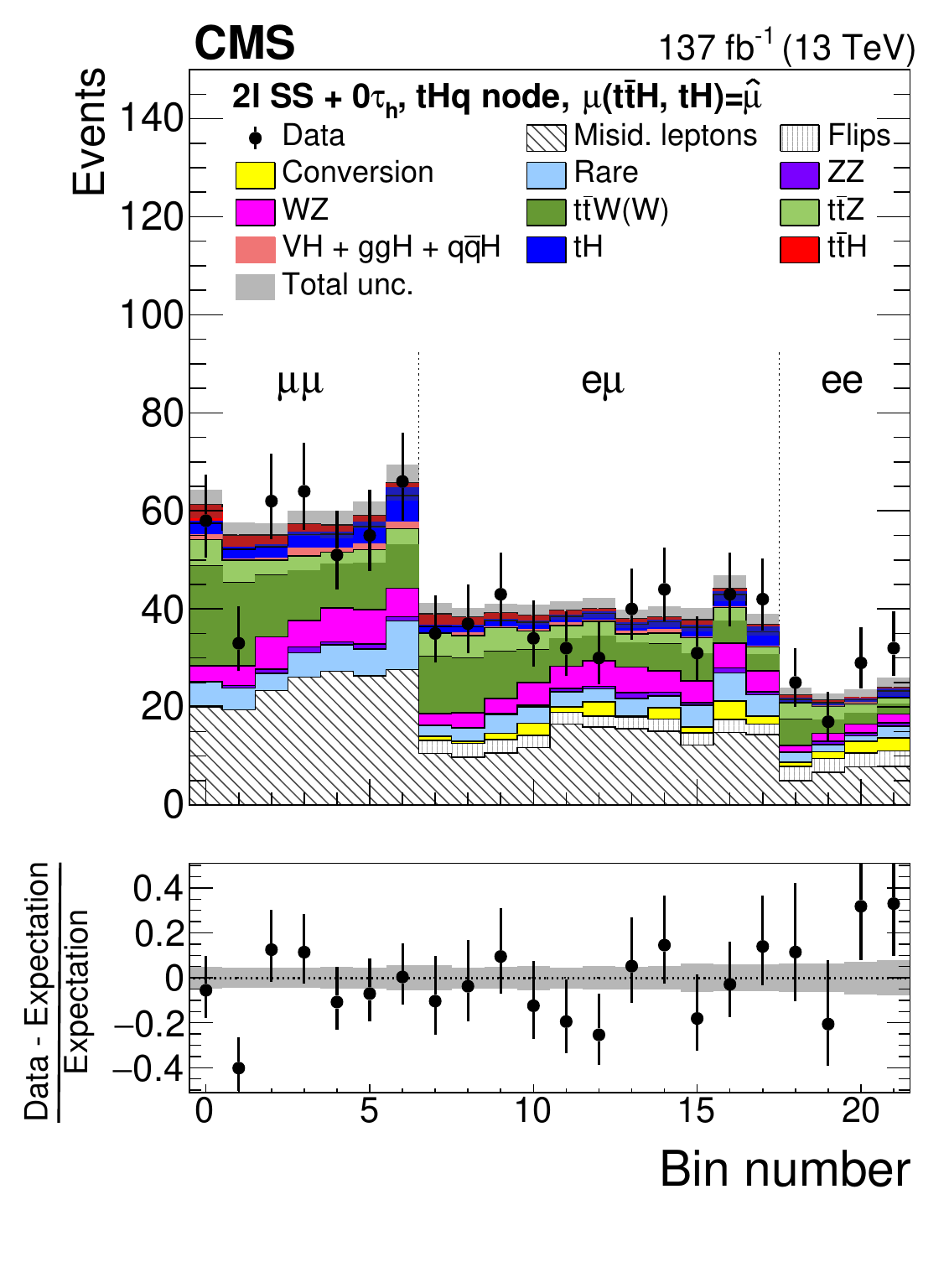}
  \centering\includegraphics[width=0.4\textwidth]{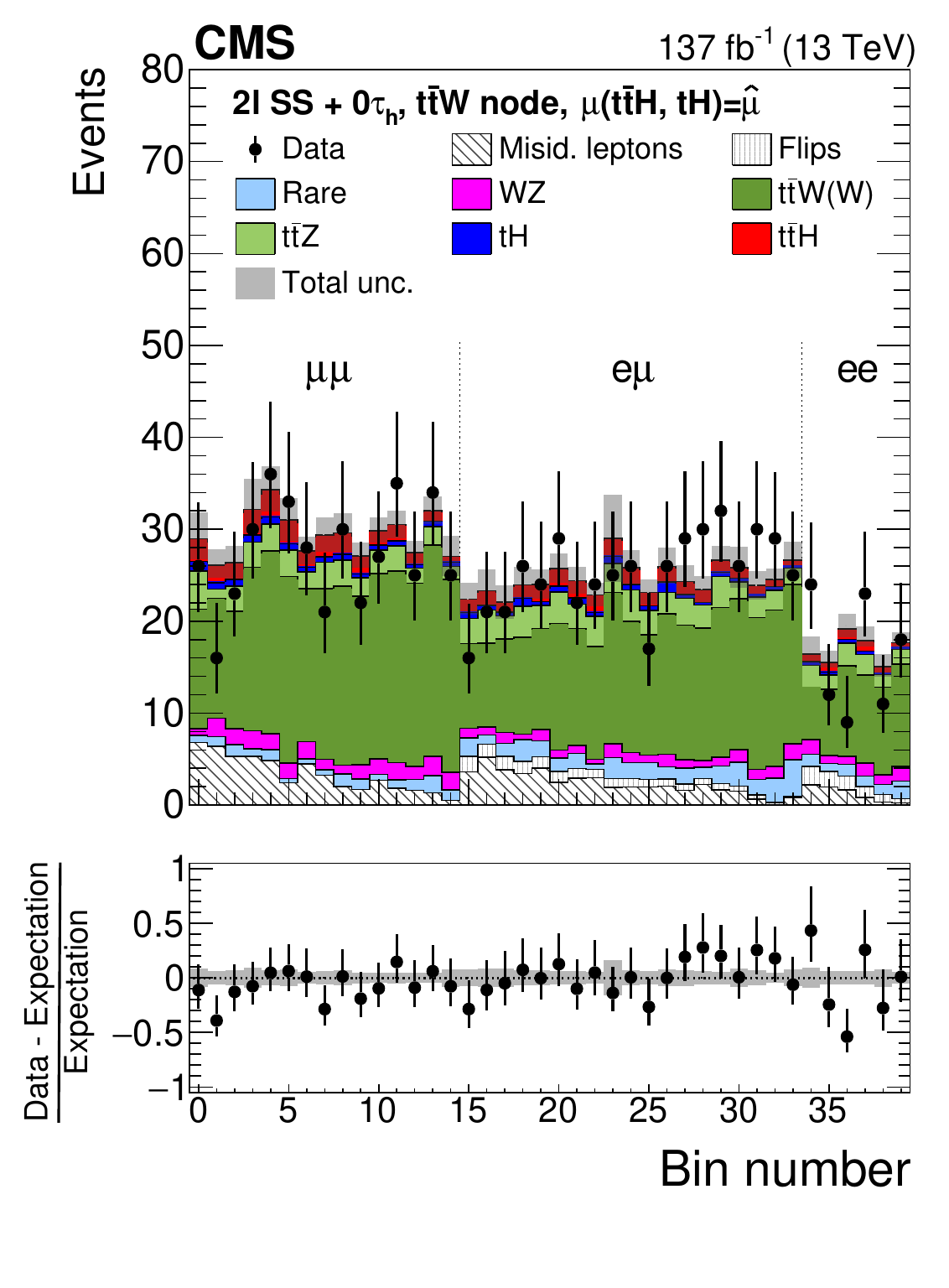}
  \centering\includegraphics[width=0.4\textwidth]{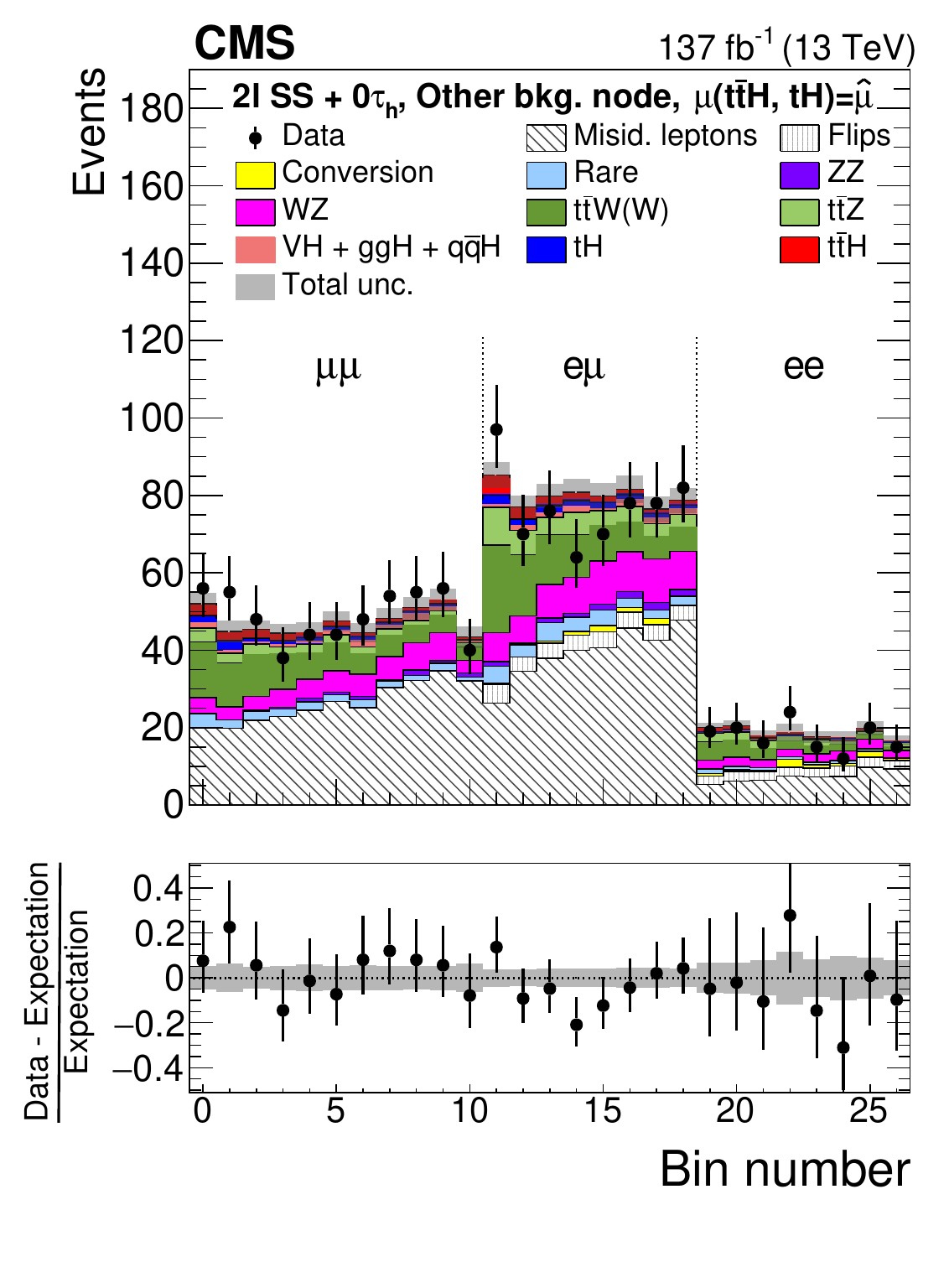}
  \caption{
    Distributions of the activation value of the ANN output node with the highest activation value 
    for events selected in the $\twoLeptonssZeroTau$ channel 
    and classified as $\ttH$ signal (\cmsTop \cmsLeft), $\tH$ signal (\cmsTop \cmsRight), $\ttW$ background (\cmsBottom \cmsLeft), and other backgrounds (\cmsBottom \cmsRight).
    The distributions expected for the $\ttH$ and $\tH$ signals and for background processes
    are shown for the values of the parameters of interest and of the nuisance parameters obtained from the ML fit.
    The best fit value of the $\ttH$ and $\tH$ production rates amounts to $\rhat_{\ttH} = 0.92$ and $\rhat_{\tH} = 5.7$
    times the rates expected in the SM.
  }
  \label{fig:postfitPlots1}
\end{figure*}

\begin{figure*}[h!]
  \centering\includegraphics[width=0.4\textwidth]{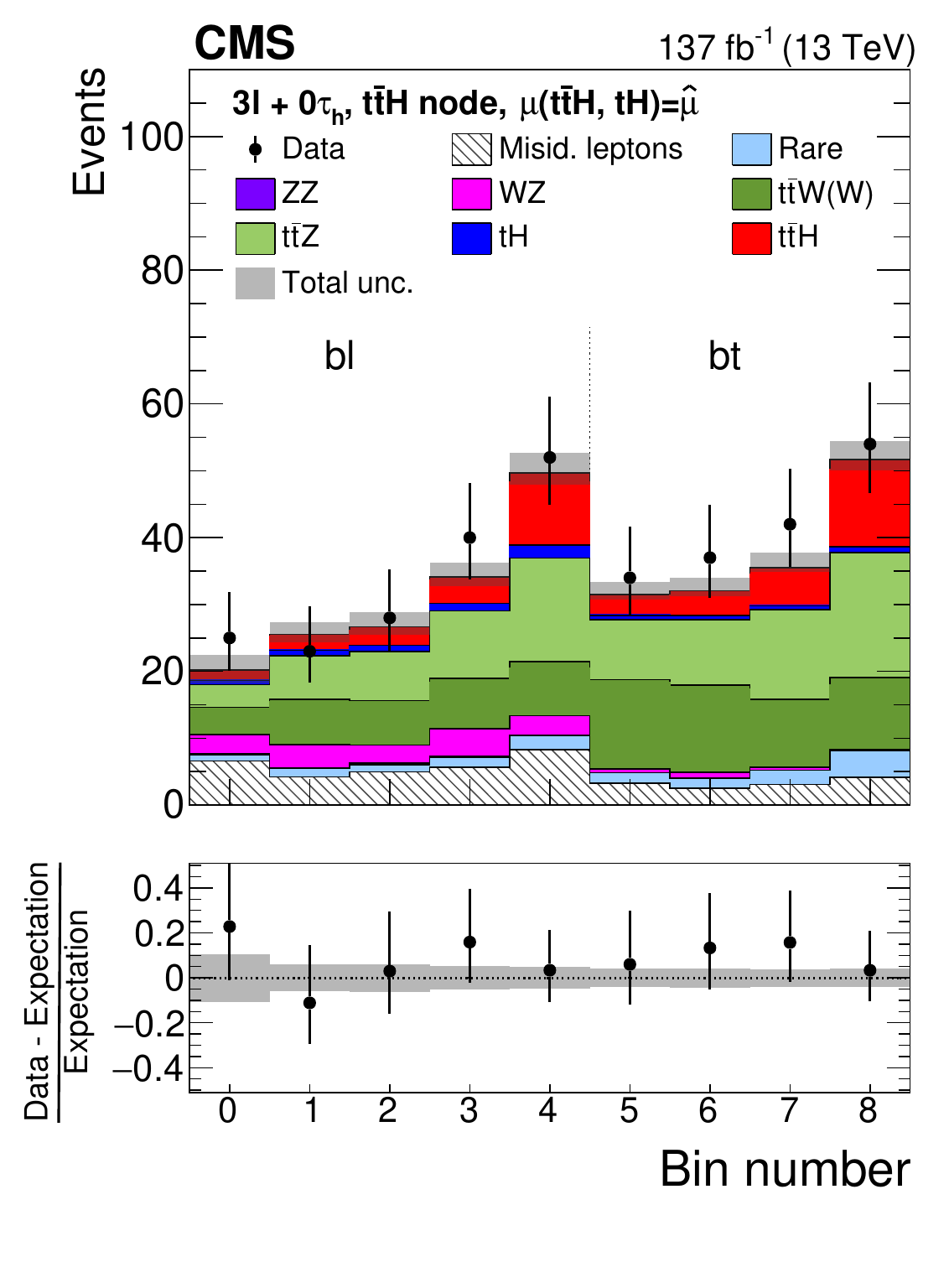}
  \centering\includegraphics[width=0.4\textwidth]{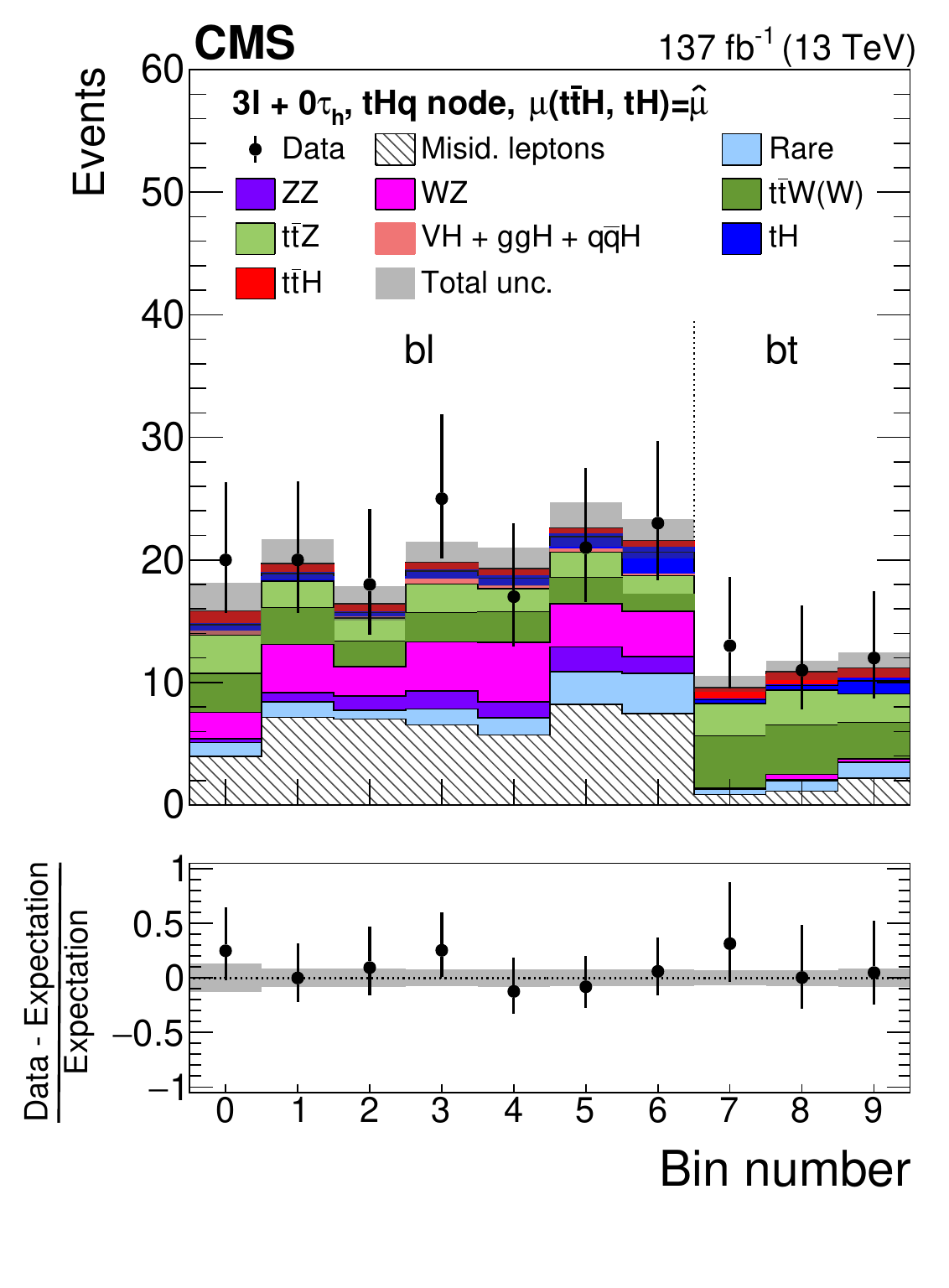}
  \centering\includegraphics[width=0.4\textwidth]{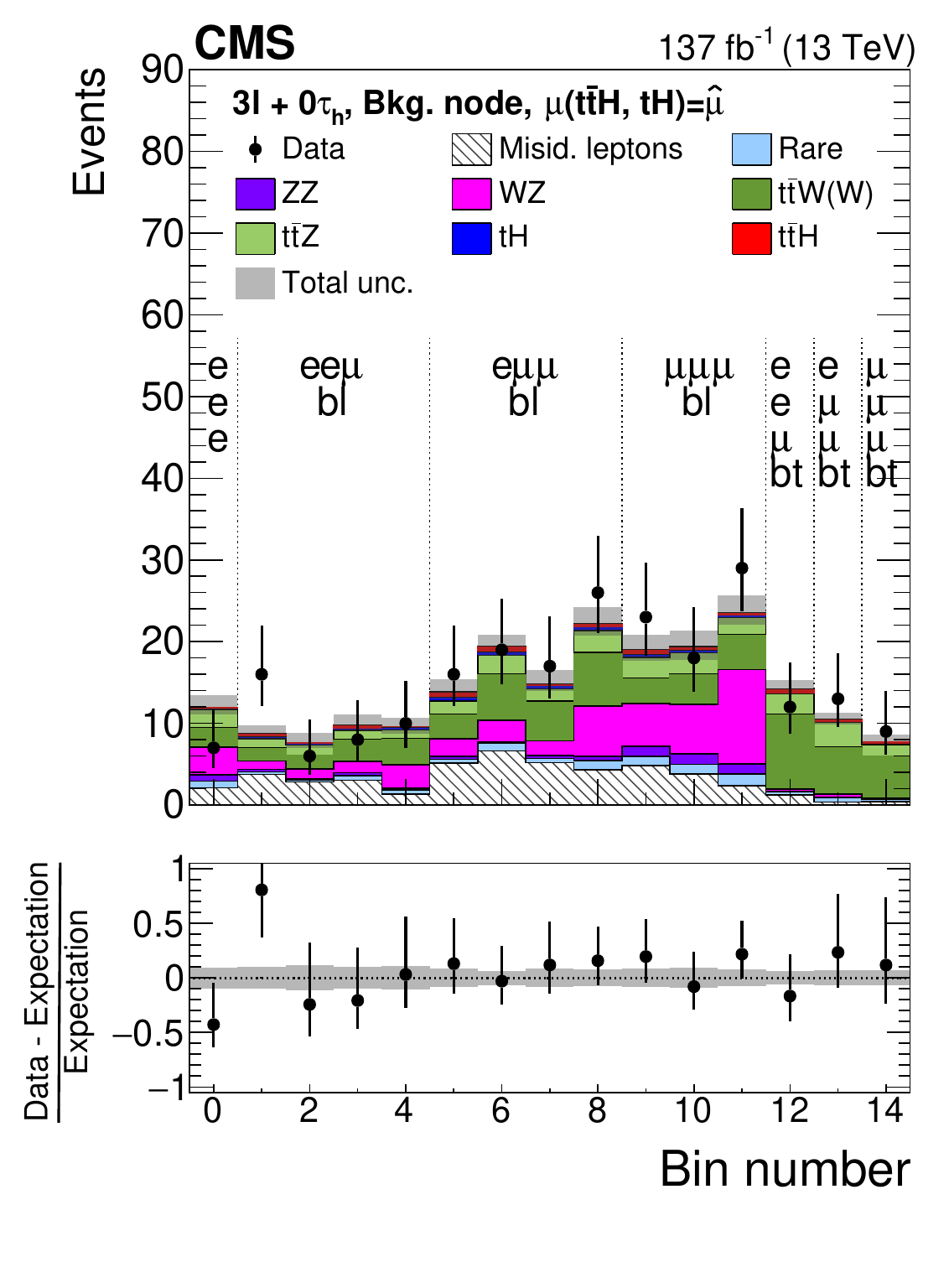}
  \centering\includegraphics[width=0.4\textwidth]{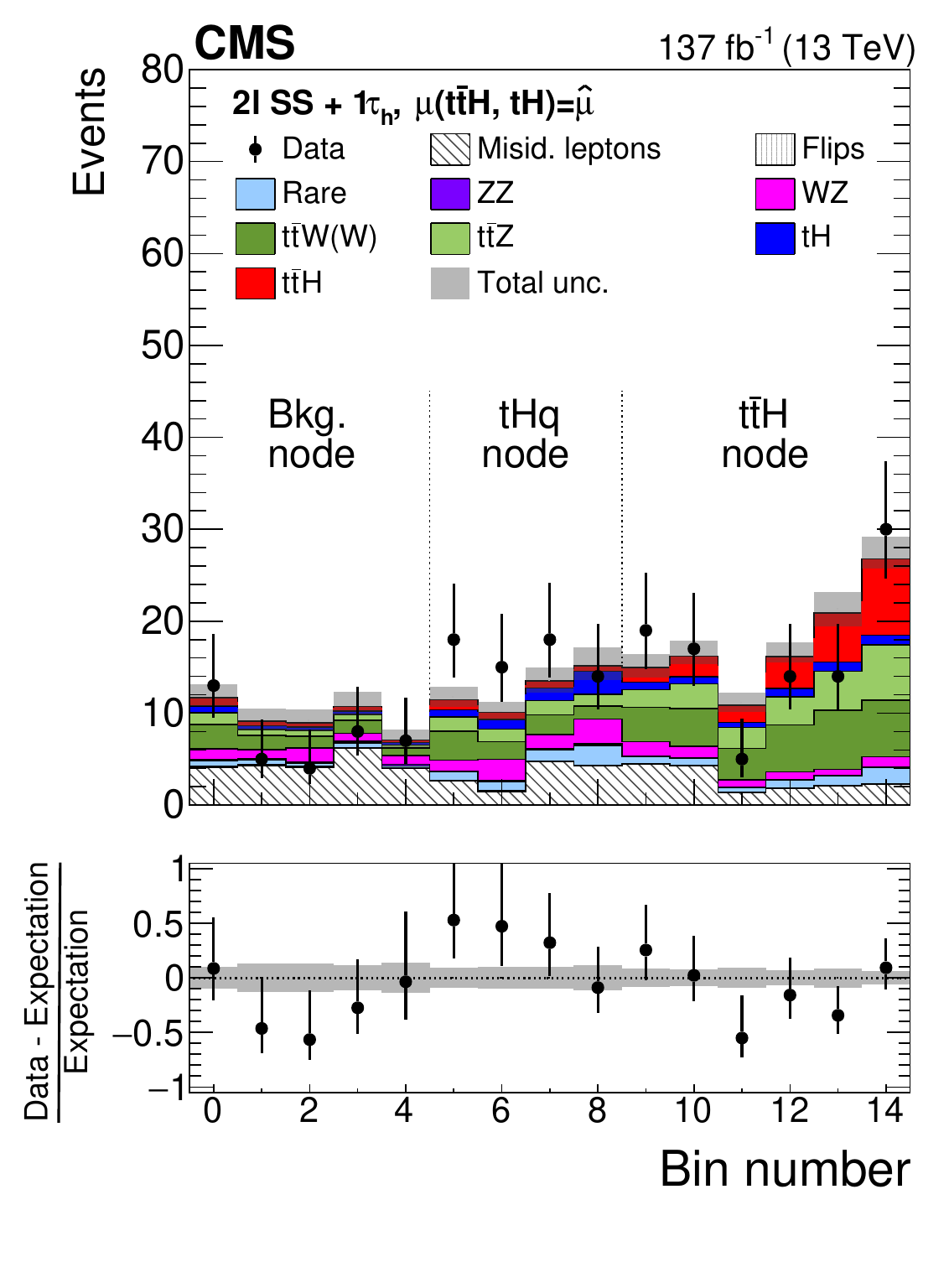}
  \caption{
    Distributions of the activation value of the ANN output node with the highest activation value 
    for events selected in the $\threeLeptonZeroTau$ channel 
    and classified as $\ttH$ signal (\cmsTop \cmsLeft), $\tH$ signal (\cmsTop \cmsRight), and background (\cmsBottom \cmsLeft),
    and for events selected in the $\twoLeptonssOneTau$ channel (\cmsBottom \cmsRight).
    In case of the $\twoLeptonssOneTau$ channel,
    the activation value of the ANN output nodes for $\ttH$ signal, $\tH$ signal, and background are shown together in a single histogram, 
    concatenating histogram bins as appropriate and enumerating the bins by a monotonously increasing number.
    The distributions expected for the $\ttH$ and $\tH$ signals and for background processes
    are shown for the values of the parameters of interest and of the nuisance parameters obtained from the ML fit.
    The best fit value of the $\ttH$ and $\tH$ production rates amounts to $\rhat_{\ttH} = 0.92$ and $\rhat_{\tH} = 5.7$
    times the rates expected in the SM.
  }
  \label{fig:postfitPlots2}
\end{figure*}

\begin{figure*}
  \centering\includegraphics[width=0.49\textwidth]{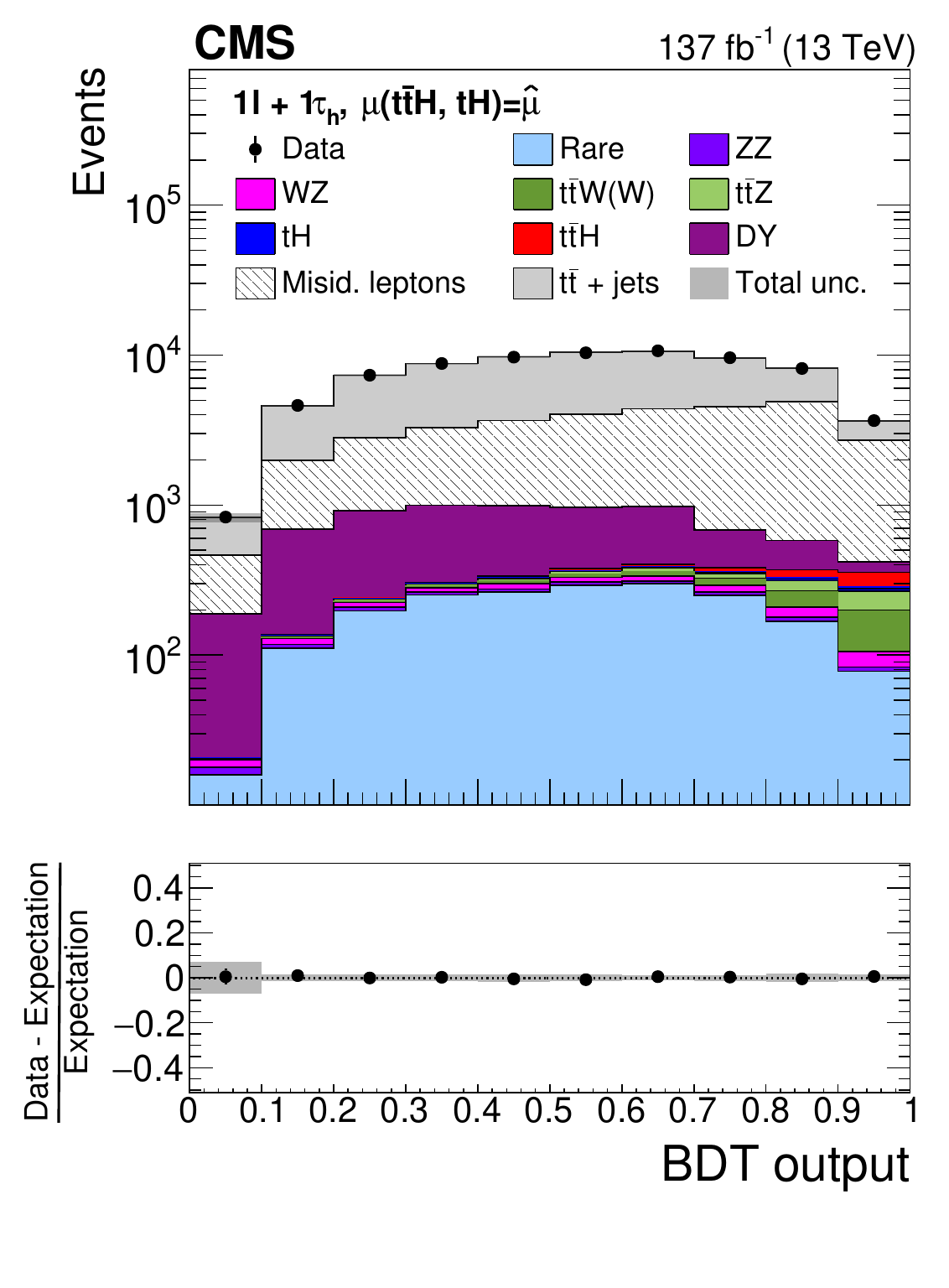}
  \centering\includegraphics[width=0.49\textwidth]{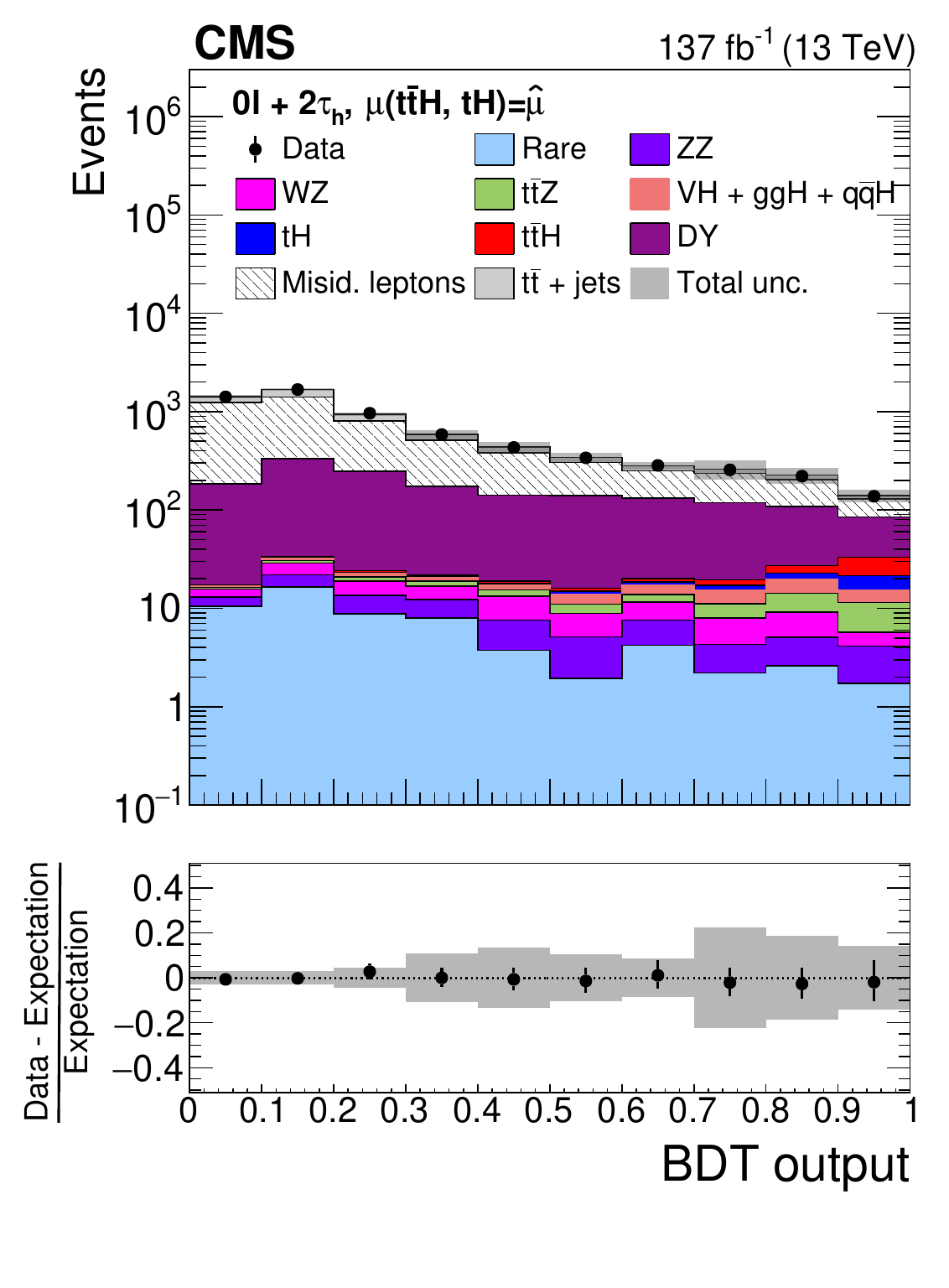}
  \centering\includegraphics[width=0.49\textwidth]{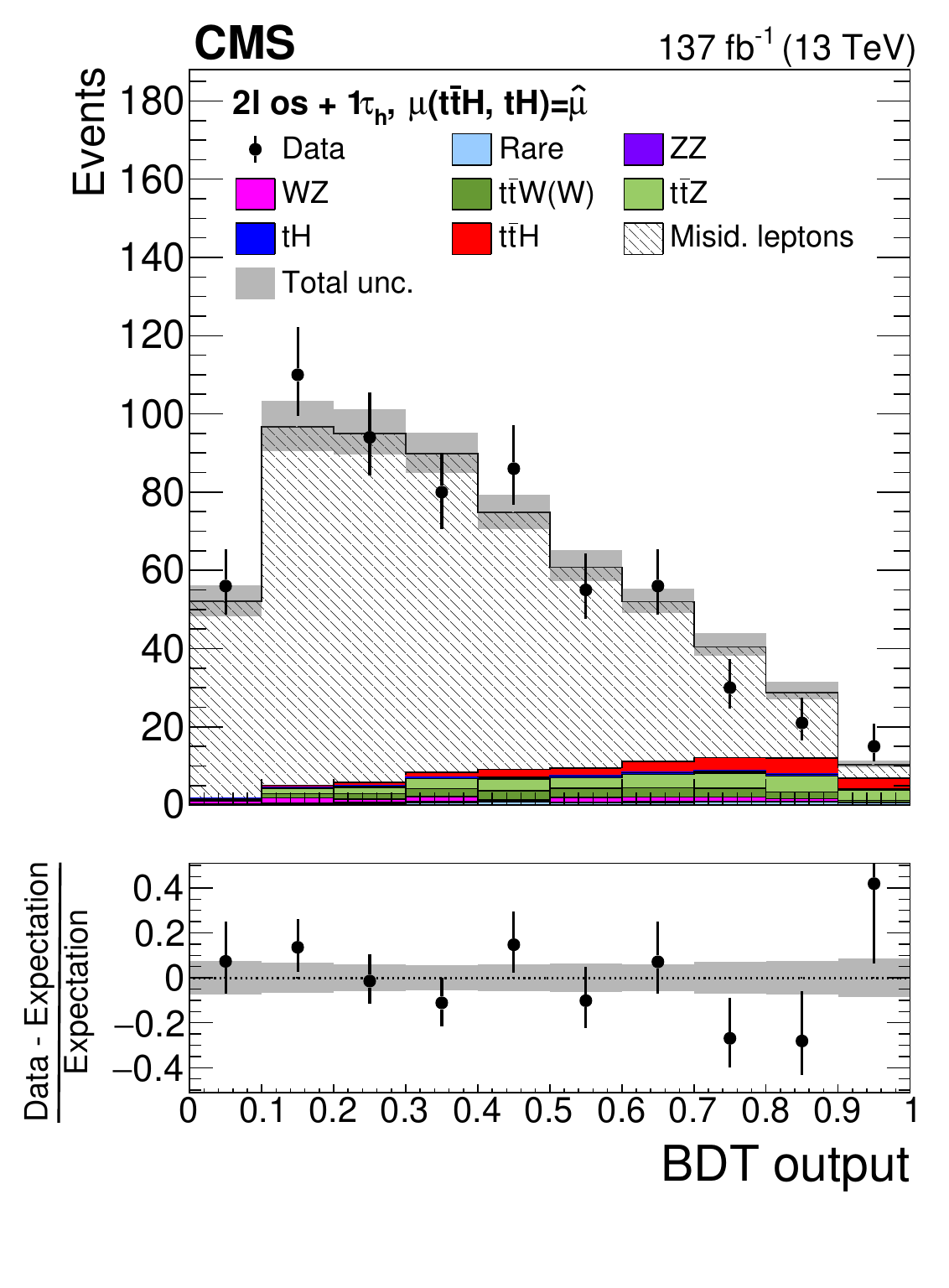}
  \caption{
    Distributions of the BDT output 
    for events selected in the $\oneLeptonOneTau$ (\cmsTop \cmsLeft), $\zeroLeptonTwoTau$ (\cmsTop \cmsRight),
    and $\twoLeptonosOneTau$ (\cmsBottom) channels.
    The distributions expected for the $\ttH$ and $\tH$ signals and for background processes
    are shown for the values of the parameters of interest and of the nuisance parameters obtained from the ML fit.
    The best fit value of the $\ttH$ and $\tH$ production rates amounts to $\rhat_{\ttH} = 0.92$ and $\rhat_{\tH} = 5.7$
    times the rates expected in the SM.
  }
  \label{fig:postfitPlots3}
\end{figure*}

\begin{figure*}
  \centering\includegraphics[width=0.49\textwidth]{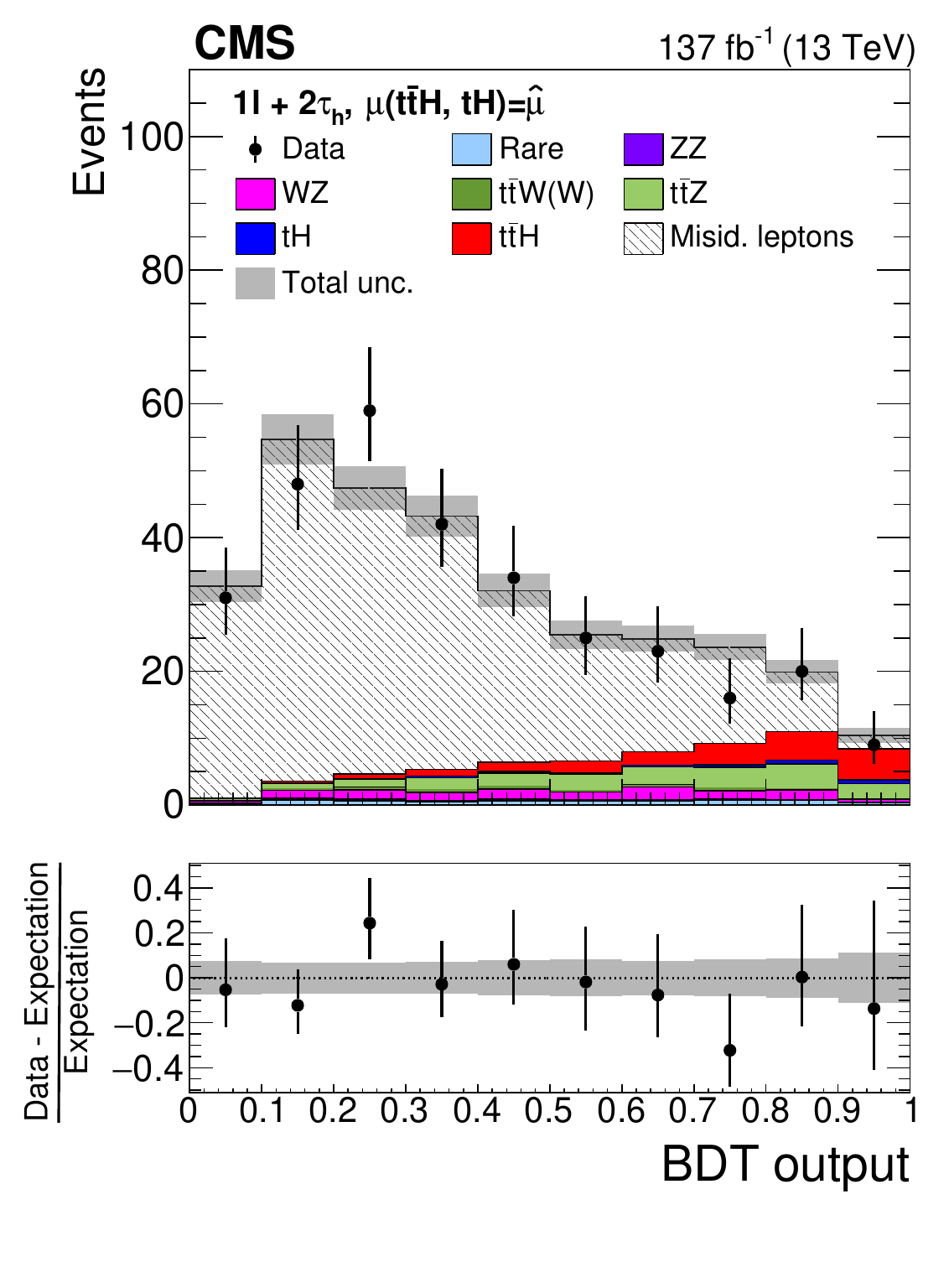}
  \centering\includegraphics[width=0.49\textwidth]{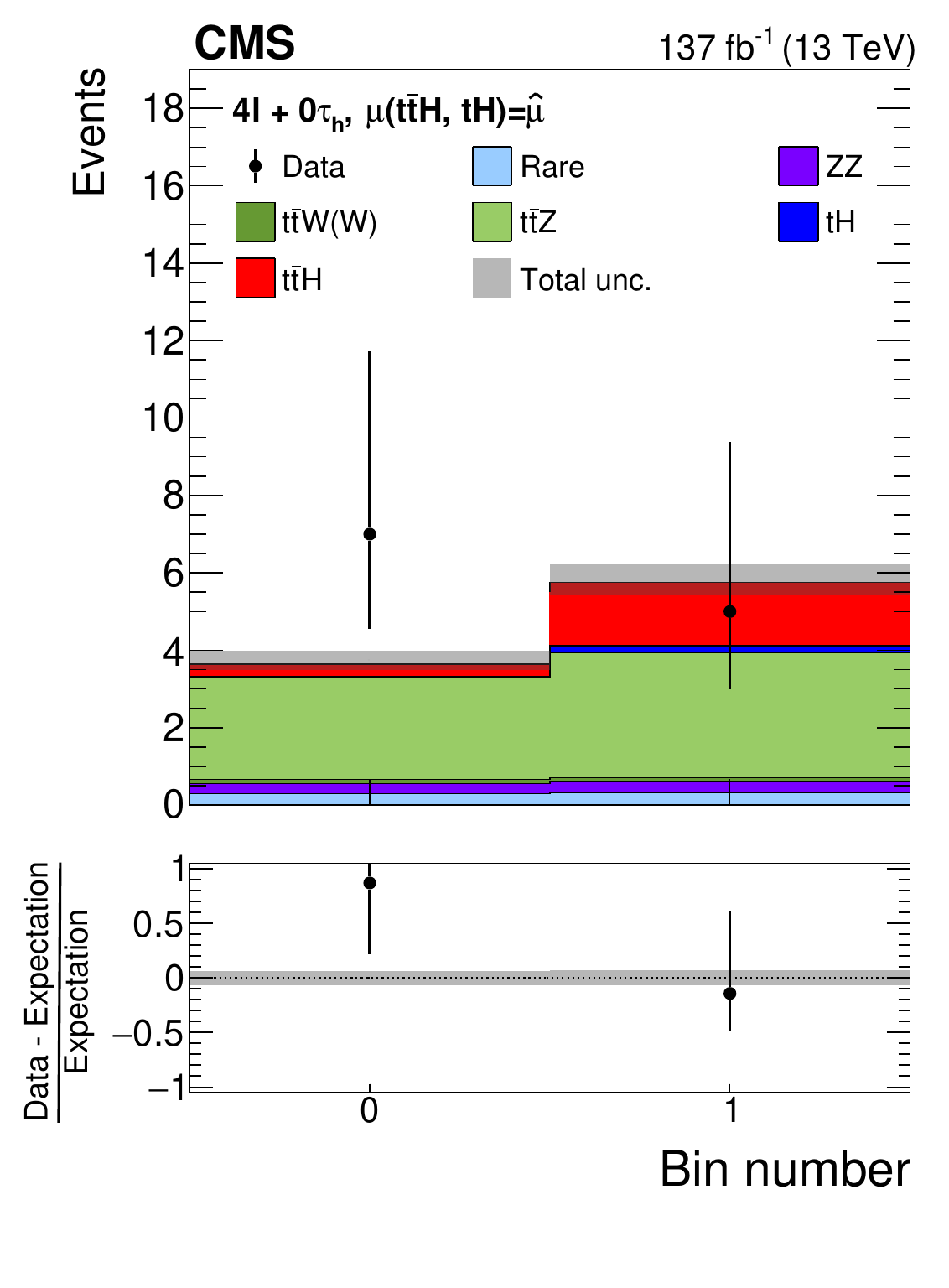}
  \centering\includegraphics[width=0.49\textwidth]{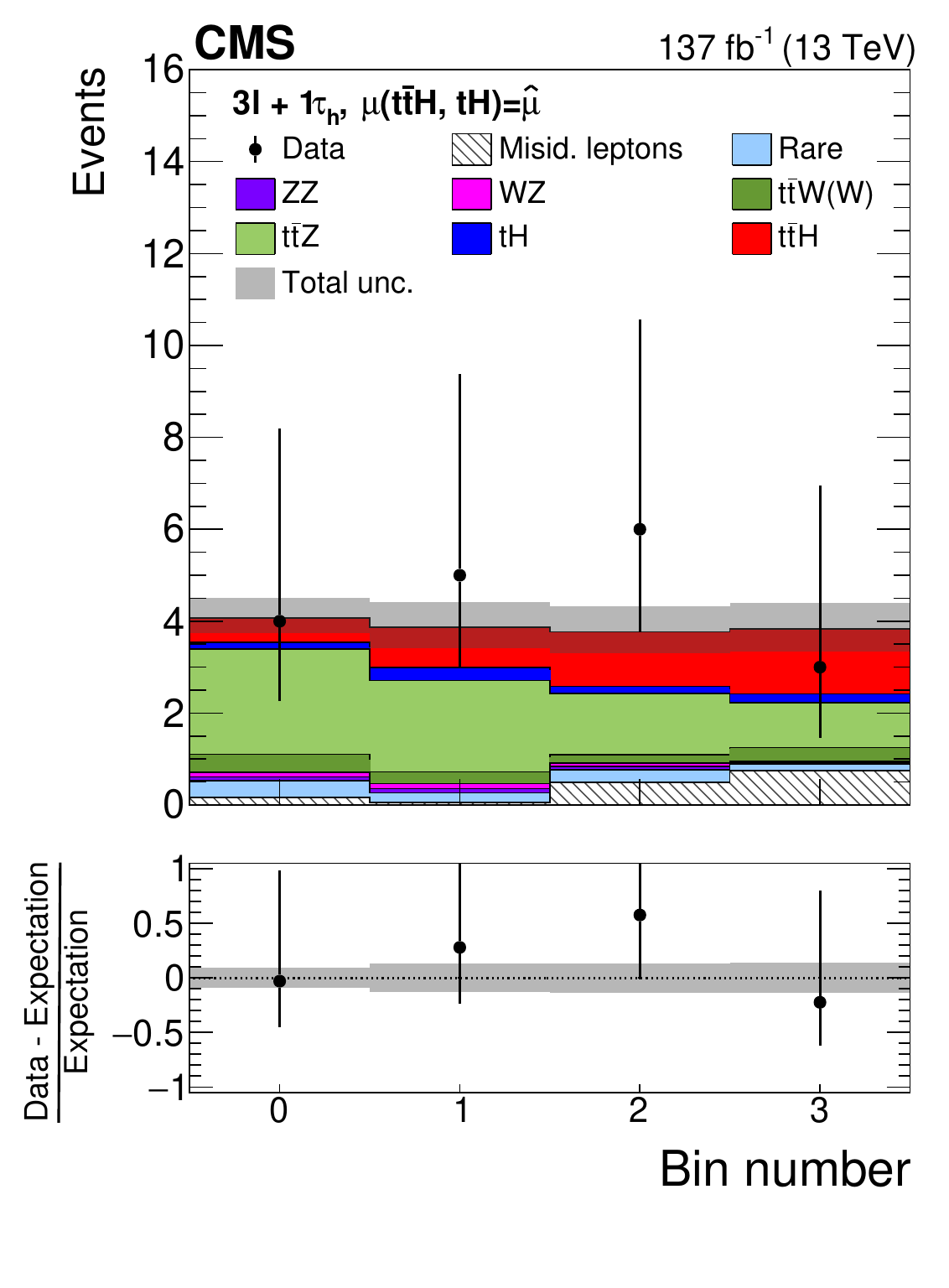}
  \centering\includegraphics[width=0.49\textwidth]{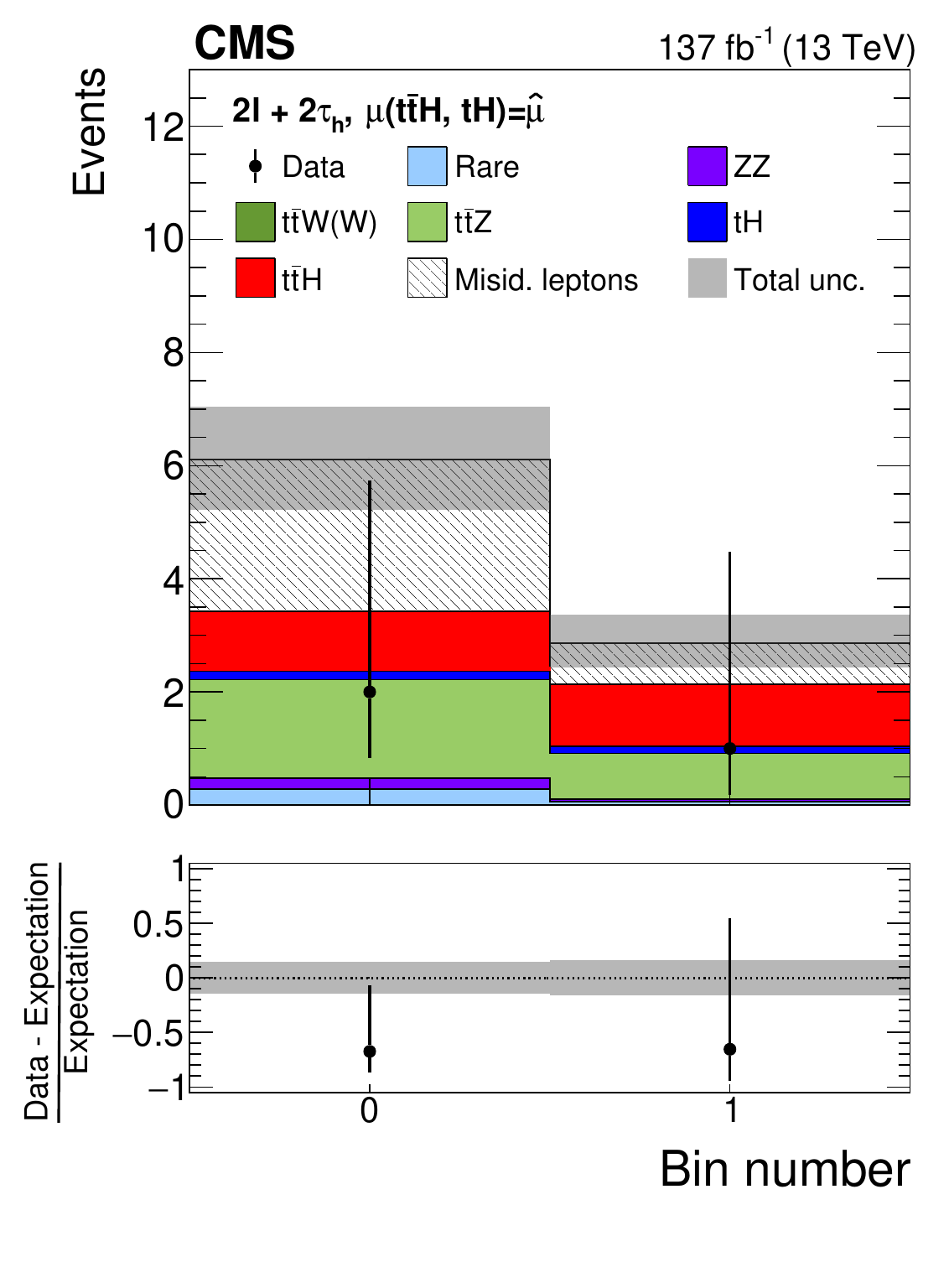}
  \caption{
    Distributions of the BDT output used for the signal extraction
    in the $\oneLeptonTwoTau$ (\cmsTop \cmsLeft), $\fourLeptonZeroTau$ (\cmsTop \cmsRight), $\threeLeptonOneTau$ (\cmsBottom \cmsLeft), and $\twoLeptonTwoTau$  (\cmsBottom \cmsRight) channels.
    The distributions expected for the $\ttH$ and $\tH$ signals and for background processes
    are shown for the values of the parameters of interest and of the nuisance parameters obtained from the ML fit.
    The best fit value of the $\ttH$ and $\tH$ production rates amounts to $\rhat_{\ttH} = 0.92$ and $\rhat_{\tH} = 5.7$
    times the rates expected in the SM.
  }
  \label{fig:postfitPlots4}
\end{figure*}

\begin{figure*}[htb]
  \centering\includegraphics[width=0.49\textwidth]{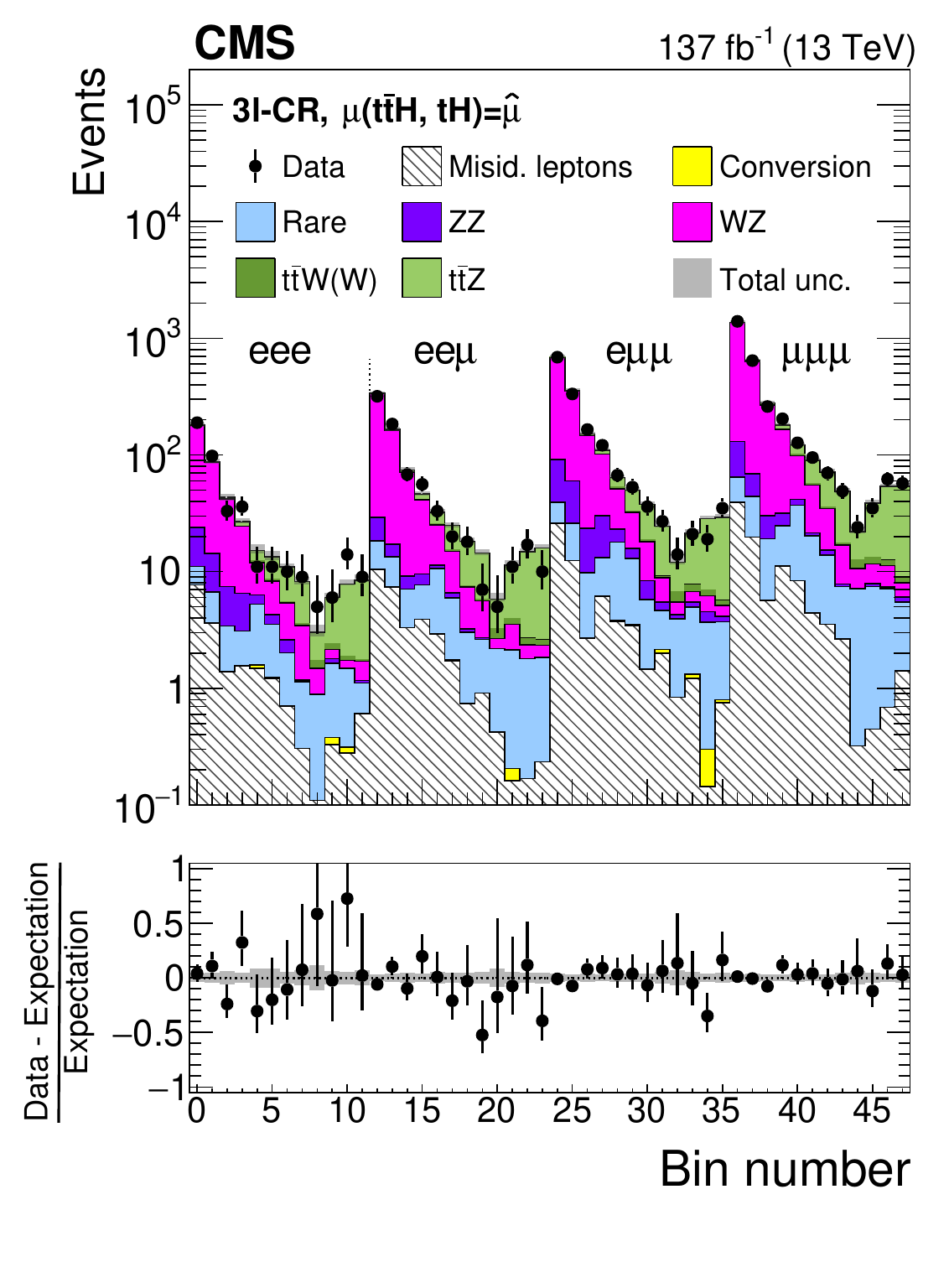}
  \centering\includegraphics[width=0.49\textwidth]{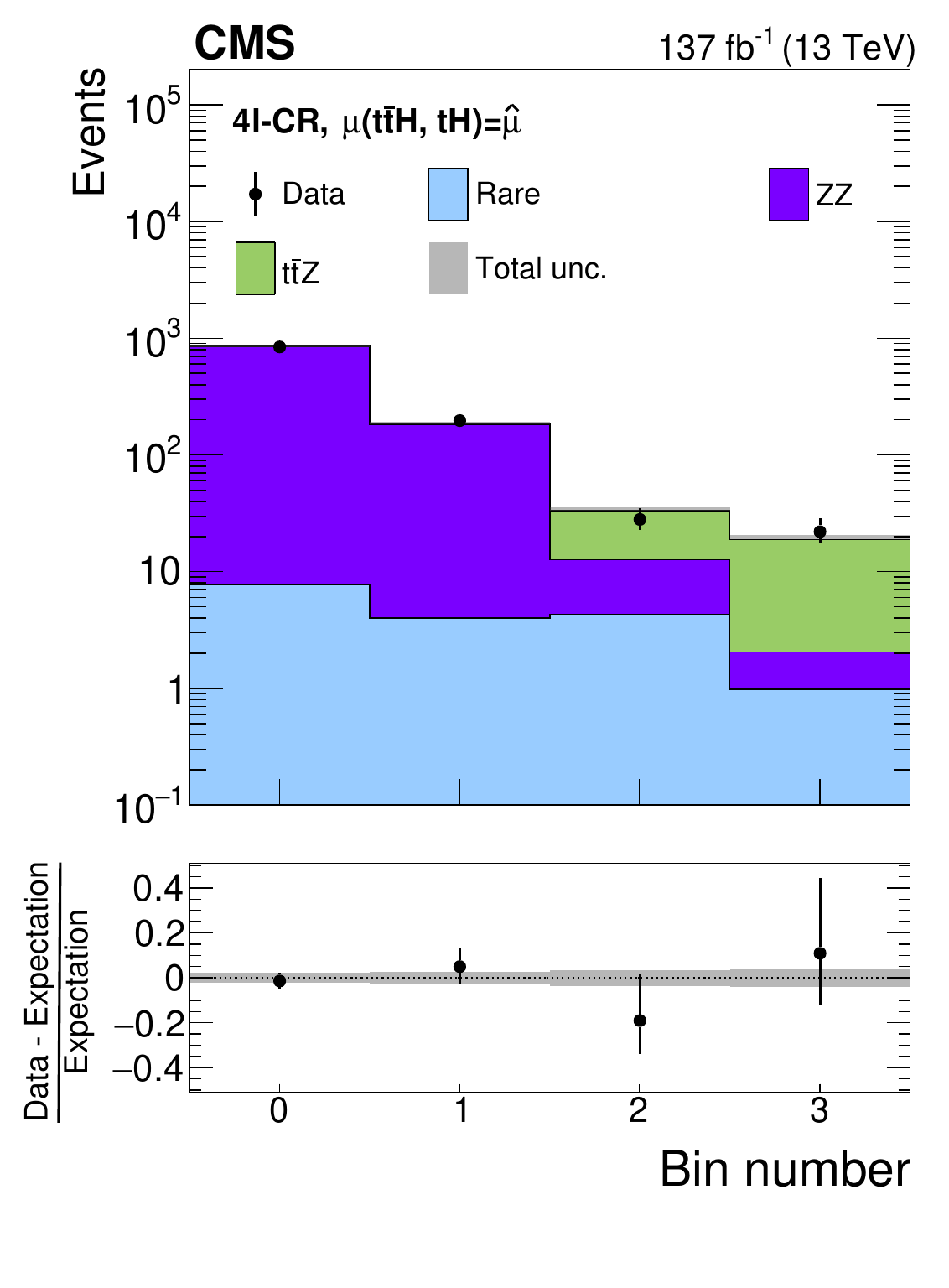}
  \caption{
    Distributions of discriminating observables
    in the $\threeLeptonZeroTau$ (\cmsLeft) and $\fourLeptonZeroTau$ (\cmsRight) control region.
    The distributions expected for the $\ttH$ and $\tH$ signals and for background processes
    are shown for the values of the parameters of interest and of the nuisance parameters obtained from the ML fit.
    The best fit value of the $\ttH$ and $\tH$ production rates amounts to $\rhat_{\ttH} = 0.92$ and $\rhat_{\tH} = 5.7$
    times the rates expected in the SM.
  }
  \label{fig:controlPlots_3leptonCR_and_4leptonCR}
\end{figure*}

Assuming the distributions of the discriminating observables for the $\tH$ and $\ttH$ signals agree with their SM expectation,
the production rate for the $\ttH$ signal is measured to be $\r_{\ttH} = 0.92 \pm 0.19 \stat ^{+0.17}_{-0.13}\syst$ times the SM expectation,
equivalent to a $\ttH$ production cross section for $\ttH$ production of $466 \pm 96\allowbreak\stat^{+70}_{-56}\syst\fb$,
and that of the $\tH$ signal is measured to be $\r_{\tH} = 5.7 \pm 2.7\stat \pm 3.0\syst$ times the SM expectation for this production rate,
equivalent to a cross section for $\tH$ production of $510 \pm 200\stat \pm 220\syst\fb$.
The corresponding observed (expected) significance of the $\ttH$ signal amounts to $4.7$ ($5.2$) standard deviations, assuming
the $\tH$ process to have the SM production rate, and that of the $\tH$ signal to $1.4$ ($0.3$) standard deviations,
also assuming the $\ttH$ process to have the SM production rate. We have estimated the agreement between the data and our
statistical model by using a goodness-of-fit test to the saturated model, obtaining a p-value of 0.097, showing no indication of
a significant difference between data and the assumed model.

The distributions that are included in the ML fit are shown in Figs.~\ref{fig:postfitPlots1} to~\ref{fig:controlPlots_3leptonCR_and_4leptonCR}.
In the $\twoLeptonssZeroTau$ and $\threeLeptonZeroTau$ channels, we show the distributions of the activation values of ANN output nodes
in the different subcategories based on lepton flavor and on the multiplicity of $\Pbottom$-tagged jets in a single histogram, 
concatenating histogram bins as appropriate, and enumerate the bins by a monotonically increasing number.
The distributions expected for the $\ttH$ and $\tH$ signals, as well as the expected background contributions, are shown
for the value of the POI and of nuisance parameters obtained from the ML fit.
The uncertainty bands shown in the figures represent the total uncertainty in the sum of signal and background contributions
that remains after having determined the value of the nuisance parameters through the ML fit.
These bands are computed by randomly sampling from the covariance matrix of the nuisance parameters as determined by the ML fit
and adding the statistical uncertainties in the background predictions in quadrature.
The data are in agreement with the sum of contributions estimated by the ML fit for the $\ttH$ and $\tH$ signals and for the background processes.
The corresponding event yields are given in Table~\ref{tab:eventYieldsPostfit}.
In the $\twoLeptonssZeroTau$, $\threeLeptonZeroTau$, and $\twoLeptonssOneTau$ channels,
the sums of events yields in all ANN output node categories are given in the table.

\begin{table*}[h!]
  \centering
  \topcaption{
    Number of events selected in each of the ten analysis channels 
    compared to the event yields expected from the $\ttH$ and $\tH$ signals and from background processes.
    The expected event yields are computed for the values of nuisance parameters and of the POI obtained from the ML fit.
    The best fit values of the POI amount to $\rhat_{\ttH} = 0.92$ and $\rhat_{\tH} = 5.7$.
    Quoted uncertainties represent the sum of statistical and systematic components. The symbol ``\NA'' indicates that
    the corresponding expected contribution is smaller than 0.1 events.
}
  \label{tab:eventYieldsPostfit}
  \resizebox{0.9\width}{!}{
    \begin{tabular}{llll}
      \hline
      Process                          & $\twoLeptonssZeroTau$ & $\threeLeptonZeroTau$ & $\twoLeptonssOneTau$ \\
      \hline                                
      $\ttH$                           & $222 \pm 51 $         & $61 \pm 15 $          & $28.9 \pm  6.4 $     \\
      $\tH$                            & $119 \pm 85 $         & $20 \pm 14 $          & $12.7 \pm  9.0 $     \\
      [\cmsTabSkip]                                            
      $\ttZ + \ttbar\Pggx$             & $322 \pm 25 $         & $145 \pm 11 $         & $29.6 \pm  3.3 $     \\
      $\ttW + \ttWW$                   & $1153 \pm 64 $        & $171.1 \pm  9.5 $     & $47.4 \pm  6.5 $     \\
      $\PW\PZ$                         & $296 \pm 31 $         & $89.7 \pm  9.7 $      & $19.4 \pm  2.9 $     \\
      $\PZ\PZ$                         & $31.2 \pm  3.3 $      & $16.2 \pm  1.6 $      & $ 1.6 \pm  0.3 $     \\
      Misidentified leptons            & $1217 \pm 91 $        & $140 \pm 11 $         & $52.0 \pm  9.6 $     \\
      Flips                            & $121 \pm 19 $         & \NA                   & \NA                  \\ 
      Rare backgrounds                 & $222 \pm 48 $         & $41.0 \pm  8.9 $      & $13.3 \pm  3.1 $     \\
      Conversion                       & $42 \pm 12 $          & $ 5.6 \pm  1.6 $      & \NA                  \\
      $\ggH + \qqH + \VH + 	\ttVH$ & $35.3 \pm  4.0 $      & $ 3.4 \pm  0.3 $      & $ 1.8 \pm  0.3 $     \\
      Total expected background        & $3517 \pm 85 $        & $627 \pm 20 $         & $179 \pm 13 $        \\
      [\cmsTabSkip]                                            
      Data                             & 3738                  & 744                   & 201                  \\
      \hline                           
    \end{tabular}
  }
 \vskip 0.8cm
 \resizebox{0.9\width}{!}{
   \begin{tabular}{lllll}
     \hline
     Process                          & $\oneLeptonOneTau$ & $\zeroLeptonTwoTau$ & $\twoLeptonosOneTau$ & $\oneLeptonTwoTau$ \\ \hline
     $\ttH$                           & $183 \pm 41 $      & $24.4 \pm  6.0 $    & $19.1 \pm  4.3 $     & $19.3 \pm  4.2 $   \\
     $\tH$                            & $65 \pm 46 $       & $16 \pm 12 $        & $ 4.8 \pm  3.4 $     & $ 2.6 \pm  1.9 $   \\
     [\cmsTabSkip]                                                                                                           
     $\ttZ + \ttbar\Pggx$             & $203 \pm 24 $      & $27.1 \pm  3.8 $    & $25.5 \pm  2.9 $     & $20.3 \pm  2.1 $   \\
     $\ttW + \ttWW$                   & $254 \pm 34 $      & $ 3.8 \pm  0.5 $    & $17.4 \pm  2.4 $     & $ 2.6 \pm  0.4 $   \\
     $\PW\PZ$                         & $198 \pm 37 $      & $42.5 \pm  8.7 $    & $ 8.4 \pm  1.6 $     & $11.8 \pm  2.2 $   \\
     $\PZ\PZ$                         & $98 \pm 13 $       & $34.2 \pm  4.8 $    & $ 1.9 \pm  0.3 $     & $ 1.8 \pm  0.3 $   \\
     DY                               & $4480 \pm 460 $    & $1430.0 \pm 220 $   & $519 \pm 28 $        & $250 \pm 16 $      \\
     $\ttbar$+jets                    & $41900 \pm 1900 $  & $861 \pm 98 $       & \NA                  & \NA                \\
     Misidentified leptons            & $25300 \pm 1900 $  & $3790 \pm 220 $     & \NA                  & \NA                \\
     Rare backgrounds                 & $1930 \pm 420 $    & $60 \pm 14 $        & $ 5.9 \pm  1.3 $     & $ 5.6 \pm  1.3 $   \\
     Conversion                       & \NA                & \NA                 & $ 0.5 \pm  0.2 $     & \NA                \\
     $\ggH + \qqH + \VH + 	\ttVH$ & $38.5 \pm  3.6 $   & $26.7 \pm  3.6 $    & $ 0.8 \pm  0.1 $     & \NA                \\
     Total expected background        & $73550 \pm 610 $   & $6290 \pm 130 $     & $584 \pm 27 $        & $295 \pm 16 $      \\
     [\cmsTabSkip]                                                                                                           
     Data                             & 73736              & 6310                & 603                  & 307                \\
     \hline
   \end{tabular}
 }
 \vskip 0.8cm
 \resizebox{0.9\width}{!}{
     \begin{tabular}{llll}
     \hline
     Process                   & $\fourLeptonZeroTau$ & $\threeLeptonOneTau$ & $\twoLeptonTwoTau$ \\ \hline 
     $\ttH$                    & $ 2.0 \pm  0.5 $     & $ 4.0 \pm  0.9 $     & $ 2.2 \pm  0.5 $   \\
     $\tH$                     & $ 0.2 \pm  0.2 $     & $ 0.8 \pm  0.6 $     & $ 0.3 \pm  0.2 $   \\
     [\cmsTabSkip]
     $\ttZ + \ttbar\Pggx$      & $ 5.9 \pm  0.4 $     & $ 6.6 \pm  0.7 $     & $ 2.5 \pm  0.3 $   \\
     $\ttW + \ttWW$            & $ 0.2 \pm  0.0 $     & $ 1.1 \pm  0.2 $     & \NA                \\
     $\PZ\PZ$                  & $ 0.6 \pm  0.2 $     & $ 0.3 \pm  0.1 $     & $ 0.2 \pm  0.0 $   \\
     Misidentified leptons     & \NA                  & $ 1.5 \pm  0.9 $     & $ 3.4 \pm  0.9 $   \\
     Rare backgrounds          & $ 0.6 \pm  0.1 $     & $ 1.0 \pm  0.3 $     & $ 0.3 \pm  0.1 $   \\
     Conversion                & \NA                  & \NA                  & \NA                \\ 
     Total expected background & $ 7.4 \pm  0.5 $     & $11.5 \pm  1.3 $     & $ 6.8 \pm  1.0 $   \\
     [\cmsTabSkip]
     Data                      & 12                   & 18                   & 3                  \\\hline
   \end{tabular}
 }
\end{table*}

The event yields of background processes obtained from the ML fit agree reasonably well with their expected production rate, given the uncertainties.
In particular, the production rates of the $\ttZ$ and $\ttW$ backgrounds
are determined to be $\r_{\ttZ} = 1.03 \pm 0.14\statsystlumi$ and $\r_{\ttW} = 1.43 \pm 0.21\statsystlumi$ times their SM expectation,
as obtained from the MC simulation.

\begin{figure*}[h!]
  \centering\includegraphics[width=0.49\textwidth]{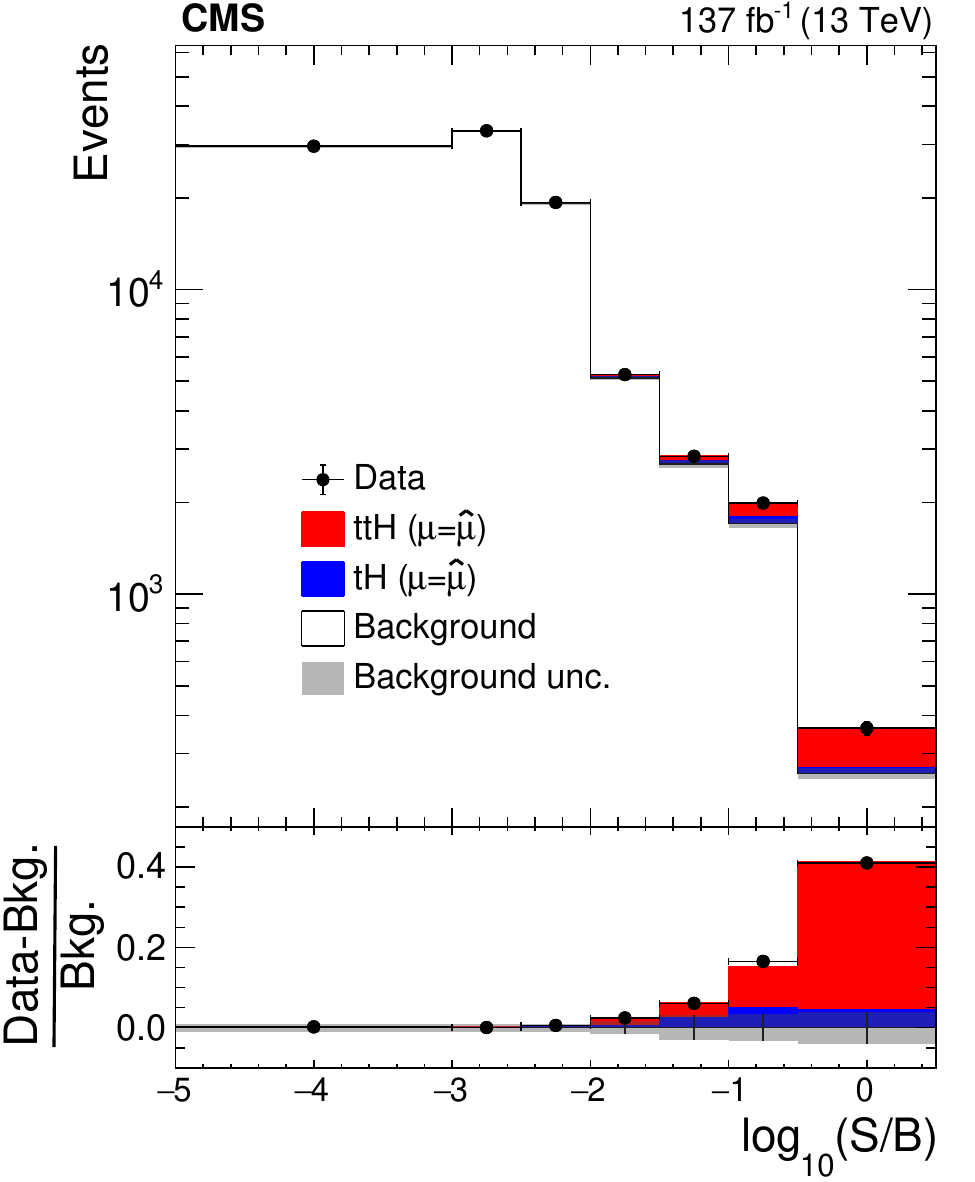}
  \caption{
    Distribution of the decimal logarithm of the ratio between the expected $\ttH+\tH$ signal and the expected sum of background contributions
    in each bin of the $105$ distributions that are included in the ML fit used for the signal extraction.
    The distributions expected for signal and background processes
    are computed for $\rhat_{\ttH} = 0.92$, $\rhat_{\tH} = 5.7$, and the values of nuisance parameters obtained from the ML fit.
  }
  \label{fig:logSoverB}
\end{figure*}

The evidence for the presence of the $\ttH$ and $\tH$ signals in the data is illustrated in Fig.~\ref{fig:logSoverB},
in which each bin of the distributions that are included in the ML fit
is classified according to the expected ratio of the number of $\ttH+\tH$ signal (S) over background (B) events in that bin.
A significant excess of events with respect to the background expectation is visible in the bins with the highest expected S/B ratio.

\begin{figure*}[h!]
  \centering\includegraphics[width=0.49\textwidth]{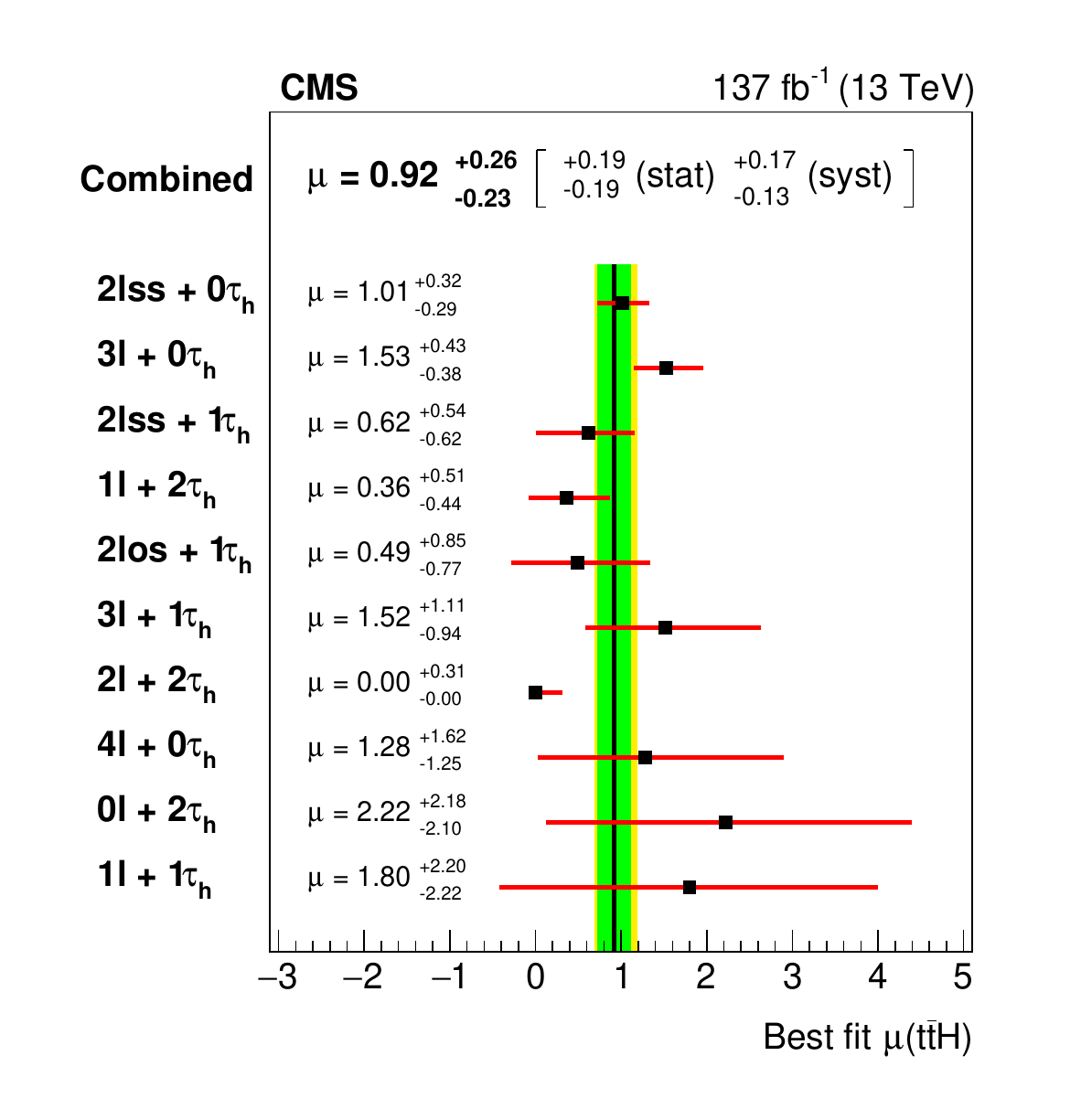}
  \centering\includegraphics[width=0.49\textwidth]{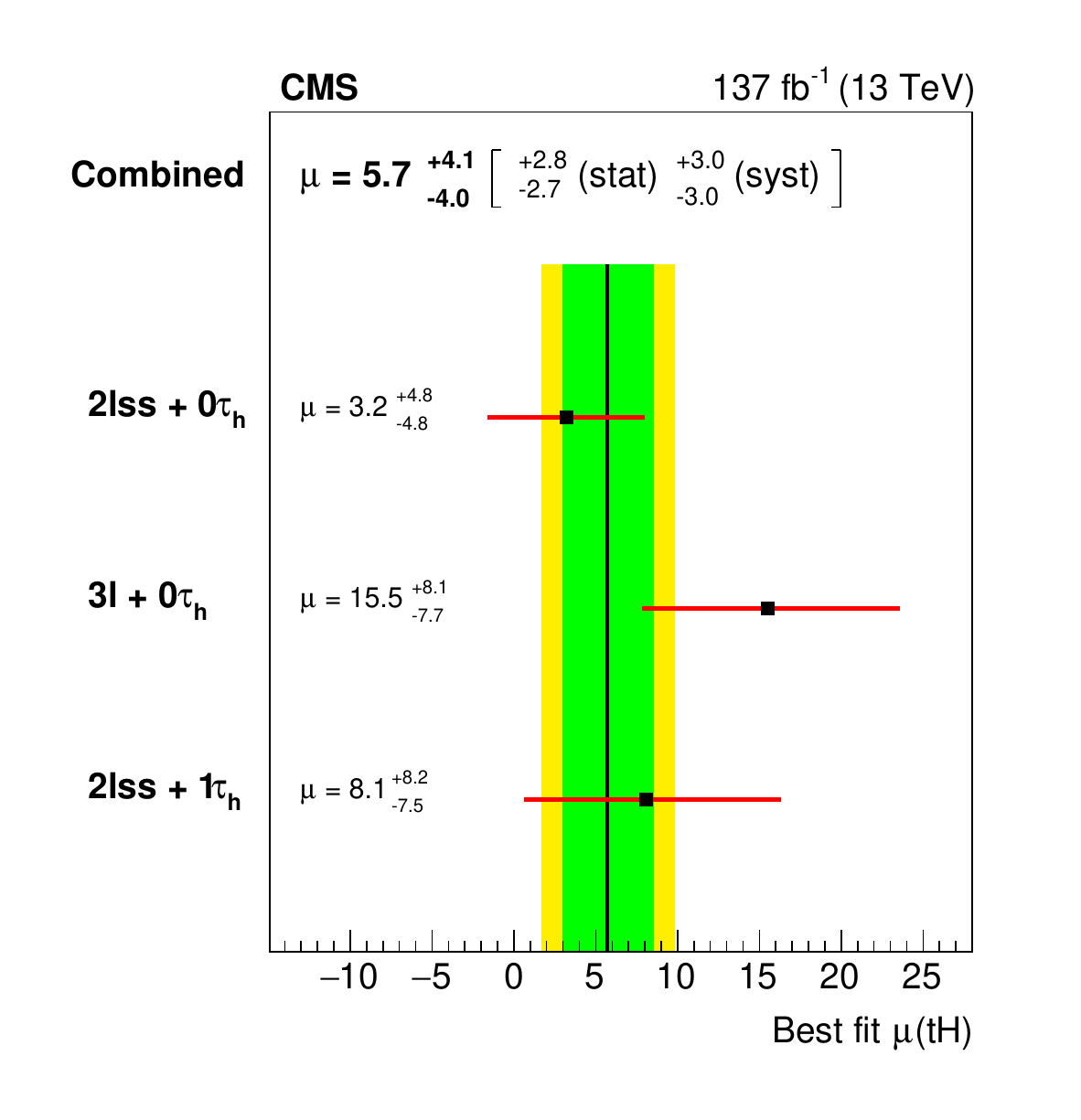}
  \caption{
    Production rate $\rhat_{\ttH}$ of the $\ttH$ signal (\cmsLeft) and $\rhat_{\tH}$ of $\tH$ signal (\cmsRight),
    in units of their rate of production expected in the SM,
    measured in each of the ten channels individually and for the combination of all channels. The central value of the signal
    strength in the \twoLeptonTwoTau is constrained to be greater than zero. 
  }
  \label{fig:signalRates}
\end{figure*}

The $\ttH$ signal rates measured in the ten individual channels are shown in Fig.~\ref{fig:signalRates}, obtained by performing a likelihood fit in which signal rates are parametrized with independent parameters, one for each channel.
The measurement of the $\tH$ production rate is only shown in the $\twoLeptonssZeroTau$, $\threeLeptonZeroTau$, and $\twoLeptonssOneTau$ channels,
which employ a multiclass ANN to separate the $\tH$ from the $\ttH$ signal.
The sensitivity of the other channels to the $\tH$ signal is small.
The $\ttH$ and $\tH$ production rates obtained from the simultaneous fit of all channels are also shown in the figure.
The signal rates measured in individual channels are compatible with each other and with the $\ttH$ and $\tH$ production rates 
obtained from the simultaneous fit of all channels.
The largest deviation from the SM expectation is observed in the $\ttH$ production
rate in the  $\twoLeptonTwoTau$ channel,
where the best fit value of the $\ttH$ signal rate is negative,
reflecting the deficit of observed events compared to the background expectation in this channel, as shown in Fig.~\ref{fig:postfitPlots4}.
The value and uncertainty shown in Fig.~\ref{fig:signalRates} are obtained after requiring   the $\ttH$ production rates in this channel to be positive.
The value measured in the $\twoLeptonTwoTau$ channel is compatible with the SM expectation at the level of $1.94$ standard deviations when constraining the signal strength in that channel to be larger than zero.
The sensitivity of individual channels can be inferred from the size of the uncertainty band in the measured signal strengths.
The channel providing the highest sensitivity is the $\twoLeptonssZeroTau$ channel, which is the channel providing the largest signal yield,
followed by the $\threeLeptonZeroTau$ and $\twoLeptonssOneTau$ channels.

\begin{figure*}[h!]
  \centering\includegraphics[width=0.49\textwidth]{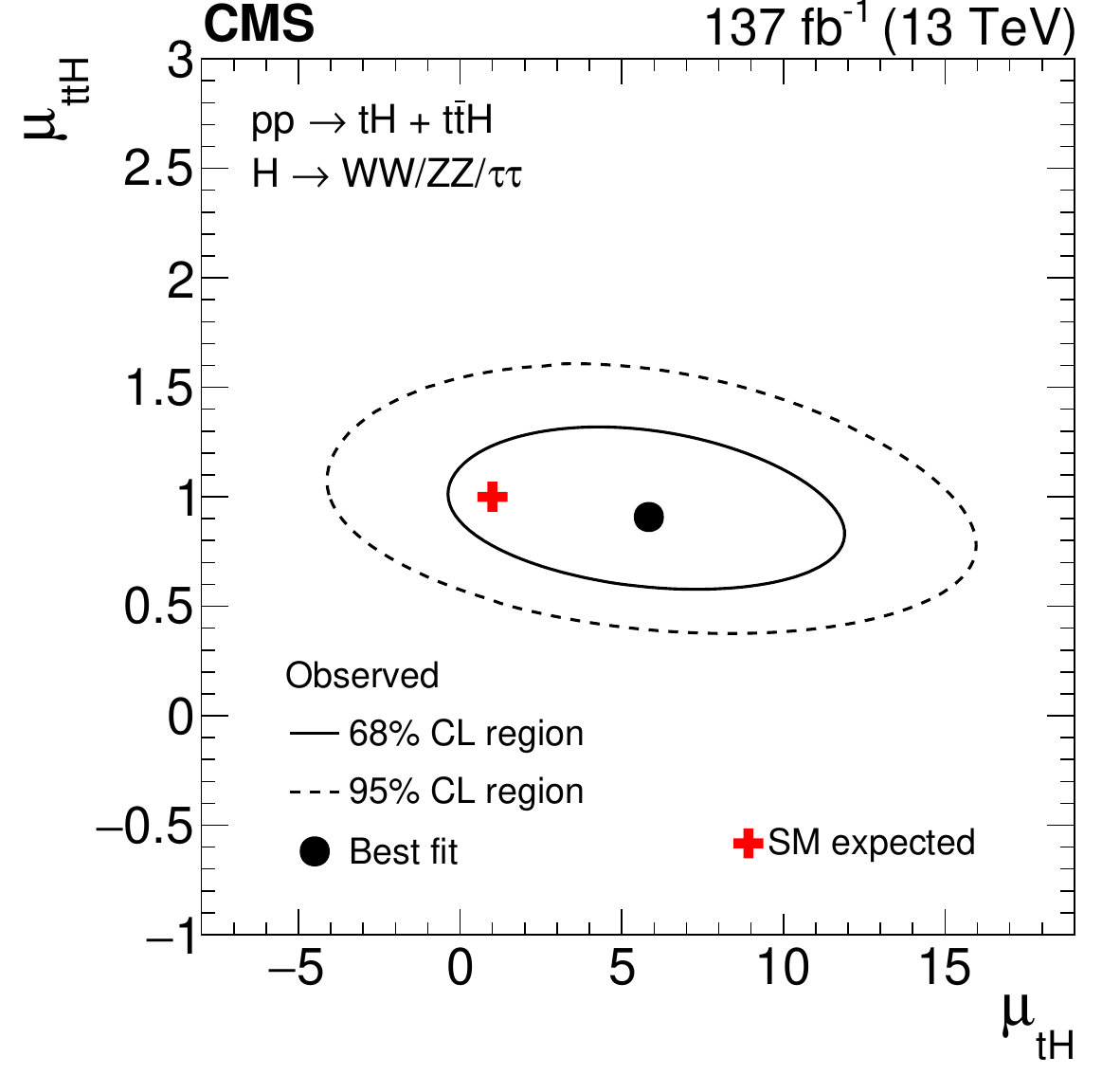}
  \centering\includegraphics[width=0.49\textwidth]{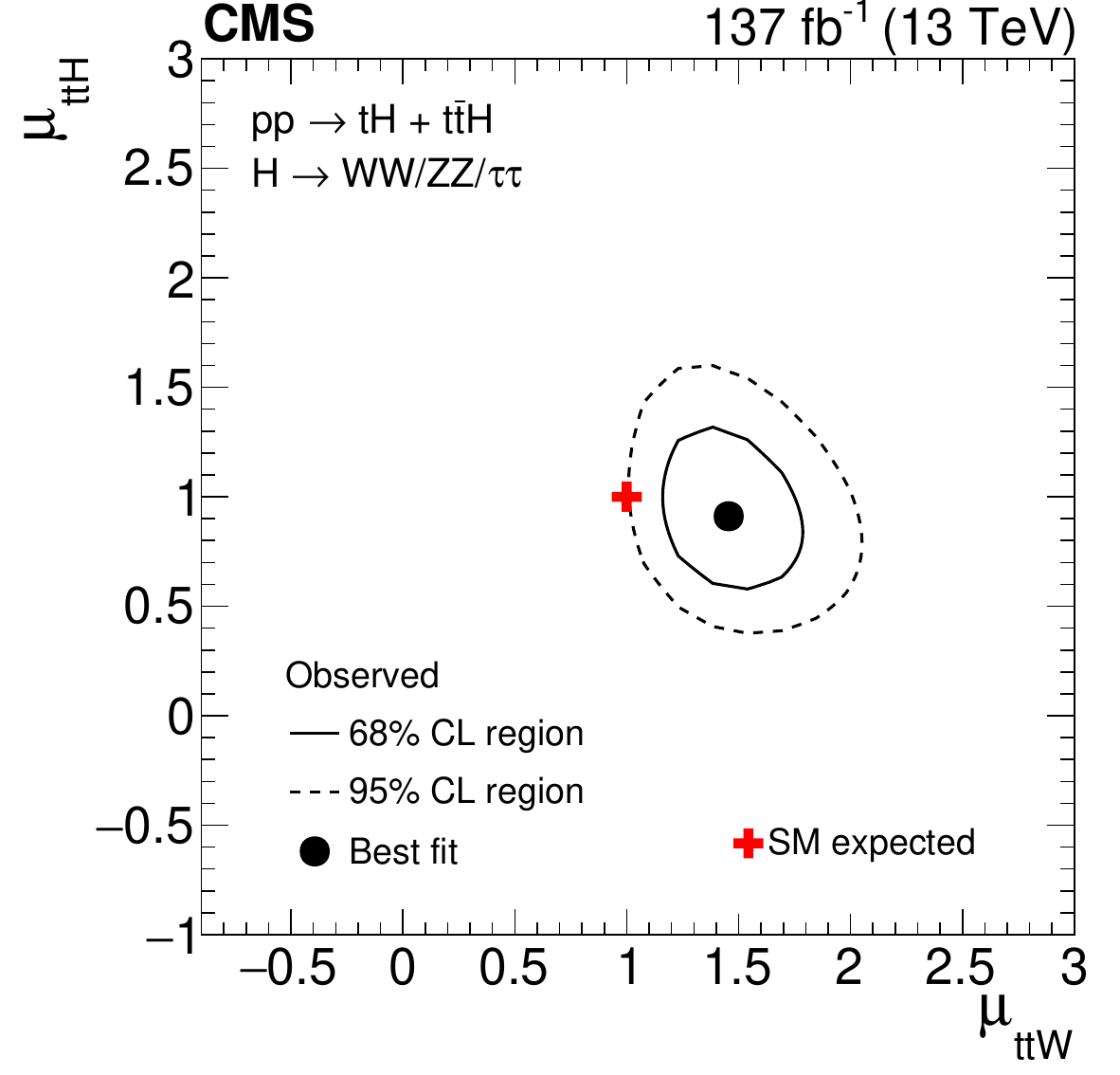}
  \centering\includegraphics[width=0.49\textwidth]{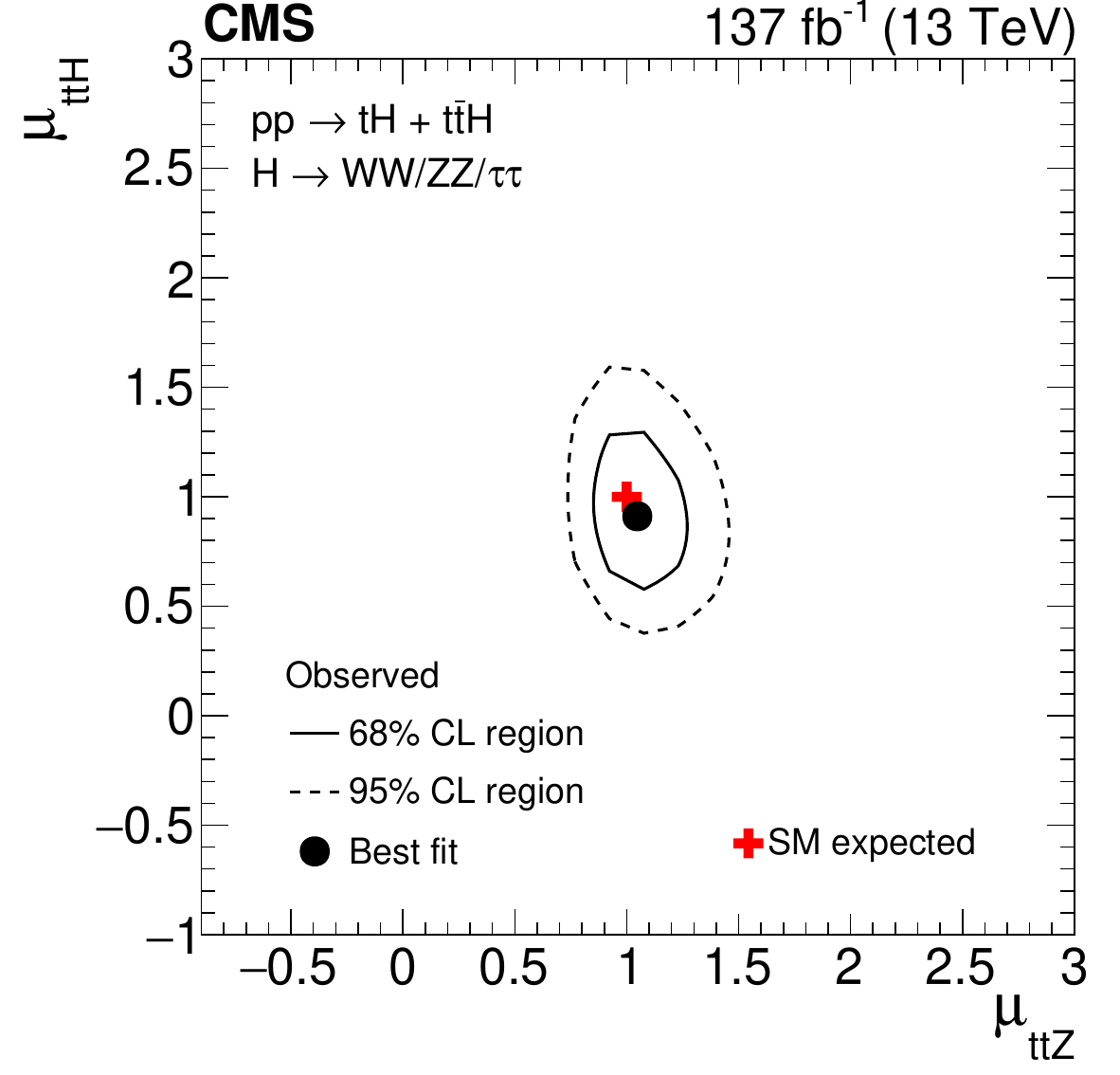}
  \centering\includegraphics[width=0.49\textwidth]{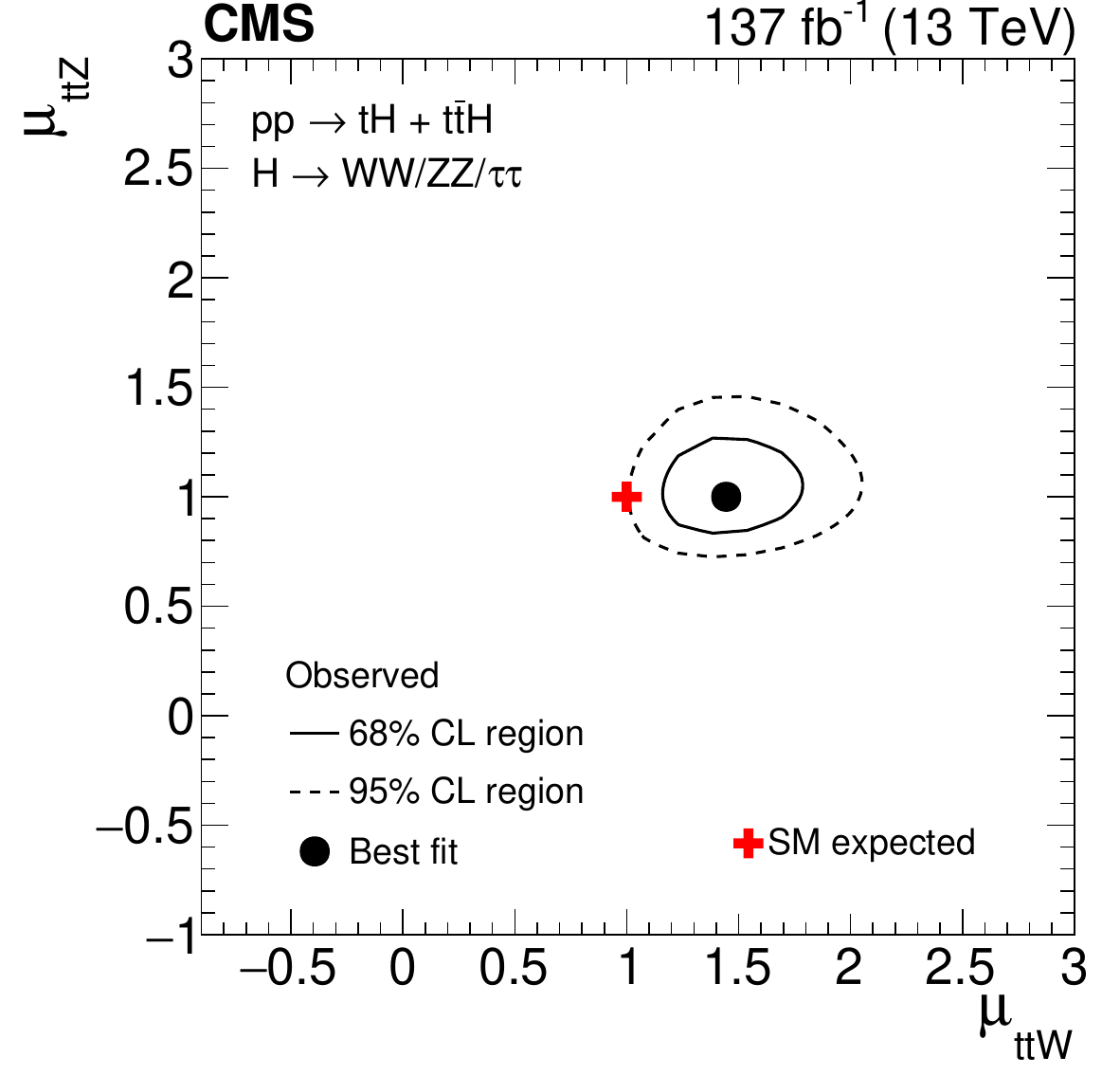}
  \caption{
    Two-dimensional contours of the likelihood function $\mathcal{L}$, given by Eq.~(\ref{eq:likelihoodFunction}),
    as a function of the production rates of the $\ttH$ and $\tH$ signals ($\r_{\ttH}$ and $\r_{\tH}$)
    and of the $\ttZ$ and $\ttW$ backgrounds ($\r_{\ttZ}$ and $\r_{\ttW}$).
    The two production rates that are not shown on either the $x$ or the $y$ axis
    are profiled such that the function $\mathcal{L}$ attains its minimum at each point in the $x$-$y$ plane.
  }
  \label{fig:signalRate_correlations}
\end{figure*}

Figure~\ref{fig:signalRate_correlations} shows the correlations between the measured $\ttH$ and $\tH$ signal rates 
and those between the signal rates and the production rates of the $\ttZ$ and $\ttW$ backgrounds.
All correlations are of moderate size, demonstrating the performance achieved by the multiclass ANN 
in distinguishing between the $\tH$ and $\ttH$ signals as well as in separating the $\ttH$ and $\tH$ signals from the $\ttZ$ and $\ttW$ backgrounds.

In the CA described in Section~\ref{sub:addchecks}, the measured production rate for the $\ttH$ signal is 
$\rhat_{\ttH} = 0.5 \pm 0.3\statsystlumi$, 
$\rhat_{\ttH} = 1.3 \pm 0.5\statsystlumi$, 
$\rhat_{\ttH} = 0.9 \pm 0.4\statsystlumi$, and
$\rhat_{\ttH} = 1.5 \pm 1.5\statsystlumi$ times the SM expectation,
in the $\twoLeptonssZeroTau$, $\threeLeptonZeroTau$, $\twoLeptonssOneTau$, and $\fourLeptonZeroTau$ channels, respectively,
while $\rhat_{\ttH} = 0.91 \pm 0.21\stat \pm 0.18\syst$ is obtained for the simultaneous ML fit of all four channels.
The $3\Plepton$- and $4\Plepton$-CRs are included in each of these ML fits.
The corresponding observed (expected) significance of the $\ttH$ signal in the CA amounts to $3.8$ ($4.0$) standard deviations.

\begin{figure*}[h!]
  \centering\includegraphics[width=0.49\textwidth]{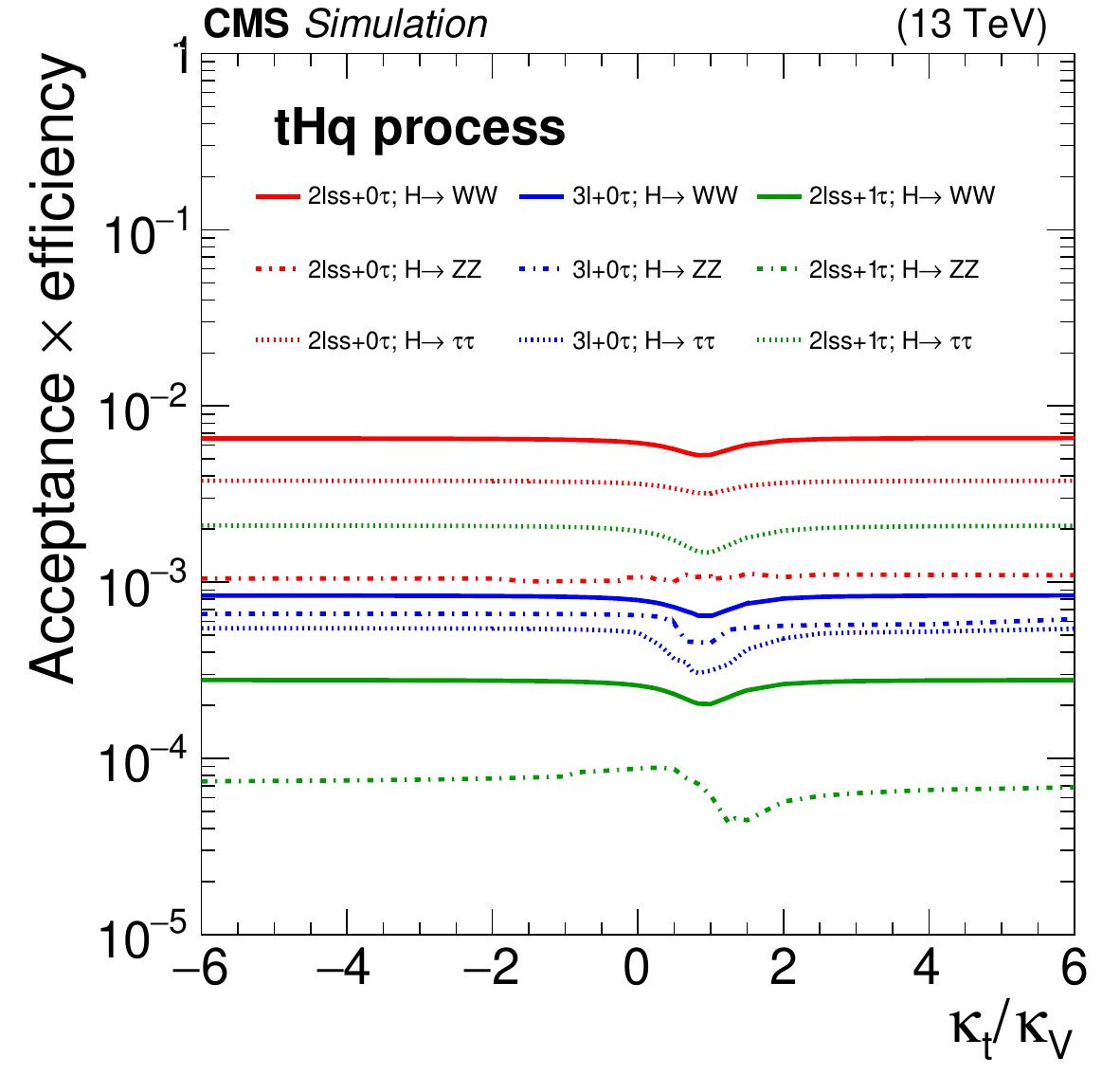}
  \centering\includegraphics[width=0.49\textwidth]{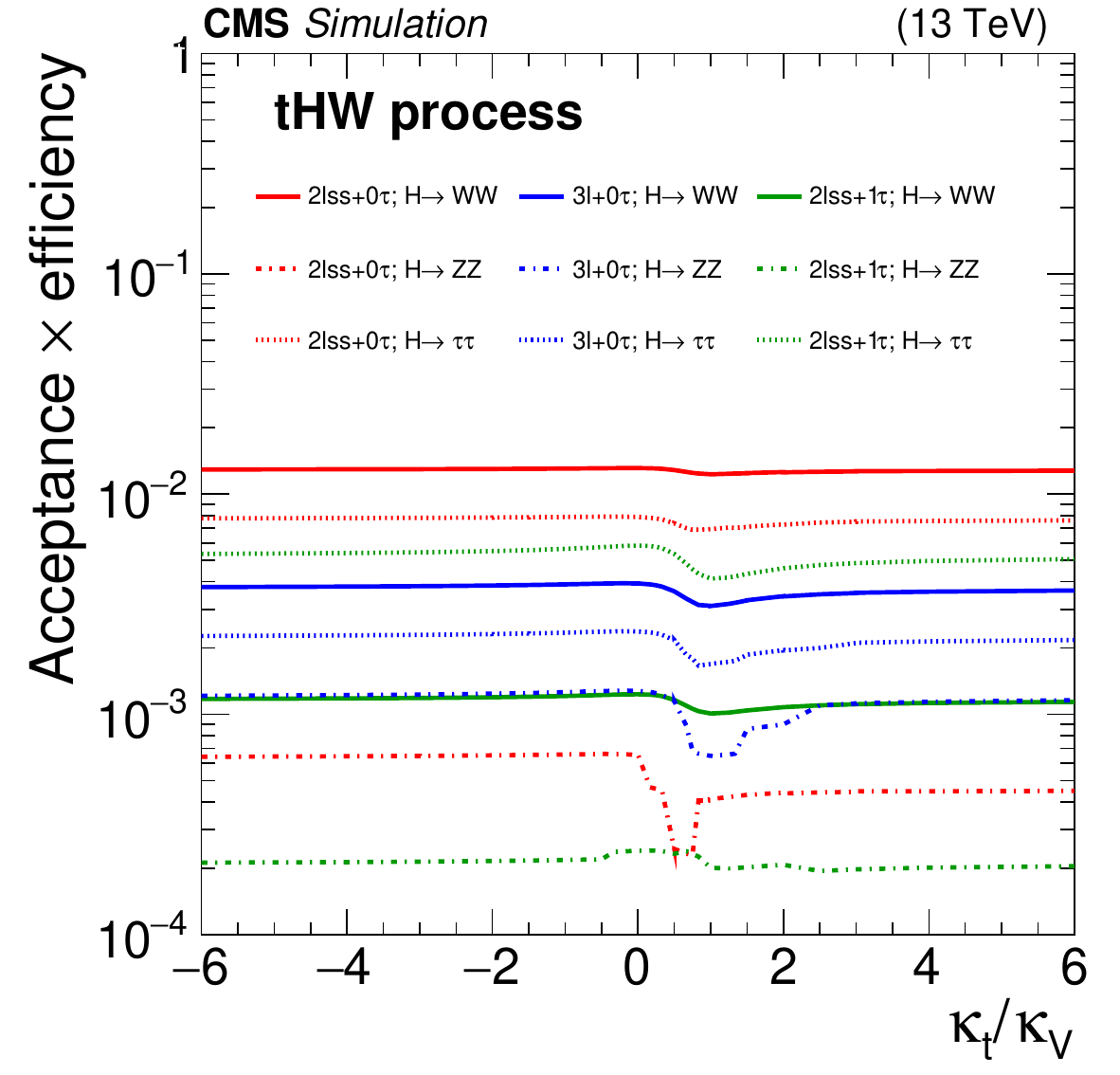}
  \caption{
    Probability for $\tH$ signal events produced by the $\tHq$ (\cmsLeft) and $\tHW$ (\cmsRight) production process
    to pass the event selection criteria for the $\twoLeptonssZeroTau$, $\threeLeptonZeroTau$, and $\twoLeptonssOneTau$ channels in each of the Higgs boson decay modes
    as a function of the ratio $\kappat/\kappaV$ of the Higgs boson couplings to the top quark and to the $\PW$ boson.
  }
  \label{fig:accTimesEff_tH}
\end{figure*}

\begin{figure*}[h!]
  \centering\includegraphics[width=0.49\textwidth]{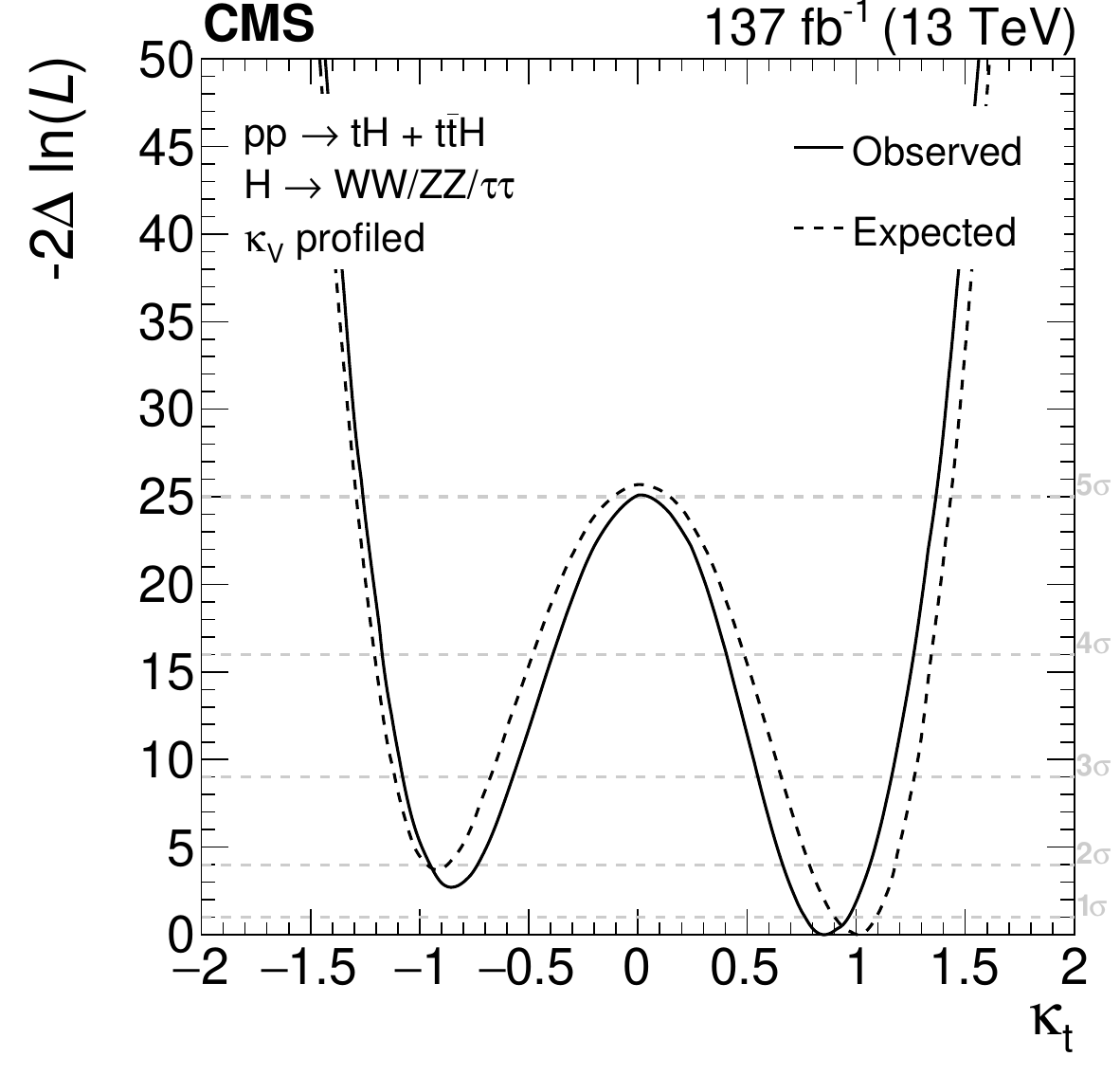}
  \centering\includegraphics[width=0.49\textwidth]{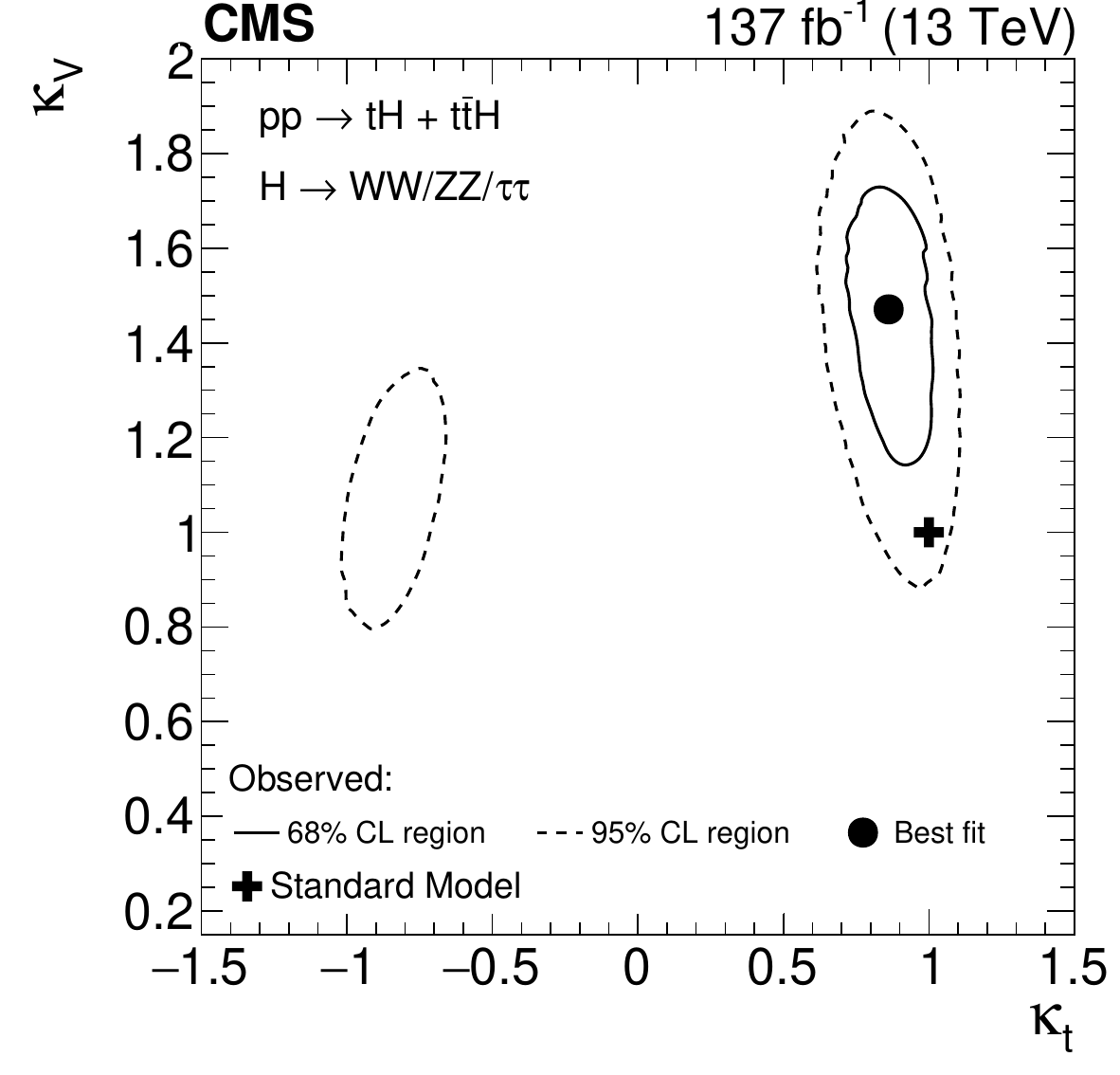}
  \caption{
    Dependence of the likelihood function $\mathcal{L}$ in Eq.~(\ref{eq:likelihoodFunction}),
    as a function of $\kappat$, profiling over $\kappaV$ (\cmsLeft), and as a function of $\kappat$ and $\kappaV$ (\cmsRight).
  }
  \label{fig:likelihoodFunction_kappat}
\end{figure*}

We now drop the assumption that the distributions of kinematic observables for the $\tH$ signal conform to the distributions expected in the SM
and determine the Yukawa coupling $\yt$ of the Higgs boson to the top quark.
We parametrize the production rates $\rhat_{\ttH}$ and $\rhat_{\tH}$ of the $\ttH$ and $\tH$ signals as a function of
the ratio of the top quark Yukawa coupling $\yt$ to its SM expectation $m_{\Ptop}/v$.
We refer to this ratio as the coupling modifier and denote it by the symbol $\kappat$.
The effect of the interference, described in Section~\ref{sec:introduction},
between the diagrams in Fig.~\ref{fig:FeynmanDiagrams_tH}
on the distributions of kinematic observables is parametrized as a function of $\kappat$ and fully taken into account,
adjusting the event yield for the $\tH$ signal as well as the distributions of the outputs of the BDTs and ANNs for each value of $\kappat$.
The changes in the kinematical properties of the event affect the probability for $\tH$ signal events to pass the event selection criteria.
The effect is illustrated in Fig.~\ref{fig:accTimesEff_tH}, 
which shows the variation of the product of acceptance and efficiency for the $\tHq$ and $\tHW$ signal contributions in each decay mode of the Higgs boson
as a function of the ratio $\kappat/\kappaV$,
where $\kappaV$ denotes the coupling of the Higgs boson to the $\PW$ boson
with respect to the SM expectation for this coupling.
The coupling of the Higgs boson to the $\PZ$ boson with respect to its SM expectation is assumed to scale by the same value $\kappaV$.
Variations of the coupling modifier $\kappaV$ from the SM expectation $\kappaV = 1$ affect the interference
between the diagrams in Fig.~\ref{fig:FeynmanDiagrams_tH} as well as
the branching fractions of the Higgs boson decay modes $\PHiggs \to \PW\PW$ and $\PHiggs \to \PZ\PZ$.
We compute the compatibility of the data with different values of $\kappat$ and $\kappaV$, as is shown in
Fig.~\ref{fig:likelihoodFunction_kappat}.
We obtain a 95\% confidence level (\CL) region on $\kappat$ consisting of the union of the two intervals $-0.9 < \kappat < -0.7$ and $0.7 < \kappat < 1.1$ at 95\% confidence level (\CL).
At 95\% \CL, both the inverted top coupling scenario and the SM expectation $\kappat = 1$ are in agreement with the data.

\section{Summary}
\label{sec:summary}

The rate for Higgs boson production in association with either one or two top quarks
has been measured in events containing multiple electrons, muons, and hadronically decaying tau leptons,
using data recorded by the CMS experiment in $\Pp\Pp$ collisions at $\sqrt{s} = 13\TeV$ in 2016, 2017, and 2018.
The analyzed data corresponds to an integrated luminosity of 137\fbinv. 
Ten different experimental signatures are considered in the analysis, differing by the multiplicity of electrons, muons, and hadronically decaying tau leptons,
and targeting events in which the Higgs boson decays via $\PHiggs \to \PW\PW$, $\PHiggs \to \Pgt\Pgt$, or $\PHiggs \to \PZ\PZ$,
whereas the top quark(s) decay either semi-leptonically or hadronically.
The measured production rates for the $\Ptop\APtop\PHiggs$ and $\Ptop\PHiggs$ signals
amount to $0.92 \pm 0.19 \stat ^{+0.17}_{-0.13}\syst$ and $5.7 \pm 2.7\stat \pm 3.0\syst$ times their respective standard model (SM) expectations.
The corresponding observed (expected) significance  amounts to $4.7$ ($5.2$) standard deviations for $\PQt\PAQt\PH$, and to $1.4$ ($0.3$) for $\PQt\PH$ production.
Assuming that the Higgs boson coupling to the tau lepton is equal in strength to the values expected in the SM,
the coupling $y_{\Ptop}$ of the Higgs boson to the top quark divided by its SM expectation, $\kappat=y_{\Ptop}/y_{\Ptop}^{\mathrm{SM}}$,  is constrained to be within $-0.9 < \kappat < -0.7$ or $0.7 < \kappat < 1.1$, at 95\% confidence level.
This result is the most sensitive measurement of the $\Ptop\APtop\PHiggs$ production rate to date.

\begin{acknowledgments}
  We congratulate our colleagues in the CERN accelerator departments for the excellent performance of the LHC and thank the technical and administrative staffs at CERN and at other CMS institutes for their contributions to the success of the CMS effort. In addition, we gratefully acknowledge the computing centers and personnel of the Worldwide LHC Computing Grid for delivering so effectively the computing infrastructure essential to our analyses. Finally, we acknowledge the enduring support for the construction and operation of the LHC and the CMS detector provided by the following funding agencies: BMBWF and FWF (Austria); FNRS and FWO (Belgium); CNPq, CAPES, FAPERJ, FAPERGS, and FAPESP (Brazil); MES (Bulgaria); CERN; CAS, MoST, and NSFC (China); COLCIENCIAS (Colombia); MSES and CSF (Croatia); RIF (Cyprus); SENESCYT (Ecuador); MoER, ERC PUT and ERDF (Estonia); Academy of Finland, MEC, and HIP (Finland); CEA and CNRS/IN2P3 (France); BMBF, DFG, and HGF (Germany); GSRT (Greece); NKFIA (Hungary); DAE and DST (India); IPM (Iran); SFI (Ireland); INFN (Italy); MSIP and NRF (Republic of Korea); MES (Latvia); LAS (Lithuania); MOE and UM (Malaysia); BUAP, CINVESTAV, CONACYT, LNS, SEP, and UASLP-FAI (Mexico); MOS (Montenegro); MBIE (New Zealand); PAEC (Pakistan); MSHE and NSC (Poland); FCT (Portugal); JINR (Dubna); MON, RosAtom, RAS, RFBR, and NRC KI (Russia); MESTD (Serbia); SEIDI, CPAN, PCTI, and FEDER (Spain); MOSTR (Sri Lanka); Swiss Funding Agencies (Switzerland); MST (Taipei); ThEPCenter, IPST, STAR, and NSTDA (Thailand); TUBITAK and TAEK (Turkey); NASU (Ukraine); STFC (United Kingdom); DOE and NSF (USA).
   
  \hyphenation{Rachada-pisek} Individuals have received support from the Marie-Curie program and the European Research Council and Horizon 2020 Grant, contract Nos.\ 675440, 724704, 752730, and 765710 (European Union); the Leventis Foundation; the A.P.\ Sloan Foundation; the Alexander von Humboldt Foundation; the Belgian Federal Science Policy Office; the Fonds pour la Formation \`a la Recherche dans l'Industrie et dans l'Agriculture (FRIA-Belgium); the Agentschap voor Innovatie door Wetenschap en Technologie (IWT-Belgium); the F.R.S.-FNRS and FWO (Belgium) under the ``Excellence of Science -- EOS" -- be.h project n.\ 30820817; the Beijing Municipal Science \& Technology Commission, No. Z191100007219010; the Ministry of Education, Youth and Sports (MEYS) of the Czech Republic; the Deutsche Forschungsgemeinschaft (DFG) under Germany's Excellence Strategy -- EXC 2121 ``Quantum Universe" -- 390833306; the Lend\"ulet (``Momentum") Program and the J\'anos Bolyai Research Scholarship of the Hungarian Academy of Sciences, the New National Excellence Program \'UNKP, the NKFIA research grants 123842, 123959, 124845, 124850, 125105, 128713, 128786, and 129058 (Hungary); the Council of Science and Industrial Research, India; the HOMING PLUS program of the Foundation for Polish Science, cofinanced from European Union, Regional Development Fund, the Mobility Plus program of the Ministry of Science and Higher Education, the National Science Center (Poland), contracts Harmonia 2014/14/M/ST2/00428, Opus 2014/13/B/ST2/02543, 2014/15/B/ST2/03998, and 2015/19/B/ST2/02861, Sonata-bis 2012/07/E/ST2/01406; the National Priorities Research Program by Qatar National Research Fund; the Ministry of Science and Higher Education, project no. 02.a03.21.0005 (Russia); the Tomsk Polytechnic University Competitiveness Enhancement Program; the Programa Estatal de Fomento de la Investigaci{\'o}n Cient{\'i}fica y T{\'e}cnica de Excelencia Mar\'{\i}a de Maeztu, grant MDM-2015-0509 and the Programa Severo Ochoa del Principado de Asturias; the Thalis and Aristeia programs cofinanced by EU-ESF and the Greek NSRF; the Rachadapisek Sompot Fund for Postdoctoral Fellowship, Chulalongkorn University and the Chulalongkorn Academic into Its 2nd Century Project Advancement Project (Thailand); the Kavli Foundation; the Nvidia Corporation; the SuperMicro Corporation; the Welch Foundation, contract C-1845; and the Weston Havens Foundation (USA).

\end{acknowledgments}

\bibliography{auto_generated}

\cleardoublepage \appendix\section{The CMS Collaboration \label{app:collab}}\begin{sloppypar}\hyphenpenalty=5000\widowpenalty=500\clubpenalty=5000\vskip\cmsinstskip
\textbf{Yerevan Physics Institute, Yerevan, Armenia}\\*[0pt]
A.M.~Sirunyan$^{\textrm{\dag}}$, A.~Tumasyan
\vskip\cmsinstskip
\textbf{Institut f\"{u}r Hochenergiephysik, Wien, Austria}\\*[0pt]
W.~Adam, T.~Bergauer, M.~Dragicevic, J.~Er\"{o}, A.~Escalante~Del~Valle, R.~Fr\"{u}hwirth\cmsAuthorMark{1}, M.~Jeitler\cmsAuthorMark{1}, N.~Krammer, L.~Lechner, D.~Liko, I.~Mikulec, F.M.~Pitters, N.~Rad, J.~Schieck\cmsAuthorMark{1}, R.~Sch\"{o}fbeck, M.~Spanring, S.~Templ, W.~Waltenberger, C.-E.~Wulz\cmsAuthorMark{1}, M.~Zarucki
% ...
\vskip\cmsinstskip
\dag: Deceased\\
1:  Also at Vienna University of Technology, Vienna, Austria\\
% ...
\end{sloppypar}
\end{document}